\documentclass [12pt]{article}
\usepackage[centertags]{amsmath}
\usepackage{amsfonts} \usepackage{amssymb}\usepackage{eufrak}
\usepackage{graphicx}
\usepackage{lscape}
\newcommand{\COMMENTO}[1]{}
\newcommand{\COMMENTOm}[1]{}

\newcommand{\II}[1]{{I(#1)}}
\newcommand{\uno}{\mathbb{I}}
\newcommand{\F}{\mathbb{F}}
\newcommand{\Z}{\mathbb{Z}}
\newcommand{\Q}{\mathbb{Q}}
\newcommand{\R}{\mathbb{R}}
\newcommand{\Co}{\mathbb{C}}
\newcommand{\C}{\mathbb{C}}
\newcommand{\Hp}{\mathbb{H}}
\newcommand{\sqa}{\sqrt{\alpha'}}
\newcommand{\alphant}{\alpha}
\newcommand{\alphat}{{ \tilde \alpha}}
\newcommand{\A}{\hat{\cal A}}
\newcommand{\B}{\hat{\cal B}}
\newcommand{\hcC}{\hat{\cal C}}
\newcommand{\D}{\hat{\cal D}}
\newcommand{\cA}{{\cal A}}
\newcommand{\cC}{{\cal C}}
\newcommand{\cE}{{\cal E}}
\newcommand{\cG}{{\cal G}}
\newcommand{\cN}{{\cal N}}
\newcommand{\cS}{{\cal S}}
\newcommand{\cW}{{\cal W}}
\newcommand{\cV}{{\cal V}}
\newcommand{\hF}{\hat{ F}}
\newcommand{\Lrpu}{{N_1}}
\newcommand{\Nu}{{\Lrpu}}

\newcommand{\Cu}{ {\cal C}_1}
\newcommand{\Cz}{ {\cal C}_0(E,\hF)}
\newcommand{\Nt}{\tilde {\cal N}_0(E)}
\newcommand{\Nop}{{\cal N}_0(E,\hF)}

\newcommand{\oh}{{\frac{1}{2}}}
\newcommand{\myexp}[1]{\exp \left(#1 \right) }

\begin{document}

\begin{titlepage}
\rightline{DFTTO-2009-67} \vskip 3.0cm
\centerline{\LARGE \bf  Open and Closed String Vertices    }
\vskip .5cm
\centerline{\LARGE \bf for branes with magnetic field and T-duality.}
\vskip .5cm
\vskip 1.0cm \centerline{\bf I. Pesando}
\vskip .6cm
\centerline{ \sl  Dipartimento di Fisica Teorica,
  Universit\`a di Torino and INFN, Sezione di Torino,} \centerline{\sl
 via P. Giuria 1,  I-10125, Torino, Italy}
 \vskip 1cm

\begin{abstract}
We discuss carefully the vertices which describe the dipole open strings and
closed strings on a D-brane with magnetic flux on a torus. 
Translation invariance along closed cycles 
forces surprisingly closed string vertices
written in open string formalism to acquire Chan-Paton like matrices.
Moreover the one loop amplitudes have a single trace for the part of gauge
group with the magnetic flux.
These peculiarities are also required by consistency of the action of
T-duality in the open string sector.
In this way we can show to all orders in perturbation theory the
equivalence of the T-dual open string theories, 
gravitational interactions included.

We provide also a new and direct derivation of the bosonic 
boundary state in presence of constant magnetic and Kalb-Ramond
background based on Sciuto-Della Selva-Saito vertex formalism.
\end{abstract}

\end{titlepage}

\section{Introduction and conclusions.}
In few years from now LHC will probably and hopefully 
have given a better picture of the physics beyond the standard model.
For example
it will probably be clear whether large extra dimensions are a feature
of Nature or not.
In the case, somewhat unexpected, that the string scale is around 1
TeV we could be able to see KK states and Regge resonances.
This makes worth studying the string interactions with KK momenta and
winding in the bottom-up approach.

In fact most of the current literature has focused in computing effective
actions for the light states which have not momentum and/or winding 
in the compact directions.
The first steps in a better understanding of the interactions among
string states with momentum and winding were taken in \cite{DiVecchia:2007dh}
where
both the open string and closed string vertices for branes with
equal magnetic field wrapped on a torus were given.
This description was up to cocycles.
The main result was that Chan-Paton factors do depend on momenta along
the directions where there is a non trivial magnetic field.
In the simplest case of a $T^2$ with coordinates $x^1,x^2$ and
adimensional magnetic field 
$2\pi \alpha' q F_{1 2}= \frac{f}{N} \uno_N \in u(N)$
this dependence is the same of a non trivial section 
of a $U(N)$ gauge bundle transforming in the adjoint.
For example the tachyonic vertex of a dipole string is given by
\begin{equation}
V_T(x; k)= :e^{i k\cdot X(x)}: ~\Lambda(k_1,k_2)
\label{intro-tach-vert}
\end{equation}
where both the compact momentum components $k_{1,2}$ have a spectrum given by 
$\frac{1}{\sqa} \frac{n_{1,2}}{N}$ ($n\in\Z$) and 
we have introduced the momentum dependent Chan-Paton matrices
\begin{eqnarray}
\Lambda_{N~I J}(\frac{1}{\sqa} \frac{n_1}{N},\frac{1}{\sqa}  \frac{n_2}{N})
=\Lambda_{N~I J}(k)
&=&
\frac{1}{\sqrt{N}} e^{-i \frac{\pi}{N} \hat h n_1 n_2}
\left( Q_N^{\hat h n_2} P_N^{-n_1}\right)_{I J}
,~~
0\le I,J < N
\nonumber\\
\end{eqnarray}
with $\hat h f \equiv -1 ~~ mod~N$ 
(we will be more precise on requirements later in eq. (\ref{fftilde}) )
and $P_{I J}\propto \delta_{I+1,J}$ and 
$Q_{I,J} \propto e^{i \frac{2\pi}{N} I} \delta_{I,J}$ are the 't Hooft
matrices.
In the same work \cite{DiVecchia:2007dh} the
transformations for the open string momenta, for the open string
metric and for other physical quantities under T-duality were discussed.

At the end of the discussion we were left with a puzzle on the true
equivalence under the full T-duality group 
of a bound state of $D2$ and $D0$ branes with a single
$D2$ branes.
In fact immediately after the discovery of the description of D-branes in
string theory it was realized that a bound state of $N$ $D2$ and $f$
$D0$ is T-dual to a single $D2$ when $gcd(N,f)=1$.
In particular in 1997 Guralnik and Ramgoolam in \cite{Guralnik:1997sy}
showed the previous equivalence using
the following chain of dualities
\begin{eqnarray}
bound(N~ D2 ~+~  f D0)
&\stackrel{T~duality~ on~y}{\longrightarrow}&
bound(N~ D1_x ~+~  f ~D1_y)
\nonumber\\
&=& G\left( bound(\frac{N}{G}~ D1_x ~+~ \frac{ f}{G} ~D1_y) \right)
\nonumber\\
&\stackrel{SL(2,\Z)~ coordinate ~rotation}{\longrightarrow}&
G ~D1_{x'}
\nonumber\\
&\stackrel{T~duality~ on~y'}{\longrightarrow}&
G ~D2
\label{Gura-Rango}
\end{eqnarray}
where $bound(\dots)$ means bound state made of $\dots$,
$G$ is equal to
$gcd(N, f)$ and ``$SL(2,\Z)$~ coordinate ~rotation'' means
that we perform a $SL(2,\Z)$ transformation on $T^2$ so that the $D1$ brane
wrapped $\frac{N}{G}$ times along $x$ and $\frac{ f}{G}$ times
along $y$ becomes a $D1$ brane wrapped once along a new $x'$ direction
and with ''perpendicular'' direction $y'$.

In \cite{DiVecchia:2007dh} we started from a tachyonic vertex for a 
single brane ($N=1$) wrapped on the $T^2$ 
\begin{equation}
V_T(x; k^t)= :e^{i k^t\cdot X(x)}: 
\label{intro-tach-vert-u1}
\end{equation}
with $k^t_{1,2}=\frac{1}{\sqa} n_{1,2}$ ($n_{1,2}\in\Z$).
Studying the action of T-duality it was possible to recover the same vertex as
in eq. (\ref{intro-tach-vert}) up to the Chan-Paton matrices, i.e. 
we found as expected that the T-duality mapped the momenta spectrum correctly.
The presence or absence of the momentum dependent Chan-Paton matrices
is not a problem as long as tree level and planar amplitudes are 
considered since the tree level contribution of these momentum
dependent Chan-Paton matrices is given by
\begin{eqnarray}
tr\left( \Lambda(k_{(1)})\dots \Lambda(k_{(M)})\right)
\propto 
\delta^{[1]}_{\sqa \sum_{o=1}^M k_{(o) 1}}
\delta^{[1]}_{\sqa \sum_{o=1}^M k_{(o) 2}}
\end{eqnarray}
where $\delta^{[M]}_{x}$ means $x \equiv 0 ~~mod~M$
and therefore these deltas are automatically satisfied upon the use of the
momentum conservation.

The problems arise when we consider mixed open-closed amplitudes and,
equivalently, non planar amplitudes.
\begin{figure}[!t]
  \begin{minipage}[t]{0.45\linewidth}{
    \includegraphics[width=0.9\textwidth,angle=-90]{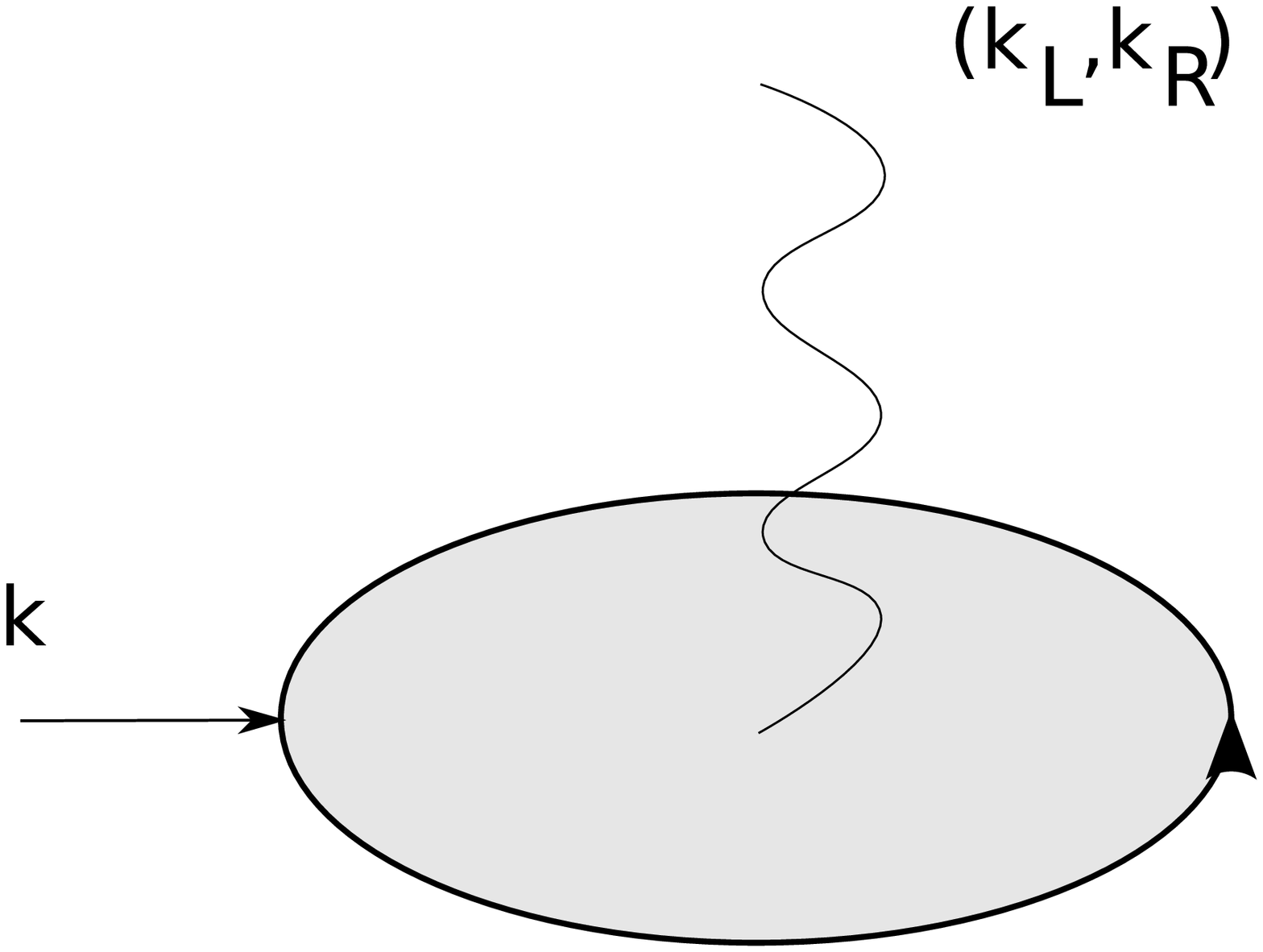}
}
\caption{The simplest mixed amplitude: one open string and one closed
  string.}
\label{figure:1open-1closed}
\end{minipage}
\hskip 1cm
  \begin{minipage}[t]{0.45\linewidth}{
    \includegraphics[width=0.9\textwidth,angle=-90]{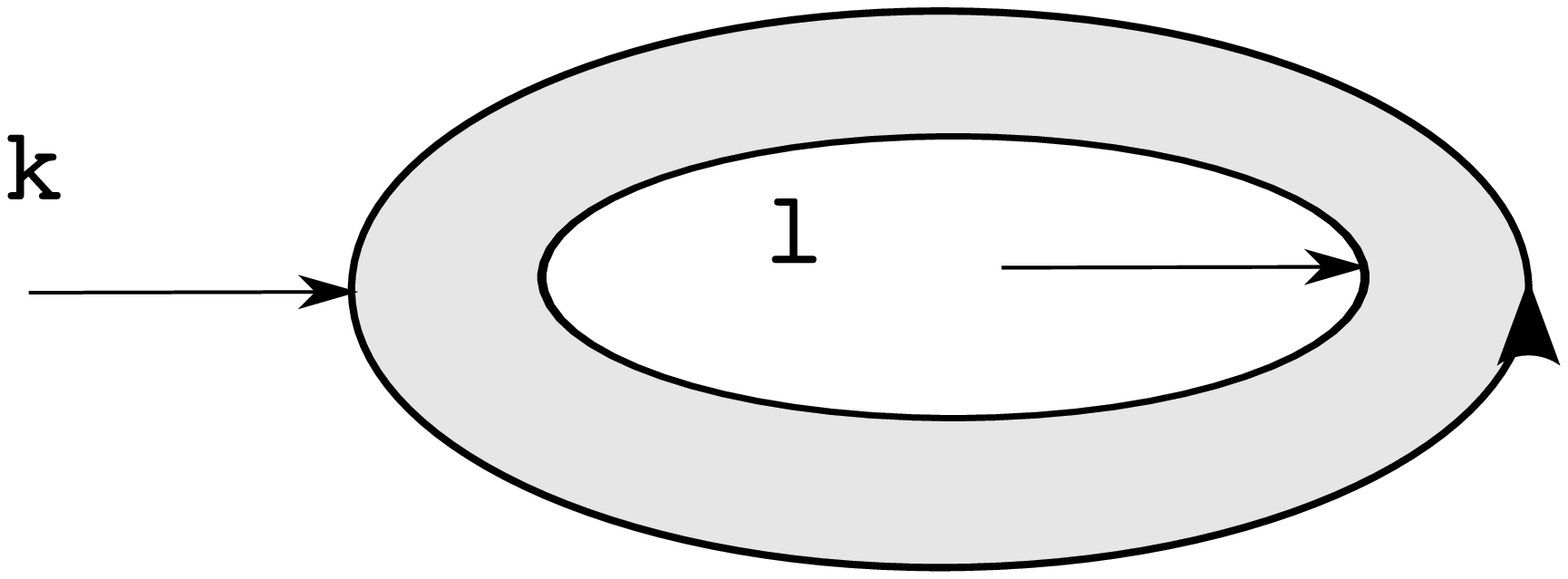}
}
\caption{The simplest non planar amplitude: one open string on one
  border and one open string on the other border of the annulus.}
\label{figure:1loop-nonplanar}
\end{minipage}

\end{figure}

Consider for example the one open string tachyon - one closed string
tachyon amplitude in fig. \ref{figure:1open-1closed}.
In the computation with the bundle we expect the amplitude to be
\begin{eqnarray}
A_{bundle}
\sim
tr\left(\Lambda(k)\right) ~\delta_{n+n_c-\hat F m_c,0}
\sim
~\delta^{[N]}_{n_{ 1}} \delta^{[N]}_{n_{ 2}}
~
\delta_{n_1+n_{c~1}-\frac{f}{N} m_{c~2},0} 
\delta_{n_2+n_{c~2}+\frac{f}{N} m_{c~1},0} 
\label{1open-1closed-bundle-naive}
\end{eqnarray}
where  $n_c$ and $m_c$ are respectively the closed string momentum and winding.
We have explicitly written the non trivial contribution from the 
Chan-Paton factor 
$tr\left(\Lambda(k)\right)\propto 
\delta^{[1]}_{\sqa k_{ 1}} \delta^{[1]}_{\sqa k_{ 2}}
=
\delta^{[N]}_{n_{ 1}} \delta^{[N]}_{n_{ 2}}
$
and the contribution from zero modes in the $T^2$ directions
$\delta_{n+n_c-\hat F m_c,0}$.

On the other side in the computation of the same diagram with a single
brane without magnetic field we expect 
\begin{eqnarray}
A^t_{single~with~\hat F^t=0}
\sim
\delta_{n^t+n^t_c,0}
\end{eqnarray}
which becomes after the T-duality
\begin{eqnarray}
A_{T-dual~of~a~single~with~\hat F = \frac{f}{N}}
\sim
~\delta_{n+n_c-\hat F m_c,0}
\sim
\delta_{n_1+n_{c~1}-\frac{f}{N} m_{c~2},0} 
\delta_{n_2+n_{c~2}+\frac{f}{N} m_{c~1},0} 
\label{1open-1closed}
\end{eqnarray}
which is the same amplitude as in
eq. (\ref{1open-1closed-bundle-naive}) {\sl without} the constraint from the
Chan-Paton factor. 
This happens since T-duality is nothing but
rewriting an amplitude using different variables.
Therefore the two amplitudes in eq.s (\ref{1open-1closed-bundle-naive})
and (\ref{1open-1closed}) are {\sl not} the same and it seems we have
two different branes: those obtained by T-duality and the magnetized ones.

A similar result holds for the non planar amplitude depicted in
fig. \ref{figure:1loop-nonplanar} where the bundle computation is
expected to give
\begin{eqnarray}
A_{bundle}
\sim
tr\left(\Lambda(k)\right) ~
tr\left(\Lambda(l)\right) ~
~\delta_{k+l,0}
\sim
~\delta^{[N]}_{n_{ 1}} \delta^{[N]}_{n_{ 2}}
~\delta^{[N]}_{m_{ 1}} \delta^{[N]}_{m_{ 2}}
~\delta_{n_1+m_1,0} 
\delta_{n_2+m_2,0} 
\label{1loop-nonplanar-bundle-naive}
\end{eqnarray}
where we have written only the Chan-Paton and zero modes contributions
from the directions where there is a non trivial bundle and we have
defined the integers $m$ as in $l= \frac{1}{\sqa}\frac{m}{N}$.
The T-dual transformation of the single brane one loop non planar
amplitude is
\begin{eqnarray}
A_{T-dual~of~a~single~with~\hat F = \frac{f}{N}}
\sim
~\delta_{k+l,0}
\label{1loop-nonplanar}
\end{eqnarray}
again missing the Chan-Paton contribution.

\begin{figure}[!t]
  \begin{minipage}[t]{0.45\linewidth}{
    \includegraphics[width=0.9\textwidth,angle=-90]{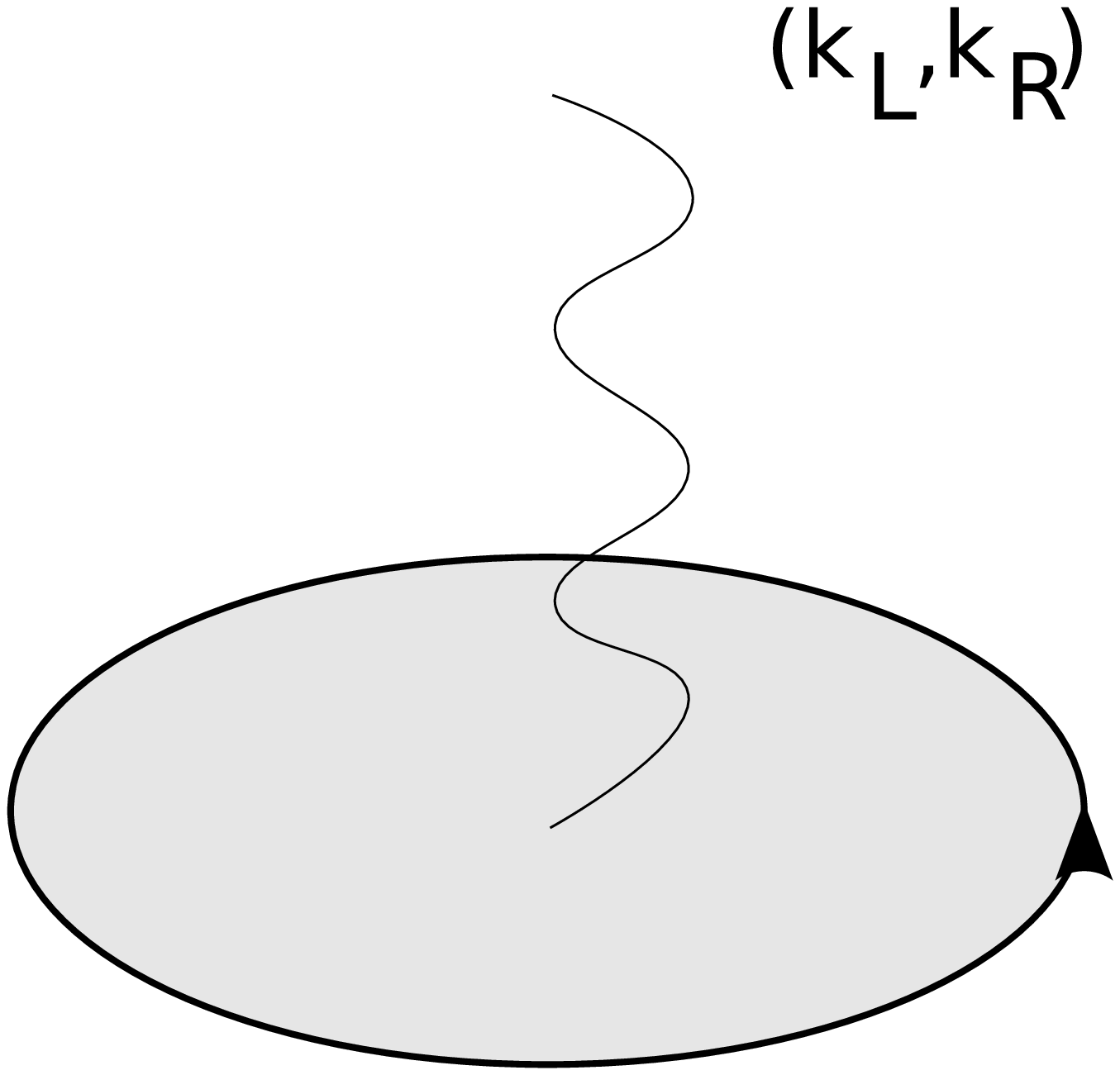}
}
\caption{The amplitude of 1 closed string on a disk, i.e. the boundary
  state.}
\label{figure:boundary}
\end{minipage}
\hskip 1cm
  \begin{minipage}[t]{0.45\linewidth}{
    \includegraphics[width=0.9\textwidth,angle=-90]{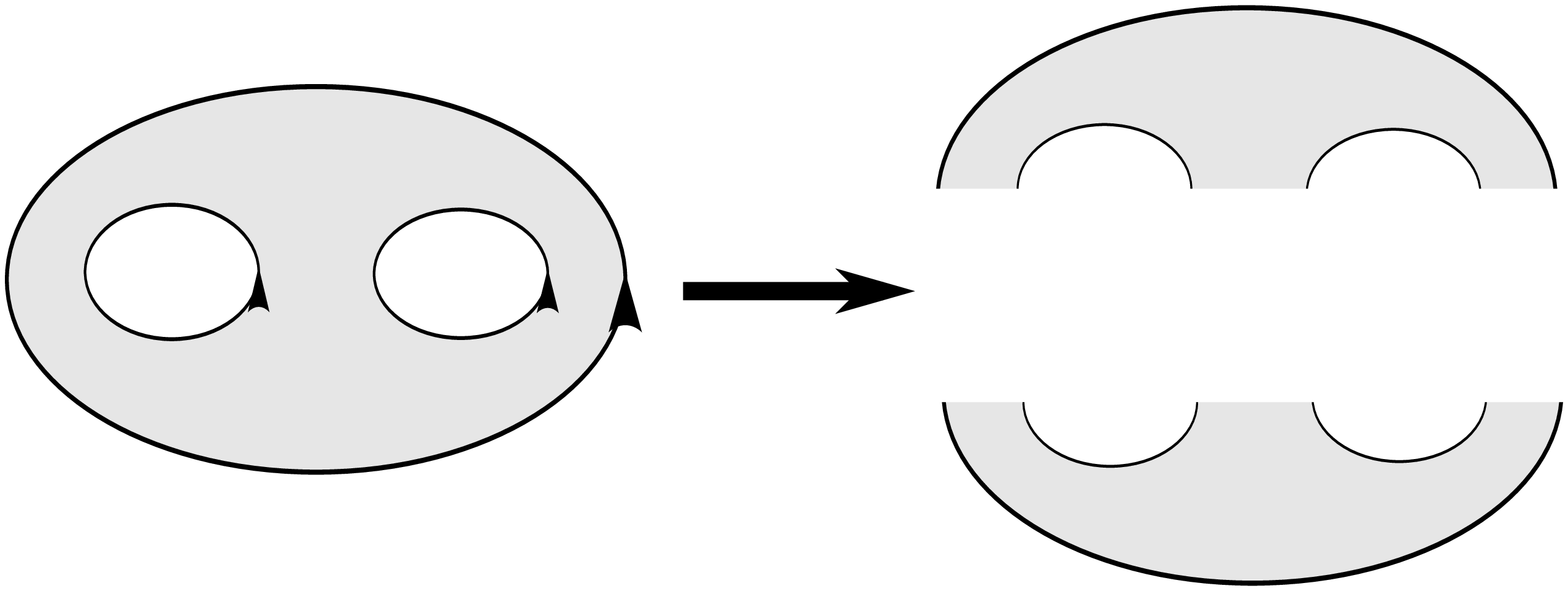}
}
\caption{Two vacuum loop factorizes into two three points tree amplitudes.}
\label{figure:2loops-factor}
\end{minipage}
\end{figure}

On the other side in (\cite{DiVecchia:2007dh},\cite{Duo:2007he}) the
boundary state for the bundle in presence of magnetic field, see
fig. \ref{figure:boundary}, was
computed directly and shown to be equal to the T-dual of the boundary
state of the single brane.
As a consequence the 2 loop vacuum amplitude is equal in both theories
since it can be computed in the closed string channel by using
boundary states.
This two loop vacuum amplitude can then be factorized 
in two three point tree amplitudes as in
fig. \ref{figure:2loops-factor} (\cite{Duo:2007he}) but this
factorization seems unique and therefore independent of whether we
compute them using the bundle picture of using the T-duality.

The aim of this article is to solve the previous puzzle.
The results are somewhat at variance with the naive expectations
derived from the common lore.
Both the computations with the bundle in
eq. (\ref{1open-1closed-bundle-naive})
and in eq. (\ref{1loop-nonplanar-bundle-naive}) are {\sl wrong}!

The 1 open string - 1 closed string in
eq. (\ref{1open-1closed-bundle-naive}) is wrong because {\sl the closed
string vertex  (in open string formalism) has a Chan-Paton like factor} 
proportional to 
$\Lambda\left( -\frac{\hat F m}{\sqa}\right)$ so that the proper
computation is
\begin{eqnarray}
A_{bundle}
&\sim&
tr\left(\Lambda(k) ~\Lambda\left( -\frac{\hat F m}{\sqa}\right) \right) 
~\delta_{n+n_c-\hat F m_c,0}
\nonumber\\
&\sim&
\delta_{n_1+n_{c~1}-\frac{f}{N} m_{c~2},0} 
\delta_{n_2+n_{c~2}+\frac{f}{N} m_{c~1},0} 
~\delta^{[N]}_{n_{ 1}-f m_{c~1}} \delta^{[N]}_{n_{ 2}+f m_{c~1}}
\nonumber\\
&\sim&
\delta_{n_1+n_{c~1}-\frac{f}{N} m_{c~2},0} 
\delta_{n_2+n_{c~2}+\frac{f}{N} m_{c~1},0} 
\label{1open-1closed-bundle}
\end{eqnarray}
which reproduces the T-dual result of the single brane 
since the Chan-Paton contribution is now compatible with momentum
conservation. 
Notice that there is no ambiguity in the ordering of the
open and the closed vertices in the amplitude with the proper cocycles
since then the open and closed vertices commute.

The 1 loop non planar amplitude in
eq. (\ref{1loop-nonplanar-bundle-naive}) is wrong too because there is a
{\sl single trace}, so that the right computation is roughly
\begin{eqnarray}
A_{bundle}
&\sim&
tr\left[\Lambda(k)~\Lambda(l)\right] ~
~\delta_{k+l,0}
\sim
\delta_{n_1+m_1,0} 
\delta_{n_2+m_2,0} 
~\delta^{[N]}_{n_{ 1}+m_{ 1}} \delta^{[N]}_{n_{ 2}+m_{ 2}}
\nonumber\\
&\sim&
\delta_{n_1+m_1,0} 
\delta_{n_2+m_2,0} 
\label{1loop-nonplanar-bundle}
\end{eqnarray}
and again the Chan-Paton contribution is compatible with momentum
conservation.
Notice however that the single trace is only for the part of gauge
group where the magnetic field  is switched on. 
If we consider a gauge group $U(NM)$ where we have a magnetic field 
embedded in a $U(N)$ subgroup the Chan-Paton has the factorized form 
$\Lambda(k)\otimes \Lambda_0$ where $\Lambda_0$  is the usual
Chan-Paton matrix then in this case the contribution is 
$ tr\left[\Lambda(k)~\Lambda(l)\right] ~ tr[\Lambda_0^{outer}] 
~tr[\Lambda_0^{inner}]$

Having the proper prescription to compute the amplitudes with dipole
strings allows to show that these amplitudes are actually the T-dual
of a system with branes without magnetic field turned on as suggested
but not proven by the chain of dualities in eq. (\ref{Gura-Rango}).

The plain of this paper is the following.
In section \ref{sect:review}
we review our conventions for open and closed string and
how we define the physical states on a bundle.
In section \ref{sect:SpectralDec}
we discuss the spectral decomposition of unity and its consequence for
the 1 loop non planar amplitude.
Then in section \ref{sect:ClosedCP}
we discuss how the request of translational invariance of
the closed string states along a cycle implies the presence of Chan-Paton like
factors for the closed string vertices in open
string formalism and we discuss how this is nevertheless true for
other simpler cases even if gone unnoticed.
To confirm the previous results 
in section \ref{sect:Cocycles} we discuss the closed string cocycles
and the open string ones and we show that everything works nicely 
when we consider the operatorial product of vertices .
To further confirm the results 
in \ref{sect:Amplitudes} we compute various amplitudes such as the
mixed $N$ open tachyons $1$ closed tachyon amplitude which we use 
to check the factorization of the 1 loop non planar in the
closed string channel. In the same section we propose a new direct
derivation of the boundary state in presence of constant magnetic 
and Kalb-Ramond background based on Sciuto-Della Selva-Saito vertex formalism.
In section \ref{sect:T-duality} we show that the amplitudes
for a dipole in presence of constant magnetic and Kalb-Ramond
background are T-dual to the amplitudes
for a dipole with a vanishing constant magnetic.
Finally in the appendices we give the details on the computations of
cocycles and of the amplitudes.


\section{A short review of closed string on torus 
and open string on non trivial bundles.}
\label{sect:review}

\subsection{Closed string conventions.}
In the following we use the notations used in
(\cite{DiVecchia:2007dh}, \cite{Di Vecchia:2006gg}).
In particular the closed string  expansion in a metric background
$E_{i j}=G_{i j}+B_{i j}$ on a generic flat space $R^{D-d}\otimes T^d$ 
 ($D=26$) is given by
\begin{eqnarray}
X^{ i}(z,\bar z) = 
\frac{1}{2} \left( \tilde X^{(c) i}_{R} (\bar z )
+ {X}^{(c) i}_{L} (z ) \right), ~~~~
\end{eqnarray}
where
\begin{eqnarray}
\tilde X^{(c) i}_{R} ( \bar z )
&=&
 x^{ i}_{R}
- 2 \alpha'  p^i_{ R} ~i~\ln(\bar z)
+
i \sqrt{2 \alpha' } \sum_{n \neq 0} \frac{1}{n}
\alphat_{n}^{i} {\bar z}^{-n}
\nonumber\\
{X}^{(c) i}_{L} (z )
&=&  x^{i}_{L}
-2\alpha'  p^i_{ L}~i~\ln(z)
+
i \sqrt{2 \alpha'} \sum_{n \neq 0}
\frac{1}{n}\alphant_{n}^{i}  z^{-n}
\end{eqnarray}
and 
${}^{(c)}$ has been added to stress that these fields are the
closed string ones,
$z=e^{2(\tau_E+i\sigma)}\in \Co$ ($0\le \sigma \le \pi$) 
and $i,j,\dots=1,2,\dots d$.
For non compact directions $\mu,\nu=0,d+1,\dots D$ we have the same
expansions with the identification $p^\mu_R=p^\mu_L= \frac{1}{2} p^\mu$.

The closed string  Hamiltonian can then be written as follows:
\begin{eqnarray}
\frac{H_c-4}{2} 
&=&
 L_0 + \tilde L_0=
 \frac{\alpha'}{4 \pi} \int_{0}^{\pi} d \sigma \left[P_{L}^{2}
  + P_{R}^{2} \right],
\nonumber\\
&=& 
N + {\tilde{N}} 
+ \frac{1}{2}
\left[G_{i j} {m}^i {m}^j
  + ( {n}_i - B_{i k} {m}^{k} ) G^{i j}
( {n}_j - B_{j h} {m}^{h} )  \right]
+ \frac{\alpha'}{2} G^{\mu\nu} k_\mu k_\nu
\nonumber\\
\label{hamilto89}
\end{eqnarray}
where
the explicit expressions of $L_0$ and ${\tilde{L}}_0$ are given by
\begin{eqnarray}
&&L_0 = 
\alpha'  G^{\mu\nu} \frac{k_\mu}{2} \frac{k_\nu}{2}
+\alpha' G_{i j} p_{R}^i  p_{R}^j+N
~~;~~ 
N=\sum_{n=1}^{\infty} 
G_{\mu\nu} \alpha_{-n}^{\mu} \alpha_{n}^{\nu}
+G_{i j} \alpha_{-n}^{i} \alpha_{n}^{j}
\nonumber\\
&&{\tilde{L}}_0 = 
\frac{\alpha'}{4} G^{\mu\nu} k_\mu k_\nu
+\alpha' G_{i j} p_{L}^i  p_{L}^j +
\tilde N
~~;~~ 
\tilde N=
\sum_{n=1}^{\infty} 
G_{\mu\nu} {\tilde{\alpha}}_{-n}^{\mu} {\tilde{\alpha}}_{n}^{\nu}
+G_{i j} {\tilde{\alpha}}_{-n}^{i } {\tilde{\alpha}}_{n}^{j }
,\label{L078}
\end{eqnarray}
The spectrum of the compact momenta  given by
\begin{eqnarray}
\left\{\begin{array}{c}
k_{L i} = G_{i j}~p^{j}_L 
= \frac{1}{2 \sqrt{\alpha'}} \left( n + E^T m\right)_i\\
k_{R i}= G_{i j}~p^{j}_R 
=\frac{1}{2 \sqrt{\alpha'}} \left( n - E m\right)_i
\end{array}
\right.
\label{pLpR89}
\end{eqnarray}
In computing amplitudes and OPEs we make use of the following non vanishing
commutators 
\begin{equation}
[x_L^i, p_L^j]=i ~G^{i,j}
~~,~~
 [ \alpha_{L n}^{i} , \alpha_{L m}^{j}] = n ~\delta_{n+m,0} ~G^{i,j}
\end{equation}
and similarly for the right movers and for the non compact directions.
The normalization of the zero modes is taken to be
\begin{equation}
\langle k_\mu, n_i,m^i | k_\mu', n_i',m^{i'} \rangle= 
(2\pi)^{D-d} \delta^{D-d}(k_\mu-k_\mu')~
(2\pi\sqa)^d \delta_{n,n'} \delta_{m,m'}
~.
\end{equation}

The vertex for a closed string state described by momentum
$(k_L,k_R)$ and quantum numbers $(\beta_L,\beta_R)$ is written as
\begin{eqnarray}
{W}^{(c)}_{\beta_L,\beta_R}(z,\bar z; k_L,k_R)
&=&
V_{\beta_L}(z;k_L)
\,\, {\tilde V}_{\beta_R}({\bar z};k_R)
\end{eqnarray}
up to cocycles which we discuss later in section \ref{clos-stri-coc-sub}
and where the right moving part of the vertex ${\tilde V}_{\beta_R}$ 
is a normal ordered functional of the closed string right moving part
$X_R^{(c)}$ and similarly for the left moving part.
The previous description is  however not completely exact for the non
compact directions since in  this case  the zero modes are common to
left and right moving sectors.
Because of this the non compact left and right part of the vertex  are not
separately normal ordered but
only the non zero modes parts are normal ordered separately while
the common zero modes are normal ordered together.

Finally for later use we write our conventions for an element of the 
T-duality group $\Lambda$ acting on the $d$ compact direction which 
is described by
\begin{eqnarray}
\Lambda &=& \left( \begin{array}{cc} \A & \B \\ \hcC & \D
\end{array}\right) \in O(d,d,\Z)
~~~~
\Lambda 
\left(\begin{array}{c c} 0 & \uno_d \\ \uno_d & 0 \end{array} \right)
\Lambda^T
=
\left(\begin{array}{c c} 0 & \uno_d \\ \uno_d & 0 \end{array} \right)
\label{lam}
\end{eqnarray}
and acts on  zero modes and operators as
\begin{equation}
\left\{\begin{array}{cc}
m^t = & \A m + \B n \\
n^t = & \hcC m + \D n \\
\end{array}
\right. 
~~~~
\left\{\begin{array}{cc}
\alpha_n^t= & \left( \A + \B E \right) \alpha_n \\
\tilde \alpha_n^t= & \left( \A - \B E^T \right) \tilde \alpha_n
\end{array}
\right. ~~~ n\in \Z^* \label{at-a-closed}
\end{equation}
and on the closed string background as 
\begin{equation}
E^{t T}= (-\hcC + \D E^T) ( \A - \B E^T)^{-1} 
~~~ E^{t }=(\hcC + \D E)(\A + \B E)^{-1} \label{Et-E}
\end{equation}
where the quantities with a ${}^t$ are the T-duality transformed ones.

\subsection{Gauge bundles.}
On the torus all quantities 
such as for example the gauge field $A_i$ have only to be periodic up to a gauge
transformation:
\begin{eqnarray}
A_i(x^j + 2\pi \sqrt{\alpha'} \delta^j_{l} )
=
\Omega_{l}(x)~A_i(x^j) ~\Omega_{l}^{-1}(x)
-i \frac{1}{q} \Omega_{l}(x) ~\partial_i \Omega_{l}^{-1}(x)
\label{Omega-main}
\end{eqnarray}
where $q$ is the gauge coupling constant and 
$\Omega_l(x) \equiv\Omega_l(x^{j \neq l})$ is the gauge transition function.
{From} now we mean by { gauge bundle} the
assignment of a background field, together
with a transition function  which fixes the periodicity property of
the  gauge field.
Notice that, if we perform  a gauge transformation
\begin{eqnarray}
A_i^\omega(x)
&=&
\omega(x)~A_i(x) ~\omega^{-1}(x)
-\frac{i}{q}\omega(x) ~\partial_i \omega^{-1}(x)
\label{gauge}
\end{eqnarray}
the transition functions transform  as
\begin{eqnarray}
\Omega^\omega_j(x)=
\omega(x^1, \dots, x^j+2\pi\sqrt{\alpha'},\dots, x^{d}) 
~\Omega_j(x) ~\omega^{-1}(x^1,\dots, x^{d})\label{ogt}.
\end{eqnarray}
Furthermore, these transition functions
have also to satisfy the {\em cocycle condition}, which
simply means that the gauge fields must be unchanged when
translated along a closed path:
\begin{eqnarray}
\Omega_{j}(x^k+2\pi\sqrt{\alpha'}\delta^k_{i}) 
\Omega_{i}(x^{k})
=
\Omega_{i}(x^k+2\pi\sqrt{\alpha'}\delta^k_{j})
\Omega_{j}(x^{k})
\label{cocycle}.
\end{eqnarray}

Some examples of the previous constructions are the followings.
\begin{enumerate}
\item
The $U(N)$ flat gauge bundle on $T^d$ 
\begin{equation}
A_i= \parallel a_i^I \delta_{I J} \parallel,~~~ \Omega_i=\mathbb{I}_{ N},
\end{equation}
for generic $a^I$ ($I=1,\dots N$ are the color indices), 
breaks the symmetry down to $U(1)^N$. 
Using this surviving symmetry we can always choose 
\begin{equation}
0\le \sqa q a_i^I < 1
\end{equation}
by a big gauge transformation given by 
$\omega=\parallel e^{i n_i^I  x^i/ \sqa} \delta_{I J} \parallel$ where 
$n_i^I\in \Z$.
The same conclusion can be reached by performing 
the gauge transformation 
$\omega=\parallel e^{i q a_i^I x^i} \delta_{I J} \parallel$
so that the previous bundle is gauge equivalent to 
\begin{equation}
A^\omega_i=0,~~~ 
\Omega^\omega_i= \parallel e^{i 2\pi \theta_i^I} \delta_{I J} \parallel
\end{equation}
where $ \theta_i^I \equiv \sqa q a_i^I$.
%
%
\item
Another less trivial $U(N)$ bundle on $T^d$ which has a
 constant magnetic field
\begin{equation}
\hat \F_{i j}= 2\pi \alpha' q \F_{i j} 
=  2\pi \alpha' q F_{i j} \mathbb{I}_{ N}
= \hat F_{i j} \mathbb{I}_{ N}
\end{equation}
is obtained for example with the choice of the gauge field
\begin{eqnarray}
A_{i} 
=-\frac{1}{2} F_{i j} x^j \mathbb{I}_{ N}
=- \frac{1}{2}\frac{2\pi}{(2\pi\sqa)^2} \frac{n_{i j}}{N} x^j \mathbb{I}_{ N}~
,
\label{A-F-main}
\end{eqnarray}
along with the gauge transition functions given by
\begin{eqnarray}
\Omega_i(x) = e^{-i \pi \sqrt{\alpha'} q F_{ i j} x^j} \omega_i
\label{Omega1}
\end{eqnarray}
where the constant matrices $\omega_i$ satisfy
\begin{equation}
\omega_i \omega_j = e^{i (2 \pi \sqa)^2 q F_{i j} } \omega_j \omega_i
~.
\end{equation}

In particular 
on a $T^2$ when $\hat F_{1 2}\equiv 2\pi \alpha' F_{1 2}=\frac{f}{N}$
we can write
\begin{equation}
\omega_1= Q_N e^{i 2\pi \theta_1},~~~~
\omega_2= P_N^{-f} e^{i 2\pi \theta_2}
\label{expl-omega-t2}
\end{equation}
and we can always choose
\begin{equation}
0\le \theta_{1,2}< \frac{1}{N}
\label{theta-range}
\end{equation}
since we can use a global transformation given by $\omega=Q_N^{k_2}
P_N^{-k_1}$ with $k_i= [ N \theta_i]$ to move the $\theta$ values into
the desired range.
\item
The previous $T^2$ case is more general than one could guess at first sight
since, given a field strength proportional to the unity in color
space,
 it is always possible to find a $SL(d,\Z)$ transformation (see
for example \cite{Duo:2007he} for a demonstration) such
that the constant magnetic field strength has only the following non vanishing 
components
\begin{eqnarray}
\hat F_{2 a-1, 2 a}= 2 \pi \alpha' q F _{2 a-1, 2 a}
~~,~~
a=1,\dots r
\label{gen-F0}
\end{eqnarray}
The block diagonal field strength (\ref{gen-F0}) and the cocycle
conditions (\ref{cocycle})
imply that we are actually working on a torus 
$T^d=\prod_{a=1}^r T^2_{(a)} \otimes T^{d-2r}$ as far the bundle is
concerned\footnote{ 
This factorized form is only true for the gauge field 
but it is not necessarily true for the
metric and the other background fields.}
since the transition functions (\ref{Omega1})
on  different torii $ T^2_{(a)}$ must
commute, for example
\begin{equation}
\omega_{2 a}\omega_{2 b}=\omega_{2 b}\omega_{2 a}
~~~~a\ne b
\end{equation}
Therefore the cocycle conditions for the transition functions 
oblige to consider the transition functions to be factorized as
$\otimes_{a=1}^{r} U(L_{(a)})\otimes U(\Lrpu)$ and the $U(N)$ bundle 
to be actually a 
$U(\prod_{b=1}^r L_{(b)} \Lrpu)$ 
bundle on $T^{2 r} \otimes T^{d -2 r}$
with $N=\prod_{b=1}^r L_{(b)} \Lrpu$
and with a constant background field   
\begin{eqnarray}
\hat F_{2 a-1, 2 a}= 2 \pi \alpha' q F _{2 a-1, 2 a}=
\frac{f_a}{L_{(a)}} \uno_{\prod_{b=1}^r L_{(b)} \Lrpu}
~~,~~
a=1,\dots r
\label{gen-F}
\end{eqnarray}
with all the other components vanishing. 
This does clearly {\sl not } happen on a
non compact surface and it is responsible for the rank reduction of
the lower dimensional effective theory from
$\prod_{b=1}^r L_{(b)} \Lrpu$ to $\Lrpu$ as we
can see from the physical states in eq. (\ref{op-str-phy-st}) and we
discuss after eq. (\ref{exam-comp-norm}). 
This fact has been used to explain the rank reduction in orientifold
theories in presence of a discrete $B$ (\cite{Pesando:2008xt}, \cite{Bachas:2008jv}).
The background field in eq. (\ref{gen-F})
is obtained  from the gauge field (up to gauge choices)
\begin{equation}
A_i = - \frac{1}{2}F_{i j} x^j \uno_{\prod_{b=1}^r L_{(b)} \Lrpu }
\label{gen-A}
\end{equation}
along with the transition functions which we take to be
\begin{eqnarray}
\Omega_1= e^{i 2\pi \theta_1} 
e^{-i \frac{f_{1}}{L_{(1)}}  \frac{x^2}{2\sqa}}
Q_{L_1}\otimes \uno_{L_2} \dots \uno_{\Lrpu} 
&~~,~~&
\Omega_2= e^{i 2\pi \theta_2} 
e^{i \frac{f_{1}}{L_{(1)}}  \frac{x^1}{2\sqa}}
P_{L_1}^{-f_1}\otimes \uno_{L_2} \dots \uno_{\Lrpu} 
\nonumber\\
\Omega_3= e^{i 2\pi \theta_3} \uno_{L_1} \otimes 
e^{-i \frac{f_{2}}{L_{(2)}}  \frac{x^4}{2\sqa}}
Q_{L_2} \dots \uno_{\Lrpu} 
&~~,~~&
\Omega_4= e^{i 2\pi \theta_4} \uno_{L_1} \otimes
e^{i \frac{f_{2}}{L_{(2)}}  \frac{x^3}{2\sqa}}P_{L_2}^{-f_2}  \dots \uno_{\Lrpu} 
\nonumber\\
&\vdots&
\nonumber\\
\Omega_{2r+1}= e^{i 2\pi \theta_{2r+1}} \uno_{L_1} \otimes \uno_{L_2} \dots \uno_{\Lrpu} 
&~~,\dots~~&
\Omega_d= e^{i 2\pi \theta_d} \uno_{L_1} \otimes
\uno_{L_2}  \dots \uno_{\Lrpu} 
\label{gen-Omega}
\end{eqnarray}
where $e^{i 2\pi \theta_i}$ are the abelian Wilson lines with 
$0\le \theta_{2a-1}, \theta_{2 a}< \frac{1}{L_{(a)}}$ and
$0\le \theta_{2r+1}, \dots \theta_{d}< 1$. 

%
%
\item 
The previous background in eq.s (\ref{gen-F}) and (\ref{gen-Omega})
can be further generalized allowing more general Wilson lines in the 
$U(\Lrpu)$ as
$e^{i 2\pi \theta_i} \uno_{\Lrpu} $ can be generalized to
$e^{i 2\pi \Theta_i}  =\parallel e^{i 2\pi  \theta_i^I} \delta_{I J}
\parallel$ (which further break $U(\Lrpu)$ down to $U(1)^\Lrpu$), 
explicitly
\begin{eqnarray}
\Omega_1= 
e^{-i \frac{f_{1}}{L_{(1)}}  \frac{x^2}{2\sqa}}
Q_{L_1}\otimes \uno_{L_2} \dots \otimes e^{i 2\pi \Theta_1}  
&~~,\dots~~&
\Omega_d= 
\uno_{L_1} \otimes
\uno_{L_2}  \dots \otimes e^{i 2\pi \Theta_d}  
\nonumber\\
\label{item-more-general-bundle}
\end{eqnarray}
where $\Theta_i$ are diagonal $\Lrpu\times\Lrpu$ matrices.

\end{enumerate}

\subsection{Dipole open strings.}

Let us now consider the open string
in a metric background given by $E_{i j}=G_{i j}+B_{i j}$
and in presence of a constant background field $\hat F_{i j}$ defined in
eq. (\ref{gen-F}) and (\ref{gen-Omega})
in the last subsection, i.e. a background where the original group
$U(N)$ is broken to $\otimes_{a=1}^{r} U(L_{(a)})\otimes U(\Lrpu)$ by
a constant magnetic field with Wilson lines.

On this background and along the compact directions
the open dipole string field expansion\footnote{
The open string is a dipole string since the magnetic field is
proportional to the unity in color space.
} is given by
\begin{eqnarray}
X^i(z,\bar z) &=&
\frac{1}{2}\left(X_L^i(z)+ X_R^i(\bar z) \right)
\label{X-open-dipole}
\end{eqnarray}
where $z=e^{\tau_E +i \sigma}\in \Hp$ ($0\le \sigma \le \pi$) and
\begin{eqnarray}
X_L^i(z)
&=&
(G^{-1} {\cal E})^i_j
\hat X_{L (0)}^j(z)
+y_0^i
\nonumber\\
X_R^i(\bar z)
&=&
(G^{-1} {\cal E}^T)^i_j
\hat X_{R (0)}^j(\bar z)
-y_0^i
\label{hatXy0Def}
\end{eqnarray}
where we have defined the following quantities
\begin{eqnarray}
{\cal{B}}_{i j} 
&=& 
B_{i j} - \hat F_{i j}
\label{calB}
\\
{\cal{E}}_{i j} 
&=& 
G_{i j} -{\cal{B}}_{i j} 
= G_{i j}
- B_{i j} + \hat F_{i j}
\label{calE}
\end{eqnarray}
and the open string metric given by
\begin{eqnarray}
{\cal{G}}_{i j} =  G_{i j} - {\cal B}_{i k} G^{k h} {\cal B}_{h j} =
{\cal{E}}^{T}_{i k} G^{k h} {\cal{E}}_{h j}
\label{openme2}
\end{eqnarray}
along with the non commutativity parameter $\Theta$ as
\begin{equation}
(\cE^{-1})^{i j}= (\cG^{-1})^{i j} - \Theta^{i j}~.
\end{equation}
Moreover we have also defined the fields $\hat X_{L (0)}^i$ and $\hat X_{R (0)}^i$
which have the usual field expansion
\begin{eqnarray}
\hat X_{L (0)}^i(z)
&=&
x^i 
-2\alpha'  p^i ~i \ln(z)
+ i \sqrt{2\alpha'} \sum_{n\ne 0} \frac{sgn(n)}{\sqrt{|n|}} a_n^i
z^{-n}
\nonumber\\
\hat X_{R(0)}^i(\bar z)
&=&
x^i 
-2\alpha'  p^i ~i \ln(\bar z)
+ i \sqrt{2\alpha'} \sum_{n\ne 0} \frac{sgn(n)}{\sqrt{|n|}} a_n^i
{\bar z}^{-n}
\label{hatXExp}
\end{eqnarray}
but have a different set of commutation relations since  the metric
$G$ is replaced by the open string metric $\cG$ and the zero modes
have a non trivial commutation relation, explicitly
\begin{eqnarray}
[x^i, x^j]= i ~2\pi\alpha' ~\Theta^{i j}
~~~~
[x^i, p^j]= i {\cal G}^{ i j}
~~~~
[a_m^i,a_n^j] = {\cal G}^{i j} ~sgn(m) ~\delta_{n+m,0}
\end{eqnarray}
The previous expansion (\ref{hatXExp}) looks as the usual one because 
we have used $x^i$ with non vanishing commutator 
and not $x_0^i$ defined as 
$x^i= x_0^i - \pi\alpha' \Theta^{i l} {\cal  G}_{l m} p^m$ 
with the usual commutator
\begin{equation}
[x_0^i, x_0^j]=0
\end{equation}
Finally $y_0^i= \sqa G^{i j} \theta_j$ are constants 
and proportional to the Wilson lines $\theta$ on the brane at $\sigma=0$
(\cite{Green:1991et},\cite{Frau:1997mq} 
and \cite{Pesando:2003ww}).
Notice the $y_0^i$ do  not enter the expansion of $X(z, \bar z)$  and
therefore they do enter the open string vertices but they do enter the
closed string vertices in the open string formalism where they are
necessary to reproduce from the open string side the phases in the
boundary state due to Wilson lines or positions.

On this background the dipole string Hamiltonian  is then given by
\begin{eqnarray}
H_o-1=L_0 
&=& 
\alpha' G^{\mu\nu} k_\mu k_\nu
+\alpha'  p^i {\cal{G}}_{i j} p^j 
+  \sum_{n=1}^{\infty} n 
{{G}}_{\mu \nu} \alpha^{\dagger \mu}_{n} \alpha^{\nu}_{n} 
+{\cal{G}}_{i j} a^{\dagger i}_{n} a^{j}_{n} 
\label{L061}
\end{eqnarray}
where the spectrum of the momentum operators in compact directions is
given in eq. (\ref{impu43}).

The spectrum in compact directions can be deduced as follows.
In order to define the physical states we need to introduce 
the conserved generalized translation operator 
${\cal T}^{(i)}_{2\pi\sqrt{\alpha'}}$
(\cite{0401040},\cite{PRD5249},\cite{DiVecchia:2007dh}) along the compact
direction $i$ by $2\pi\sqrt{\alpha'}$
so that the physical states can be defined to satisfy
\begin{equation}
\forall i~~~~
{\cal T}^{(i)}_{2\pi\sqrt{\alpha'}} |phys\rangle = |phys\rangle
\end{equation}
The action of these operators is given by\footnote{
With respect to (\cite{DiVecchia:2007dh},\cite{Pesando:2008xt}) 
we have slightly changed notation as 
$| I,J\rangle_{here}=| J,I\rangle_{there}=| I\rangle_{\sigma=0}
~|J\rangle_{\sigma=\pi} $. 
}
\begin{eqnarray}
{\cal T}^{(i)}_{2\pi\sqrt{\alpha'}} |\Phi; \, I , J\rangle
=
 e^{2\pi \sqrt{\alpha'}\,i\, \cG_{i j} p^j 
} 
({\omega_i}^\dag)_{M I} 
~(\omega_i)_{ J L} 
|\Phi;\,M , L \rangle 
 \label{bst1}
 \end{eqnarray}
where 
$| I,J\rangle=| I\rangle_{\sigma=0}~|J\rangle_{\sigma=\pi}$  
( with hermitian conjugate
$\langle I,J|= {}_{\sigma=0}\langle I|~{}_{\sigma=\pi}\langle J|$ )
is an
element of basis for the color indexes (see (\cite{DiVecchia:2005vm})
for more details).
The color index $I$ is actually a collection of indices
corresponding to the different factors into which the $U(N)$ bundle
transition functions split
$\otimes_{a=1}^{r} U(L_{(a)})\otimes U(\Lrpu) \subset U(N)$
as $I=(I_1,I_2,\dots I_{r+1})$ 
($1\le I_a \le L_{(a)}$, $1\le I_{r+1} \le \Lrpu$ )
and similarly for $J$.

Since $\omega_{2 a-1}^{L_{(a)}}=\omega_{2
  a}^{L_{(a)}}=\uno_{L_{(a)}}$ we can iterate eq. (\ref{bst1}) $L_i$
times and get the spectrum of $p_i$;
the compact momenta  have therefore spectrum\footnote{
\label{foot:GenBundMom}
If we consider the more general bundle given in eq.
(\ref{item-more-general-bundle}) of the previous section, we find that
the momenta depend on the the Wilson lines, i.e. they depend on both the
starting brane and the ending one.
For a string with color $I_{r+1}$ at $\sigma=0$ and
color $J_{r+1}$ at $\sigma=\pi$ we get 
$p^i_{I_{r+1} J_{r+1}} = {\cal{G}}^{i j} \frac{1}{\sqa} 
(\frac{n_i}{L_i}
+\theta_i^{I_{r+1}}
-\theta_i^{J_{r+1}})
$
.
In the main text we consider only the case with $\theta=0$ while in
appendix we consider the more general case.
} 
\begin{eqnarray}
p^i = {\cal{G}}^{i j} \frac{1}{\sqa} \frac{n_i}{L_i}
\label{impu43}
\end{eqnarray}
where we have defined $L_{ 2 a}= L_{2 a -1}=L_{(a)}$ 
for $1\le a \le r$ and $L_i=1$ for $2 r< i \le d$.

It then follows that the normalized string states are given by
(\cite{Di Vecchia:2006gg},\cite{Pesando:2008xt})
\begin{eqnarray}
|\chi; k_\mu, n_i; u\rangle
&=&
\frac{1}{(2\pi\sqa)^{d/2}}
|\chi \rangle \otimes
|k_\mu\rangle 
\nonumber\\
&&
\otimes 
\Lambda_{L_{(1)}; I_1 J_1}(n_1,n_2) 
~|\frac{n_1}{\sqa L_{(1)}},\frac{n_2}{\sqa L_{(1)}} \rangle_p
~|I_1, J_1\rangle
\nonumber\\
&&
\otimes 
\Lambda_{L_{(2)}; I_2 J_2}(n_3,n_4)
~|\frac{n_3}{\sqa L_{(2)}},\frac{n_4}{\sqa L_{(2)}} \rangle_p
~ |I_2, J_2\rangle
\dots
\nonumber\\
&&
\otimes
~T_{u~\Lrpu; I_{r+1} J_{r+1}}
~| \frac{ n_{2 r+1}}{\sqa},\dots \frac{ n_{d}}{\sqa}\rangle_p 
~|I_{r+1}, J_{r+1}\rangle
\nonumber\\
\label{op-str-phy-st}
\end{eqnarray} 
where $|\chi \rangle$ is the collective name for the quantum numbers
associated with the non zero modes,
$|\frac{n_1}{\sqa L_{(1)}},\frac{n_2}{\sqa L_{(1)}} \rangle_p$ is the momentum
eigenvector in the first torus of coordinates $x^1,x^2$ and similarly
for the others.
The physical meaning of 
writing $\Lambda_{L_{(1)}; I_1 J_1}(n_1,n_2) ~|I_1, J_1\rangle$
is that for a given momentum 
$\left(\frac{n_1}{\sqa L_{(1)}},\frac{n_2}{\sqa L_{(1)}}\right)$
not all the possible $L_{(1)}^2$  $~|I_1, J_1\rangle$ color index
combinations are possible, as it is usual with the trivial bundle, but
only one.

In the eq. (\ref{op-str-phy-st}) 
$T_u$ are the usual $\Lrpu^2$ hermitian $u(\Lrpu)$ generators which
are normalized to unity as $tr(T_u ~T_v)= \delta_{u,v}$ 
and can be traded for the $\Lrpu^2$ color states $|I_{r+1}, J_{r+1}\rangle$.

The $\Lambda$ are the hermitian momentum dependent Chan-Paton matrices
given by\footnote {We use a double notation for the dependence of
  $\Lambda$ on momenta: by giving either 
  their true values $k=( \frac{n_1}{L \sqa},\frac{n_2}{L \sqa})$ 
or by the integers associated with them $(n_1,n_2)$. We can easily
pass from one notation to the the other.
Notice however that in presence of Wilson lines the $\Lambda_L$s still
depend on the integers associated with the momenta and do not depend
on Wilson lines.}
\begin{eqnarray}
~\Lambda_{L; I J}(n_1,n_2)
&=&\Lambda_{L; I J}(\frac{n_1}{L \sqa},\frac{n_2}{L \sqa)}
=\Lambda_{L; I J}(k)
\nonumber\\
&=&
\frac{1}{\sqrt{L}} e^{-i \frac{\pi}{L} \hat h n_1 n_2}
\left( Q_L^{\hat h n_2} P_L^{-n_1}\right)_{I J}
,~~
0\le I,J < L
\label{LambdaDef}
\end{eqnarray}
where $n_1$ and $n_2$ are two arbitrary momenta not necessarily in
directions $1$ and $2$
and\footnote{
\label{foot:fftilde}
 We can always choose
$ f \tilde f$ even since it $f$ were odd we can shift $\hat
h\rightarrow \hat h +L$ and $\tilde f \rightarrow \tilde f +f $ in
order to get $\tilde f$ even.
}
\begin{eqnarray}
P_{I,J}=\delta_{I+1,J}, ~~~~Q_{I,J}= e^{i \frac{2\pi}{L} I} \delta_{I,J}
\nonumber\\
\hat h f =-1 +\tilde f L \equiv -1 ~~ mod~L
\nonumber\\
f \tilde f\in 2 \Z
\label{fftilde}
\end{eqnarray}
 which have the following properties:
\begin{itemize}
\item commutation with non abelian Wilson lines
\begin{eqnarray}
\omega_i \Lambda_L(n_1,n_2) \omega_i^\dagger
=
 e^{i \frac{2\pi}{L} n_i} \Lambda_L(n_1,n_2)
~~~\Rightarrow
\omega_i^\dagger \Lambda_L(n_1,n_2) \omega_i
=
 e^{-i \frac{2\pi}{L} n_i} \Lambda_L(n_1,n_2)
\label{Lambda-transl}
\end{eqnarray}
\item Chan-Paton matrices algebra
\begin{eqnarray}
\Lambda_L(n_1,n_2) \Lambda_L(m_1,m_2)
&=&
e^{-i \pi \frac{ \hat h}{L}(n_1 m_2- n_2 m_1)}
\frac{1}{\sqrt{L}}
\Lambda_L(n_1+m_1,n_2+m_2)
\nonumber\\
&=& 
e^{-i \pi  \alpha' (\frac{n_1}{\sqa L},\frac{n_2}{\sqa L}) \Theta_{C P} 
(\frac{m_1}{\sqa L},\frac{m_2}{\sqa L})^T
}
\frac{1}{\sqrt{L}}
\Lambda_L(n_1+m_1,n_2+m_2)
~~~~
\end{eqnarray}
with 
\begin{equation}
\Theta_{C P}
= L {\hat h} \epsilon
= L {\hat h} 
\left(\begin{array}{cc}
0& +1 \\
-1 & 0
\end{array}\right),
\label{ThetaCP}
\end{equation} 
and
\begin{equation}
\Theta_{CP} \hF = (1- \tilde f L) ~\uno= - \hat h f ~\uno
\label{ThetaF}
.\end{equation}
\item hermitian conjugation 
\begin{equation}
\Lambda_L^\dagger(n_1,n_2)=\Lambda_L(-n_1,-n_2)
\end{equation}
\item normalization
\begin{equation}
tr\left(\Lambda_L^\dagger(n_1,n_2) \Lambda_L(m_1,m_2) \right)
= \delta_{n,m}
\label{LambdaNorm}
\end{equation}
\end{itemize}
To show that the states in eq.  (\ref{op-str-phy-st}) are normalized
we perform computations like 
\begin{eqnarray}
&&
\langle M, K |
~{}_p\langle\frac{n_1}{\sqa L},\frac{n_2}{\sqa L} |
~\Lambda^*_{L; M K}(n_1,n_2) 
~\Lambda_{L; I J}(m_1,m_2) 
~|\frac{m_1}{\sqa L},\frac{m_2}{\sqa L} \rangle_p
~| I, J\rangle
\nonumber\\
&&
=
(2\pi\sqa)^2 \delta_{n,m}
~ \delta_{K,J}\delta_{M,I}
~\left(\Lambda^\dagger_{L}\right)_{ K M}(n_1,n_2) 
~\Lambda_{L; I J}(m_1,m_2) 
\nonumber\\
&&
=
(2\pi\sqa)^2 \delta_{n,m}
~tr\left(
\Lambda_{L}^\dagger(n_1,n_2) 
~\Lambda_{L}(m_1,m_2)
\right)
=
(2\pi\sqa)^2 \delta_{n,m}
\label{exam-comp-norm}
\end{eqnarray}

We notice that even if we start with $\prod_{b=1}^r L_{(b)} \Lrpu$
branes the number of massless states $k_\mu^2=0$ is only $\Lrpu^2$. 
This does not mean that we have not $(\prod_{b=1}^r L_{(b)} \Lrpu)^2$
states as we naively would expect but that some of them become
massive, with a mass of order $\frac{1}{L}$:
it is essentially a Scherk-Schwarz reduction mechanism
and is the key idea of the explanation of the rank reduction.

Finally  the vacuum is given by
\begin{equation}
|0\rangle= |0\rangle_p ~
\sum_I |I,I \rangle
\label{open-vacuum}
\end{equation}
therefore the physical state (\ref{op-str-phy-st}) in the case of the tachyon 
(and disregarding a possible cocycle which we discuss later) is
associated to the $\sigma=0$ vertex and to the $\sigma=\pi$ vertex by
\begin{equation}
|T_o\rangle 
= \lim_{x\rightarrow 0^+} V_T(x;k) |0\rangle
= \lim_{y\rightarrow 0^-} V_T(y;k) |0\rangle
\end{equation}
The $\sigma=0$ vertex of a tachyon with polarization $t_u(k)$ is given by
\begin{equation}
V_T(x;k)
=
t_u :e^{i k^T X(x)}:
~\Lambda_{L_{(1)}; I_1 J_1}(n_1,n_2)
\otimes\dots
~T^u_{ I_{r+1} J_{r+1}}
\end{equation}
or, in a more unconventional way which makes clear how it acts on indices
(again up to possible cocycles)
\begin{eqnarray}
V_T(x;k)
&=&
t_u :e^{i k^T \hat X_{L(0)}(x)}:
~|I_1\rangle_{\sigma=0}  ~\Lambda_{L_{(1)}; I_1 J_1}(n_1,n_2)  
~{}_{\sigma=0}\langle J_1|
\otimes\dots
\nonumber\\
&&
\hspace{4em}\otimes
~|I_{r+1}\rangle_{\sigma=0}  ~T^u_{ I_{r+1} J_{r+1}}  ~{}_{\sigma=0}\langle J_{r+1}|
\label{T-vertex-wo-coc}
\end{eqnarray}
where we have written $X_{L(0)}(x)$ and not $X(x)$ to stress the T-duality invariant
nature of this expression 
(\cite{Pesando:1999hm}).
The analogous vertex for the emission from the $\sigma=\pi$ boundary ($y<0$)
is then given by
\begin{eqnarray}
V_T(y;k)
&=&
t_u :e^{i k^T \hat X_{L(0)}(y)}:
~|I_1\rangle_{\sigma=\pi}  ~(\Lambda_{L_{(1)}})^T_{ I_1 J_1}(n_1,n_2)  
~{}_{\sigma=\pi}\langle J_1|
\otimes\dots
\nonumber\\
&&
\hspace{4em}\otimes
~|I_{r+1}\rangle_{\sigma=\pi}  ~(T^{u ~T})_{ I_{r+1} J_{r+1}}
~{}_{\sigma=\pi}\langle J_{r+1}|
\label{T-vertex-wo-coc-sigmapi}
\end{eqnarray}
where the transpose is necessary because is the $\sigma=\pi$ boundary
is traveled in the opposite direction of the $\sigma=0$ one.

\section{Spectral decomposition of unity and unique trace in one loop amplitudes.}
\label{sect:SpectralDec}
Since we know that the physical states are given by eq. (\ref{op-str-phy-st}) 
we can write the spectral decomposition of the unity.

Here and in the following we consider the case where 
the magnetic field is switched on only on a $T^2$, i.e. $r=1$ since it is easier
to write the formulae and it is easy to consider the more general
case $r>1$.

\begin{figure}[!t]
  \begin{minipage}[t]{0.4\linewidth}{
    \includegraphics[width=0.9\textwidth,angle=-90]{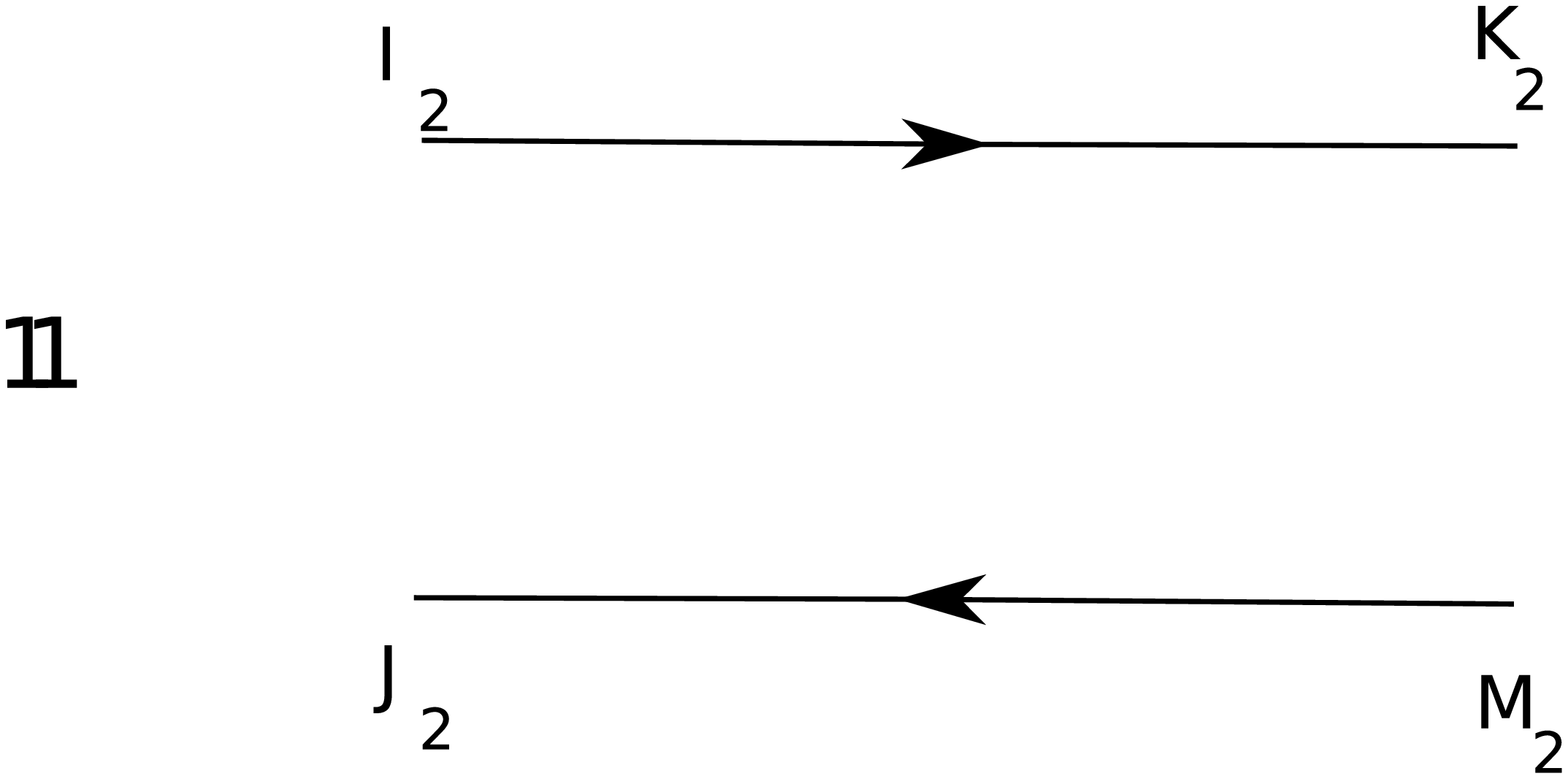}
}
\caption{The spectral decomposition of unity 
with the ``free'' color indices made explicit.
}
\label{figure:Unity-true}
\end{minipage}
\hskip 5mm
\begin{minipage}[t]{0.4\linewidth}{
    \includegraphics[width=0.9\textwidth,angle=-90]{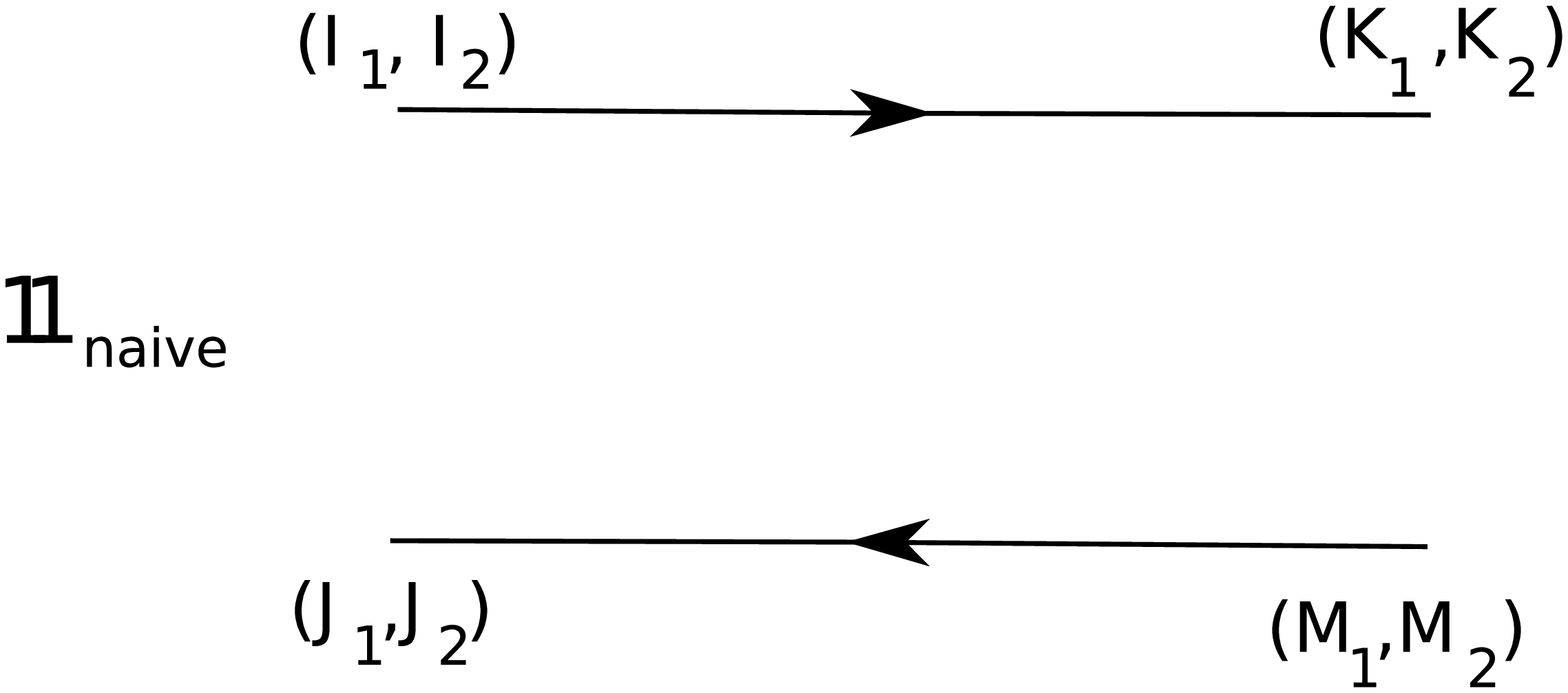}
}
\caption{The usual color spectral decomposition of unity 
with the color indices made explicit
which is not the right one in presence of a magnetic background.}
\label{figure:Unity-naive}
\end{minipage}
\end{figure}

The spectral decomposition of the unity is then given by %
\begin{eqnarray}
\uno 
&=&
\int \frac{d^{D-d}k_\mu}{ (2\pi)^{D-d} ~(2\pi\sqa)^d}
\sum_{\chi,n_i,u}
|\chi; k_\mu, n_i; u\rangle
~\langle \chi; k_\mu, n_i; u|
\nonumber\\
&=&
\int \frac{d^{D-d}k_\mu}{ (2\pi)^{D-d} ~(2\pi\sqa)^d}
\sum_{\chi,n_i,u}
|\chi \rangle \otimes
|k_\mu\rangle 
\otimes 
\Lambda_{L; I_1 J_1}(n_1,n_2) 
~|\frac{n_1}{\sqa L},\frac{n_2}{\sqa L} \rangle_p
~|I_1, J_1\rangle
\nonumber\\
&&
\hspace{6em}
\otimes
~T_{u~\Lrpu; I_{2} J_{2}}
~| \frac{ n_{3}}{\sqa},\dots \frac{ n_{d}}{\sqa}\rangle 
~|I_{2}, J_{2}\rangle
~~
\langle M_2, K_2|
~{}_p\langle\frac{n_3}{\sqa},\dots \frac{n_d}{\sqa} |
~T^*_{u~\Lrpu; K_{2} M_{2}}
\nonumber\\
&&
\hspace{6em}
\langle K_1, M_1|
~{}_p\langle\frac{n_1}{\sqa L},\frac{n_2}{\sqa L} |
~\Lambda^*_{L; K_1 M_1}(n_1,n_2) 
\otimes
\langle k_\mu|
\otimes 
\langle\chi|
\nonumber\\
\end{eqnarray}
where we have simplified the notation setting $L=L_{(1)}$
and $I_1,I_2,K_1,K_2$ are the color indices at $\sigma=0$ boundary
while
$J_1,J_2,M_1,M_2$ are the color indices at $\sigma=\pi$ boundary.
As previously noticed the only color indices which are ``free'' and can 
be summed in the spectral decomposition of the unity 
are those associated with the unbroken part of the gauge group.
The structure of the ``free'' color indices is pictured in
fig. \ref{figure:Unity-true} and is different from what we naively
we would expect and is pictured in fig. \ref{figure:Unity-naive}.
The very reason of this phenomenon is nothing but the fact that a 
momentum eigenstate is a unique fixed superposition of strings with
different colors (in the part of the group which feels the magnetic moment).

It is then obvious to define the trace for a generic operator $O$ as
\begin{eqnarray}
Tr(O) 
&=&
\int \frac{d^{D-d}k_\mu}{ (2\pi)^{D-d} ~(2\pi\sqa)^d}
\sum_{\chi,n_i,u}
~\langle \chi; k_\mu, n_i; u|
~O
|\chi; k_\mu, n_i; u\rangle
\end{eqnarray}
This definition has proven the right one to compute the normalization
of the Moebius amplitude necessary for the tadpole cancellation when a
discrete $B$ is present in an orientifold (\cite{Pesando:2008xt}).

Let us now see some consequences of this definition.  In order to
stress the main point, i.e. the presence in the one loop amplitudes of
a unique trace for the part of the gauge algebra where the magnetic
field is turned on, we sketch now the annulus and the simplest non
planar one loop amplitudes whose generalization we
compute in detail in sec \ref{ssec:1loopNP}.

The simplest one loop amplitude is  the annulus which is given by
\begin{eqnarray}
Z&\propto&
\int \frac{d\tau}{\tau} 
~Tr_{nzm}e^{-\tau L_{0,nzm}^{X+b c} }
~\int \frac{d^{D-d}k_\mu}{ (2\pi)^{D-d} ~(2\pi\sqa)^d}
~e^{-\tau \alpha' G^{\mu\nu} k_\mu k_\nu }
~\delta^{D-d}(0)
\nonumber\\
&&
\sum_{n_i}
e^{-\tau \left[
\left( \frac{n_1}{L_1}, \frac{n_2}{L_2},n_3,\dots n_d \right)
\cG^{-1}
\left( \frac{n_1}{L_1}, \frac{n_2}{L_2},n_3,\dots n_d \right)^T
\right]
}
tr\left(\Lambda_L^\dagger(n_1,n_2) \Lambda_L(n_1,n_2) \right)
\sum_u tr( T_u T^\dagger_u)
\nonumber\\
&\propto&
\Nu^2
\int \frac{d\tau}{\tau} 
~Tr_{nzm}e^{-\tau L_{0,nzm}^{X+b c} }
~\int \frac{d^{D-d}k_\mu}{ (2\pi)^{D-d} ~(2\pi\sqa)^d}
~e^{-\tau \alpha' G^{\mu\nu} k_\mu k_\nu }
\sum_{(n_i)}
e^{-\tau
\sum
{\cal{G}}^{i j}  
\frac{n_i}{L_i} \frac{n_j}{L_j}
}
\nonumber\\
\end{eqnarray}
where we have used the $\Lambda$ normalization given in
eq. (\ref{LambdaNorm}) and we have not supposed the metric to be factorized.
This result clearly shows how the amplitude is proportional only to the
product of traces of the part of the group left unbroken by the
magnetic field.

This point becomes even clearer in
the simplest non planar one loop amplitude pictured in
fig. \ref{figure:1loop-nonplanar}. 
This amplitude in the case of two tachyon vertices given in
eq.s (\ref{T-vertex-wo-coc}) and (\ref{T-vertex-wo-coc-sigmapi}) becomes 
\begin{eqnarray}
&&Tr(V_T(x=1;k_1)~\Delta ~V_T(y=-1;k_2) ~\Delta)
=
\nonumber\\
&&\hspace{4em}
= t_u t_v
\int \frac{d\tau_1}{\tau_1} 
\int \frac{d\tau_2}{\tau_2} 
~\int \frac{d^{D-d}k_\mu}{ (2\pi)^{D-d} ~(2\pi\sqa)^d}
\sum_{(n_i)}
\nonumber\\
&&\hspace{5em}
\times
\langle k, \frac{n}{\sqa L}|
~Tr_{nzm}( 
e^{i k_1^T X(e^{-\tau_1})}
e^{i k_2^T X(-e^{-(\tau_1+\tau_2))}}
)
~|k, \frac{n}{\sqa L}\rangle
\nonumber\\
&&\hspace{5em}
\times
tr(\Lambda^\dagger(n) \Lambda(k_1) \Lambda(n) \Lambda(k_2))
~~tr(T_u)~ Tr(T_v)
\end{eqnarray}

\section{Translational invariant operators and Chan-Paton factors for
  closed string vertices.}
\label{sect:ClosedCP}
We want now check that closed string vertices are invariant for a
translation along a closed curve on the torus.
 
Since the closed string vertices cannot be applied directly to the
open string vacuum to generate open string states 
we need an explicit expression for the conserved generalized
translation operator ${\cal T}^{(i)}_{2\pi\sqrt{\alpha'}}$ along the
$x^i$ torus cycle  which can be written as
\begin{eqnarray}
{\cal T}^{(i)}_{2\pi\sqrt{\alpha'}} 
=
 e^{2\pi \sqrt{\alpha'}\,i\, \cG_{i j} p^j 
} 
({\omega_i}^\dag)_{M I} 
~(\omega_i)_{ J L} 
~
|M , L \rangle 
\langle I , J |
\label{gen-T-oper}
\end{eqnarray}
so that any operator $O$ which is invariant along the cycle
parametrized by $x^i$ satisfies
\begin{equation}
{\cal T}^{(i)}_{2\pi\sqrt{\alpha'}}  
~O 
~{\cal T}^{(i)~-1}_{2\pi\sqrt{\alpha'}} 
=O
~~~~\forall i
\label{DefTranInvOp}
\end{equation}
It is then natural to assume that all physical operators are invariant
for a translation along all cycles.
We can now see what happens if we apply the previous translation
operators to closed string vertices written in open string formalism.
The {\sl naive} closed string vertex operators can be written in open string
formalism as (\cite{Ademollo:1974fc}, \cite{Frau:1997mq},
\cite{Pesando:2003ww}) and up to cocycles as
\begin{equation}
W_{T_c,naive}(z, \bar z; k_L, k_R)
=
:e^{i k_L^T X_L(z)}:
~
:e^{i k_R^T X_R(\bar z)}:
~
\sum_{I,J} |I,J\rangle~\langle I,J|
\end{equation}
where $X_L(z)$ and $X_R(\bar z)$ are the open string fields in
eq. (\ref{X-open-dipole}) and the left and right moving part are
separately normal ordered.

If we apply the translation operator along the $x^i$ cycle 
on the previous naive vertex we get
\begin{eqnarray}
{\cal T}^{(i)}_{2\pi\sqrt{\alpha'}}  
~ W_{T_c,naive}(z, \bar z; k_L, k_R)
~{\cal T}^{(i)~-1}_{2\pi\sqrt{\alpha'}}
&=& 
e^{i 2\pi \sqa
( 
k_L^T G^{-1} {\cal E}
+ k_R^T G^{-1} {\cal E}^T
)_i }
~
W_{T_c,naive}(z, \bar z; k_L, k_R) 
\nonumber\\
&=&
e^{i 2\pi (n -\hF m)_i }
~
W_{T_c,naive}(z, \bar z; k_L, k_R) 
\label{T-w-wrong-phase-W}
\end{eqnarray}
which is clearly not translational invariant when $\hF_{i j}\in \Q$.

To cure this problem we propose to define
the closed string vertex operators up to cocycles as
\begin{eqnarray}
W_{T_c}(z, \bar z; k_L, k_R)
&=&
:e^{i k_L^T X_L(z)}:
~
:e^{i k_R^T X_R(\bar z)}:
\nonumber\\
&&
~\times\sqrt{L}
\sum |(I_1,I_2),J\rangle~\Lambda_{L; I_1 K_1}\left(-L~{\hF m}\right) 
~\langle (K_1,I_2),J|
\label{Main-W-wo-coc}
\end{eqnarray}
Since $\hF_{1 2}=\frac{f}{L}$ the action of the matricial part of the
generalized translation operator ${\cal T}^{(i)}_{2\pi\sqrt{\alpha'}}$ is now
\begin{eqnarray}
({\omega_i}^\dag)_{M_0 I_0} ~(\omega_i)_{ J_0 L_0} ~|M_0 , L_0 \rangle 
\langle I_0 , J_0 |
&\times&
\sum |I_1,J\rangle~\Lambda_{L; (I_1)_1 (I_2)_1}\left(-L~{\hF m}\right) ~\langle I_2,J|
\nonumber\\
&\times&
({\omega_i}^T)_{M_3 I_3} ~(\omega_i^*)_{ J_3 L_3} ~|M_3 , L_3\rangle 
\langle I_3 , J_3 |=
\nonumber\\
=
|M_0 , L_0 \rangle 
&&
\left( {\omega_i}^\dag \Lambda_{L}\left(-L~{\hF m}\right)
     {\omega_i}\right)_{(M_0)_1 (I_3)_1}
\left( {\omega_i}  {\omega_i}^\dagger\right)_{M_0 J_3}
\langle I_3 , J_3 |
\nonumber\\
\Rightarrow
\omega_i^\dagger~ 
\Lambda_L\left(- L~{\hF m}\right) 
\omega_i
&=&
\omega_i^\dagger~ 
\Lambda_L\left(-{f m^2},{f m^1}\right) 
\omega_i
\nonumber\\
&=& e^{- i \frac{2\pi}{L} (-L~{\hF m})_i}
\Lambda_L\left(-{f m^2},{f m^1}\right) 
\end{eqnarray}
as follows from eq. (\ref{Lambda-transl}).
The additional phase then cancels exactly the one due to the
operatorial part in eq. (\ref{T-w-wrong-phase-W}).
While this is contrary to the common lore we will show that when
cocycles are taken into account everything works as expected, i.e.
for example
two closed string vertices have the same OPE as in a pure closed
string theory. 
Before discussing the cocycles which are more technical we would like
to make a series of comments:
\begin{itemize}
\item
the previous vertex is not in contrast with the derivation of
the same from the open string $\sigma$ model (\cite{Pesando:2003ww})
since only the massless
states  which have vanishing winding $m=0$ couple to the world-sheet
action and in this case we have $\sqrt{L} \Lambda_L(0,0)=\uno_L$;
\item
one could have canceled the unwanted phase in
eq. (\ref{T-w-wrong-phase-W}) by using a matrix on the $\sigma=\pi$
boundary.
 A reason to prefer the solution given is that
$y_0$ which is defined in eq. (\ref{hatXy0Def}) enters the vertex 
as $e^{i m^T G y_0/ \sqa}$ and 
is identified (up to constant) with the commuting part of the Wilson
lines on the brane at $\sigma=0$   and therefore we can
associate with $\Lambda_L\left(-L~{\hF m}\right)$ the non abelian part
of the Wilson lines coded into $\omega_i$;
\item
finally we notice that a dependence on Chan-Paton in closed string
vertices is present in other systems even if not clearly stated. 
Consider a flat $U(2)$ bundle on $S^1$ defined on two branes 
with Wilson lines turned on as 
$\Omega_1= \left(\begin{array}{cc} 
e^{i 2\pi \theta_1^1} & \\
& e^{i 2\pi \theta_1^2} 
\end{array}
\right)$. 
Then any closed string vertex operator contains a (trivial) matrix
dependence
$\left(\begin{array}{cc} 
e^{i 2\pi \theta_1^1 m^1} & \\
& e^{i 2\pi \theta_1^2 m^1} 
\end{array}
\right)$ associated to the fact that the factor  
$e^{i m^T G y_0/\sqa}$ feels
two different Wilson lines of the two branes.
In fact consider the coupling of a closed string state described by $W$
to the system made by the two branes, then the coupling is given very
roughly by
the sum $\langle W \rangle_{brane ~1}+\langle W \rangle_{brane ~2}$.
If we make explicit the dependence on $y_0$, i.e. on the Wilson lines
we can write
$\langle e^{i m^T \theta_1} W \rangle_{brane ~1}+\langle e^{i m^T
  \theta_2} W \rangle_{brane ~2}$
But then if we want to describe the previous sum as due to a unique
system described by a bundle we must write
$ Tr \langle 
\left( \begin{array}{cc} e^{i 2\pi \theta_1^1 m^1} & \\
& e^{i 2\pi \theta_1^2 m^1} \end{array}\right) W \rangle$.
\end{itemize}

\section{Cocycles for a well defined open/closed string theory.}
\label{sect:Cocycles}
Until now we have discussed the vertices without explicitly  writing
the cocycles. The cocycles are necessary to have a well-defined open
and closed theory especially when Chan-Paton like factors are
introduced for closed string vertices.

\subsection{Cocycles for the closed string theory and framing.}
\label{clos-stri-coc-sub}
The first step is to determine the cocycles in the closed string
formalism since we want to reproduce in open string formalism the
closed string result for the product of two vertices.
A well defined closed strings (which is a good CFT plus a proper
treatment of zero modes) must satisfy at least the
following criteria
\begin{enumerate}
\item closed string vertices commute;
\item a proper behavior under Hermitian conjugation.
\end{enumerate}
The first request is necessary in order to ensure the mutual locality of closed
vertices (\cite{PolBook}), or in other words, that vertices obey the
 spin-statistics theorem.
For implementing the wanted properties we look for a solution of the
form (\cite{PolBook})
\begin{eqnarray}
\label{W-closed}
{\cal W}^{(c)}_{\beta_L,\beta_R}(z,\bar z; k_L,k_R)
&=&
c(k_L,k_R;p_L,p_R) V_{\beta_L}(z;k_L)
\,\, {\tilde V}_{\beta_R}({\bar z};k_R)
\end{eqnarray}
where the cocycles are given by
\begin{eqnarray}
\label{c_closed}
c(k_L,k_R;p_L,p_R)&=&
e^{i\pi \left(
n_i A_c^{i j} \hat n_j
+n_i B_{c~j}^i \hat m^j
+m^i C_{c~ i}^{~~j} \hat n_j
+m^i D_{c~i j} \hat m^j
\right)}
%
~~
\label{cocycl-closed}
\end{eqnarray}
We choose the  coefficients $A_c,B_c,C_c,D_c$ to be matrices in order to be
general.
They must be determined, as said
before, so that two arbitrary vertices are mutually local,
i.e. commute.
This is also equivalent to the fact that the radial
ordering of a product of vertices is given by a unique expression
which can be derived by 
analytically continuing whichever
particular ordering of the vertices is chosen to
perform the computation.

In order to compute these matrices we consider the ordering of
the product
of two arbitrary vertices as
\begin{eqnarray}
\label{OPE-closed}
{\cal W}^{(c)}_{\beta_L,\beta_R}(z,\bar z; k_{L },k_{R })
{\cal W}^{(c)}_{\alpha_L,\alpha_R}(w,\bar w;  l_{L},l_{R })
&=&
e^{i\Phi_{(c)}(k,l)}
c(k_{L}+l_{L },k_{R }+l_{R }; p_L,p_R)
\nonumber\\
&\times&
V_{\beta_L}(z;k_{L })
V_{\alpha_L}(w;l_{L })
\,\, \, \,
{\tilde V}_{\beta_R}({\bar z};k_{R })
{\tilde V}_{\alpha_R}({\bar w};l_{R })
\nonumber\\
\end{eqnarray}
where we have defined the phase
\begin{equation}
\label{OPE-phase-closed}
e^{i \Phi_c(k,l) }
=
e^{-i\pi\left[
n^T_l A_{c}  n_k
+n^T_l B_c  m_k
+m^T_l C_c  n_k
+m^T_l D_{c}  m_k
\right]}
%
\end{equation}
In the previous expression  we have used the compact momenta 
\begin{equation}
k_L=\frac{1}{2 \sqrt{\alpha'}}\left(n_k +E^T m_k\right),~~~~
k_R=\frac{1}{2 \sqrt{\alpha'}}\left(n_k -E m_k\right)
\end{equation}
and similarly for $l$.
The general solution of the constraints listed at the beginning of the
section is discussed in appendix (\ref{Closed-Cocycles})
 and is given by
\begin{eqnarray}
A_c &=& Z_{A~A} + 2[ Z_{A 0}]_S+ 2 Z_A,
~~~~
Z_{A~A}^T=-Z_{A~A},
\nonumber\\
D_c &=& Z_{D~A}  + 2 [Z_{D 0}]_S+2 Z_D,
~~~~
Z_{D~A}^T=-Z_{D~A},
\nonumber\\
B_c&=& \frac{1}{2} \uno +Z_{B}
\nonumber\\
C_c&=&B^T_c-\uno+2 Z_C =  -\frac{1}{2} \uno +Z_{B}^T + 2 Z_C
\label{gen-sol-coc.closed}
\end{eqnarray}
where all matrices are integer valued and we have defined
$2[M]_S=M+M^T$\footnote{
This last definition implies therefore that 
$2[ Z_{A 0}]_S$ is symmetric matrix with even diagonal entries
and arbitrary integer off diagonal entries and similarly for the other cases.
}
We notice that there is a kind of ``gauge invariance'':
 matrices $A_c, B_c, C_c$ and $D_c$ are in the same class of
equivalence when they yield the same phase (\ref{OPE-phase-closed}) 
and this happens for any choice of the  $Z_A, Z_D$ and $Z_C$ matrices.
Moreover because the phase (\ref{OPE-phase-closed})  is actually only
a sign we can always take a representative of the previous matrices
with all entries in $\Z_2$ and we can always choose $[A_c]_A$ and
$[D_c]_A$ as we want since $-1 \equiv 1~~mod~2$. 
This last possibility turns out to be
fundamental for having a consistent theory of open and closed string
interaction. 

In particular the hermiticity condition implies
\begin{equation}
e^{i \Phi_c(k,k) }
=1
\end{equation}

For future reference we write the effect of a T-duality transformation 
defined by a $\Lambda \in O(d,d,\Z)$ matrix in eq. (\ref{lam}) 
 on the cocycles matrices as 
\begin{eqnarray}
M_c=
\left( \begin{array}{cc} D_c & C_c \\ B_c & A_c \end{array} \right)
&=& \Lambda^T
\left( \begin{array}{cc} D_c^t & C_c^t \\ B_c^t & A_c^t \end{array} \right)
\Lambda
\label{A-At-main}
\end{eqnarray}

As it were first discussed in (\cite{Pesando:2003ww}) neither the previous
values are T-duality invariant nor the amplitudes are,
even if it was shown that different solutions differ by momentum
dependent phases which are the same at all loops and therefore the
transition probabilities are the same.
In (\cite{Hellerman:2006tx}, \cite{Duo:2007he}) 
in the framework of the Reggeon approach
it was proposed to keep fixed the the Reggeon vertex but change the
states represented by bra vectors by a phase necessary to compensate
the change of the cocycle.
This approach suffers from an asymmetry since one must fix a
default cocycle and then keep this cocycle fixed for
the Reggeon while changing the states. 
Despite of this we will show in sec. \ref{sect:T-duality} that this
procedure extends correctly to the open string sector.
\COMMENTOm{
Here we propose to introduce the concept of framing, i.e. in order to
specify completely the amplitudes we must give both $E=G+B$ and the
``frame'' given by the matrices  $A_c,B_c,C_c,D_c$.
The reason is that in this way we are completely symmetric and there
is not any preferred frame even if different framings differ by the
choice of the matrix $M_c$ and therefore have amplitudes which differ
by an overall momentum dependent phase but have the same transition functions
}

We will see that open string formalism has not any preferred frame but
has a preferred gauge.

There is actually a further constraint we can impose. 
We would like the naive condition 
\begin{equation}
|B(\hF)\rangle= tr( P e^{i \oint A}) |B\rangle
\label{BF-B-rel}
\end{equation}
to be true. We have used the adjective naive because the previous
equation is what can be deduced from path integral and usually the
results from path integral can suffer from some subtleties.
In any case, as we show in section (\ref{BouStaSec}) 
it is possible to show that eq. (\ref{BF-B-rel}) holds
if we restrict the possible values of the cocycles to
\begin{equation}
[M_c]_S
=
\left( \begin{array}{cc} 2[Z_{A 0}+Z_A]_S & 2(Z_B^T+Z_C) 
\\ 
2(Z_B+Z_C^T) & 2[Z_{D 0}+Z_D]_S  \end{array} \right)
\equiv
0~~~ mod ~4
\Rightarrow
e^{\frac{i}{2} \Phi_c(k,k)}=1
\label{halfPhi1}
\end{equation} 
The constraint implies $e^{\frac{i}{2} \Phi_c(k,k)}=1$ but  it is not
true the converse nevertheless the previous choice is T-duality invariant and
is therefore better.
A particular choice satisfying both the previous constraints and the
ones from the open string construction given in
eq. (\ref{ConstClosStriOpenForm})  is
\begin{equation}
A_c=0,~~~~ B_c=-C_c= \oh \uno,~~~~ D_c= f \tilde f ~\epsilon
~\leftrightarrow~
M_{c 0}=
\left( \begin{array}{cc} f \tilde f ~\epsilon & -\oh \uno \\ 
\oh \uno & 0 \end{array} \right)
\label{Mc0ff}
\end{equation}
and corresponds to the closed string tachyonic vertex
\begin{equation}
\hspace{-2em}
{\cal W}_{T_c}(z,\bar z; k_L, k_R)
=
e^{-i\frac{1}{2}\pi (n_i \hat m^i -m^i \hat n_i) 
+i \pi~ f \tilde f ( m^1 \hat m^2- m^2 \hat m^1) }
:e^{ i \left[\frac{1}{2}k_\mu X^\mu_{L (c)}(z) +k_{L i} X_{L (c)}^i(z) \right]} :
:e^{ i \left[\frac{1}{2}k_\mu \tilde X^\mu_{R (c)}(\bar z)
              +k_{R i} \tilde X_{R (c)}^i(\bar z) \right]} :
\end{equation}
which does depend on the magnetic field in a way which may or may be
not be removed  by a ``gauge'' choice of $Z_D$. 
This gauge transformation is possible when $f \tilde f\in 2
\Z$ which can always be chosen and we have chosen in
eq. (\ref{fftilde}) therefore we can always choose $M_c$ to be gauge
equivalent to
\begin{equation}
M_{c 0}\equiv
\left( \begin{array}{cc} 0 & -\oh \uno \\ 
\oh \uno & 0 \end{array} \right)
\label{Mc0}
\end{equation}

\subsection{Cocycles for the open string theory.}
\label{sec-coc-vert-open-str}
We consider now the open-closed string theory.
In particular a well defined  string theory, which amounts to a proper
treatment of zero modes besides a well behaved CFT, must satisfy
the following 5 constraints:
\begin{enumerate}
\item
the open string vertices at $\sigma=0$ and $\sigma=\pi$ commute;
\item
the open string emission vertex from $\sigma=0$ commutes with the
closed string vertices;
\item
in a similar way the open string emission vertex from $\sigma=\pi$
commutes with the closed string vertices;
\item
the closed string vertices in open string formalism must have the same
product (OPE) as in closed string formalism, as a consequence
we want the closed string vertices to commute;
\item a proper behavior under Hermitian conjugation.
\end{enumerate}

To these constraints we want to add
\begin{enumerate}
\item 
the open closed string mixed amplitudes are sensitive to the value of
$y_0$ which is periodic since it can be identified with the Wilson lines
therefore these amplitudes, but not necessarily the the vertexes, 
must be invariant under this periodic identification;
\item
the naive relation 
$|B(\hF)\rangle= tr( P e^{i \oint A}) |B\rangle$
holds.
\end{enumerate}

On a compact space the previous constraints are not at all trivial and
they are discussed in depth in appendix \ref{app-cocyles}.
Here we summarize the non trivial results in the simplest case as
before where the magnetic field is turned on only within a 2-torus.

Since the cocycles depend only on momenta the ones  in
tachyonic vertex operators are the same of all the other vertices,
therefore we write only the tachyonic vertices which read:
\begin{eqnarray}
{\cal V}_{(0) T}(x; k)
&=&
: e^{ i k_M X^M(x) } :
~\Lambda_{L}(n_1,n_2)\otimes T_u
\COMMENTO{
\nonumber\\
&=&
: e^{ i k_M \hat X^M_{L (0)}(x) } :
~\Lambda_{L}(n_1,n_2)\otimes T_u
\nonumber\\
&=&
: e^{ i \left(  
k_\mu X^\mu_{L }(x) +(G \cE^{-T}  k)_i (X^i_{L}(x)-y_0^i) 
\right)} :
~\Lambda_{L}(n_1,n_2)\otimes T_u
}
\label{OpenStringVertex-sigmazero}
\\
\nonumber\\
{\cal V}_{(\pi) T}(y; k)
&=&
: e^{ i k_M X^M(y) } :
~\Lambda_{L}^T(n_1,n_2)\otimes T_u^T
\COMMENTO{
\nonumber\\
&=&
 e^{i~2 \pi \alpha' k_M (-\uno+\Theta \cG)^M_N p^N}
~: e^{ i k_M \hat X^M_{L (0)}(y)} :
~\Lambda_{L}^T(n_1,n_2)\otimes T_u^T
\nonumber\\
&=&
 e^{i~2 \pi \alpha' k_M (\Theta \cG)^M_N p^N}
\nonumber\\
&&\times
~: e^{i ~k_\mu X^\mu_{L }(x) +i (G\cE^{-T}k)_i (X^i_{L}(y)-y_0^i) } :
~\Lambda_{L}^T(n_1,n_2)\otimes T_u^T
}
\nonumber\\
\label{OpenStringVertex-sigmapi}
\\
{\cal W}_{T_c}(z,\bar z; k_L,k_R,y_0)
&=&
e^{\frac{i}{2 } \Phi_{(c)}(k,k)}
e^{-i \pi \alpha'~ k_{R M} (\cE^{-1} \cG \cE^{-1})^{M N} k_{L N}}
\times
\sqrt{L}
\Lambda_L\left(-{f m^2},{f m^1}\right) 
\otimes\uno_{N_1}
\nonumber\\
&&
\times
: e^{ i k_{L M}  X_{L }^M(z)} :
: e^{ i k_{R M}  X_{R}^M(\bar z)} :
~
\COMMENTO{
\nonumber\\
&=&
e^{\frac{i}{2 } \Phi_{(c)}(k,k)}
~e^{-i \pi \alpha'~ k_{R M} (G^{-1} \cE \Theta \cE G^{-1})^{M N} k_{L N}}
~e^{i m^M G_{M N} \frac{y_0^N}{\sqa}}
\times
\sqrt{L}
\Lambda_L\left(-L \hat F m\right) 
\otimes\uno_{N_1}
\nonumber\\
&&
\times
: e^{ i (\cE^T G^{-1}k_L)_{M}  \hat X_{L (0) }^M(z)} :
: e^{ i (\cE G^{-1} k_R)_{M}  \hat X_{R (0)}^M(\bar z)} :
~
}
\label{ClosedStringVertex}
\end{eqnarray}
where $x=|x| e^{i 0}$, $y=|y| e^{i \pi}$ and $z$ in the upper complex plane.
Other expressions for the vertices which are more useful in the
computations are given in eq.s (\ref{OpenStringVertices1}).
\COMMENTO{
In the previous equations we have written the open string vertices in
three different forms since we want to show that the vertices can be
written 
either in the naive way (first form)
or in a T-duality invariant way (\cite{Pesando:1999hm}),
in this latter case they can be written 
using either $\hat X_{L(0)}$ which has the
usual OPE with the open string metric (second form) 
or using $X_{L(0)}$ which has the usual OPE with the closed string
metric (third form) .
}

A certain number of comments are worth doing.
\begin{enumerate}
\item
It is important to stress that the previous vertices satisfy all the
wanted constraints iff we use the following gauge for the closed
string cocycle 
\begin{eqnarray}
[A_c]_A=0
~~~~
[D_c]_A=\tilde f  ~L \hF=f \tilde f \epsilon,
~~~~
C_c-B_c^T=\uno
\label{ConstClosStriOpenForm}
\end{eqnarray}
where we have defined $\tilde f$ in eq. (\ref{fftilde}).
\item
The phase in the closed string vertex $e^{-i \pi \alpha'~ k_R^T 
  \cE^{-1} \cG \cE^{-1} k_L^T}$ is needed to get a vertex independent
on the ordering of the left and right part since it let us to perform
the substitution $\ln(z-\bar z)
\rightarrow \ln|z -\bar z|$ in the amplitudes.
\item
The closed string vertices are determined up to  signs.
In fact from the hermiticity of the closed string vertices we know
$e^{i \Phi_{(c)}(k,k)}=1$ therefore the associated phase in open string formalism
is a sign
$e^{\frac{i}{2 } \Phi_{(c)}(k,k)}\in \{-1,1\}$
which can depend on $k$ nevertheless the choice $e^{i \oh \Phi_{(c)}(k,k)}=1$
is always possible and it is the best one.
\end{enumerate}

In the case of the more general gauge bundle described in eq.
(\ref{item-more-general-bundle}) where the Wilson lines are turned on
to break the gauge group to $U(1)^\Lrpu$ the previous vertices must be
supplemented by the effect of the change of the Wilson lines. 
Consider to the emission of
an open string from the $\sigma=0$ boundary with color $J$ if the
momentum of the emitted string is 
$\sqa k_{i,J L} = \left(\frac{n_i}{L_i}
+\theta_i^{J}
-\theta_i^{L} \right)$ 
then the final string has color $L$ on the $\sigma=0$ boundary 
and the change in the Wilson lines implies that
\begin{equation}
\Delta y_0^i= -2\pi \alpha' G^{i j} (\theta_j^J-\theta_j^L)
\end{equation}
since $G \frac{y_0}{\sqa}= \theta$
(see appendix \ref{app-cocyles} for further details).
In particular the previous results reduce to what found in the simpler
case $B=F=0$ in \cite{Pesando:2003ww}.

\section{Some examples of amplitudes.}
\label{sect:Amplitudes}
In this section we would like to compute some amplitudes which can be
used to check the picture proposed in the previous sections.
In particular the computation in \ref{ssec:OpenOPEs} comments on the

The computation in sec. \ref{ssec:1Cl} is a warming up to the one in
sec. \ref{BouStaSec} which shows that the momentum dependent
Chan-Paton-like matrix in the closed vertices written in open string formalism
is necessary in order to
reproduce the boundary state from the open string point of view. 
This Chan-Paton-like matrix gives 
raise to a ``trivial'' momentum dependent sign which
has been shown (\cite{Duo:2007he}) 
essential for the correct factorization of the open
string  two loops vacuum amplitude as shown in
fig. \ref{figure:2loops-factor} and the same sign 
turns out to be fundamental for the T-duality to hold as discussed in
sec. \ref{sect:T-duality}.
In sec. \ref{ssec:NOp1Cl} we compute the
one closed string and $N$ open string tachyonic amplitudes where the
would-be closed string Chan-Paton gives again a momentum dependent sign.
Using these results we can finally check that
the one loop non planar amplitudes factorize correctly in the
closed string channel in sec. \ref{ssec:1loopNP}.

\subsection{Product of two open string vertices.}
\label{ssec:OpenOPEs}
It is easy to check that the OPEs of two open string vertices are
given by
\begin{eqnarray}
{\cal V}_{(0) T}(x_1; k)
{\cal V}_{(0) T}(x_2; l)
&=&
\frac{1}{\sqrt{L}} 
e^{ - i \pi \alpha' ~ k_M ( \Theta + \Theta_{C P})^{ M N} l_N}
(x_1 - x_2)^{2 \alpha' k_M \cG^{ M N} l_N}
\nonumber\\
&& 
: e^{ i   k_M X^M(x_1) + i   l_M X^M(x_2) } :
~\Lambda_{L}(k+l)\otimes T_u T_v
\label{ProdOpVer}
\end{eqnarray}
and
\begin{eqnarray}
{\cal V}_{(\pi) T}(y_1; k)
{\cal V}_{(\pi) T}(y_2; l)
&=&
\frac{1}{\sqrt{L}} 
e^{  i \pi \alpha' ~ k_M ( 2 \cG^{-1}+\Theta + \Theta_{C P})^{ M N} l_N}
(y_1 - y_2)^{2 \alpha' k_M \cG^{ M N} l_N}
\nonumber\\
&& 
e^{  -i \pi \alpha' ~ (k+l)_M \cG^{ M N} (k+l)_N}
: e^{ i   k_M X^M(y_1) + i   l_M X^M(y_2) } :
~\Lambda_{L}^T(k+l)\otimes (T_v T_u)^T
\nonumber\\
\label{OpenOPEsigma=pi}
\end{eqnarray}
where despite the apparent asymmetry due to the  phase 
$e^{i 2  \pi\alpha' k_M \cG^{M N} l_N}$ the amplitudes are completely symmetric.

Notice how the nice feature of being able of writing the OPEs in function of
$\Theta_{tot}=\Theta+\Theta_{CP}$ is not anymore true when Wilson
lines are turned on as discussed in section \ref{appssec:OpenOPEs}.

\subsection{1 closed string emission amplitude from the disk.}
\label{ssec:1Cl}
We start by computing the amplitude given in
fig. \ref{figure:boundary} which must give the result already found in
the closed string channel (\cite{DiVecchia:2007dh},\cite{Duo:2007he}).
The complete amplitude for a closed tachyon is then given by
\begin{eqnarray}
{\cal A}(1 T_c)=
\Cz ~ \Nt~ \frac{1}{2\pi}~\langle 0|~{\cal W}_{T_c}(z,\bar
z;k_L,k_R)~|0\rangle~~|_{z=i}
\label{1ptCloAmp}
\end{eqnarray}
where $\Cz$ is the normalization of the disk amplitude, $\Nt$ is the
normalization of the closed string vertices and is independent of the
open string background,  
$\frac{1}{2\pi}$ is the left over of the $SL(2,\R)$ gauge fixing
at $z=i$,
$|0\rangle$ is the opens string vacuum given in eq.
(\ref{open-vacuum}) 
and finally
${\cal W}_{T_c}$ is the closed string vertex in open string formalism 
given in eq. (\ref{ClosedStringVertex}).
An easy computation gives
\begin{eqnarray}
{\cal A}(1 T_c)&=&
\Cz ~ \Nt~ \frac{1}{2\pi}
\times
N_1 L~ e^{i \frac{\pi}{L}\hat h f^2 m^1 m^2}
~\delta^{[L]}_{-L \hF m}
\nonumber\\
&&
\times
e^{\frac{i}{2}\Phi_{(c)}(k,k)}
e^{i m^T G \frac{y_0}{\sqa}}
\delta(k_\mu)
(2\pi\sqa)^d \delta_{\cE^T G^{-1} k_L+\cE G^{-1} k_R,0}
|z-\bar z|^{2\alpha' k_R^T \cE^{-1} \cG \cE^{-1} k_L}
\nonumber\\
\end{eqnarray}
where in the first line we have the Chan-Paton contribution and in the
second the operatorial one.
It is worth noticing that the
absolute value  $|z-\bar z|$ which ensures a well defined phase for the
amplitude is due to the non operatorial cocycle in the closed string vertex
exactly as in the case without magnetic field (\cite{Pesando:2003ww}).
The conservation of the compact momentum 
$\delta_{\cE^T G^{-1} k_L+\cE G^{-1} k_R,0}= \delta_{(n-\hF m)/\sqa,0}$
can be rewritten as
$\delta_{n,\hF m}$ which implies that $\hF m=\frac{f}{L}(m^2,-m^1)$ is
actually integer as also imposed  by the Chan-Paton trace 
$\delta^{[L]}_{-L \hF m}=\delta^{[L]}_{ f  m^1}\delta^{[L]}_{ f m^2}$. 
The exponent of $|z-\bar z|$
becomes 
${2\alpha' k_R^T \cE^{-1} \cG \cE^{-1} k_L}=
-{2\alpha' k_L^T G^{-1}  k_L}=-2$
due to the
mass shell condition $\alpha' ( G^{00} k_0^2+ k_L^T  G^{-1} k_L)=1$
therefore we get
\begin{eqnarray}
{\cal A}(1 T_c)&=&
\Cz ~ \Nt~ \frac{1}{2\pi}
N~ 
e^{\frac{i}{2}\Phi_{(c)}(k,k) +i \pi \frac{\hat h f^2}{L} m^1 m^2}
e^{i m^T G \frac{y_0}{\sqa}}
\nonumber\\
&&
\times
(2\pi)^{D-d}\delta^{D-d}(k_\mu)
(2\pi\sqa)^d \delta_{n,\hF m}
\times \frac{1}{|z-\bar z|^2}
\label{1ptCloTacFin}
\end{eqnarray}
Given our choice $e^{\frac{i}{2}\Phi_{(c)}(k,k)}=1$ (\ref{halfPhi1})
it is then obvious that,
if we can identify $G_{i j} \frac{  y_0^{j}}{\sqa}= 2\pi \sqa q a_{0 ~i}$,
we reproduce exactly the closed string result, phases
included obtained with the boundary state formalism.
With our choices (\ref{fftilde}) and (\ref{Mc0ff})
we have both
$\frac{\hat h f^2}{L}= \frac{f}{L} (-1+\tilde f)\equiv -\frac{f}{L}~~mod~2$ and 
$e^{\frac{i}{2} \Phi_c(k,k)}=1$ so that we can reproduce the amplitude
obtained from the boundary state.

\subsection{Boundary state.}
\label{BouStaSec}

We want to generalize the previous computation to all the closed
string states, i.e we want to derive the generating function in the
closed string Hilbert space of all
the one point closed string coupling to the disk: 
this is nothing else but the boundary state.
We compute therefore the amplitude
\begin{eqnarray}
\langle B(F); V_L,V_R|
&=&
\frac{ \Cz ~ \Nt}{2\pi}
\nonumber\\
&&
\langle x_L=x_R=0; 0_{a_{(c)}}, 0_{\tilde a_{(c)}} |
~e^{-\frac{i}{2 } \Phi_{(c)}(G p_{(c)},G p_{(c)})}
~e^{-i \pi \alpha'~ p_{R}^M (\cE^{T} \cG^{-1} \cE^{T})_{M N} p_{L}^N}
\nonumber\\
&&
\times
Tr\left( \sqrt{L}
\Lambda_L\left(-L \hat F m\right) 
\otimes \uno_{\Lrpu}
\right)
~\times
{}_p\langle 0| \cS_L(z;V_L) ~\cS_R(\bar z;V_R) |0\rangle_p
\nonumber\\
\label{BouGen}
\end{eqnarray}
where we have introduced the closed string state $\langle x_L=x_R=0|$ 
normalized as $\langle x_L=x_R=0| k_L, k_R\rangle=1 $
and the Sciuto-Della Selva-Saito vertex 
$\cS_L(z;V_L)$ ($\cS_R(\bar z;V_R)$)
as discussed in (\cite{DiVecchia:1988cy}, \cite{DiVecchia:1986jv})
which acts on both the open string Hilbert space and the left (right)
closed string Hilbert space.
These vertices are given for arbitrary local $SL(2,\C)$  
coordinates $V_L(u;z)$ and $V_R(u;\bar z)$ defined as
\begin{eqnarray}
V_L(u;z)
&=&
\frac{a_L u + b_L z}{c_L u +b_L},
~~~~
b_L(a_L - c_L z)=1,
~~~~
V_L(0;z)=z
\label{defVL}
\end{eqnarray}
and similarly for $V_R$.

Performing explicitly the computation 
the amplitude (\ref{BouGen}) can then be written as
\begin{eqnarray}
\langle B(F);V_L, V_R | 
&=&
N~ \frac{ \Cz ~ \Nt}{2\pi}
\nonumber\\
&&\hspace{-5em}
~\langle k_\mu=0|
~\sum_{s\in\Z^d} \frac{1}{(2\pi \sqa)^d} \langle n=L~ \hF ~s, m= L~ s|
e^{-\oh i \Phi_{(c)}(k,k) +i \pi \frac{\hat h f^2}{L} m^1 m^2}
~e^{i ~m^M ~G_{M N} ~\frac{y^N_0}{\sqa} } 
\nonumber\\
&\times&
\langle 0_a, 0_{\tilde a} |~
\myexp{- \sum_{n,m=0}^\infty a^N_{(c) n} (\cE \cG^{-1} \cE)_{N M} \tilde a^M_{(c) m}
D_{n m}(U_L V_R)
}
\label{1cl-from-SDS}
\end{eqnarray}
where $D_{n m}$ is a (pseudo)representation of the $SL(2,\C)$ group as
explained in appendix \ref{app:BounRegg}.
This expression allows to compute any  one point closed string amplitude by
\begin{eqnarray}
{\cal A}(\{\beta_L,\beta_R\})
&=&
\langle B(F); V_L,V_R|
\left( \left. \frac{d V_L}{ d u} \right|_{u=0} \right)^{-
\Delta_{\beta _L}}
~\left( \left.\frac{d V_R}{ d u} \right|_{u=0} \right)^{-
\Delta_{\beta _R}}
 |\beta_L,\beta_R\rangle
\nonumber\\
\label{1ptGenAmp}
\end{eqnarray}
where $\Delta_{\beta _L}$ is the conformal weight of the closed string
left moving state
$|\beta_L\rangle$ and similarly for the right moving part.
This result is independent on $V_L$ and $V_R$ for the physical states 
since any closed string physical  state
$|\beta_L,\beta_R\rangle$ is annihilated by the  $SL(2,\C)$ generators.

The previous expression for the boundary state (\ref{1cl-from-SDS}) is
nevertheless not physical, i.e.
\begin{eqnarray}
\langle B(F);V_L, V_R | (L_n - \tilde L_{-n}) \ne 0
\end{eqnarray}
   
We can now fix the local coordinates in such a way we get the usual
boundary, i.e. we require $D_{n m}(U_L V_R)=\delta_{n,m}$ and therefore 
we impose
\begin{eqnarray}
u&=&
U_LV_R(u)
=
\frac{1}{b_L} 
\frac{(a_L c_R -a_R c_L) u + b_R(a_L -c_L \bar z)}
{(a_R -c_R z) u - b_R (z-\bar z)}
\nonumber\\
&\Rightarrow&
V_L(u;z)= \frac{c \bar z ~u + z}{c u +1},
~~~~
V_R(u;\bar z)= \frac{ \frac{1}{c} z ~u +  \bar z}{ \frac{1}{c} u +1},
\label{SpecVLVR}
\end{eqnarray}
so that 
 we get the final result
\begin{eqnarray}
\langle B(F) | 
&=&
N~ \frac{ \Cz ~ \Nt}{2\pi}
~\langle k_\mu=0|
\nonumber\\
&&~\sum_{s\in\Z^d} \frac{1}{(2\pi \sqa)^d} \langle n=L~ \hF ~s, m= L~ s|
e^{-\oh i \Phi_{(c)}(k,k)+i \pi \frac{\hat h f^2}{L} m^1 m^2}
~e^{i ~m^M ~G_{M N} ~\frac{y^N_0}{\sqa} } 
\nonumber\\
&&
\langle 0_a, 0_{\tilde a} |
e^{- \sum_{n=1}^\infty a^N_{(c) n} (\cE \cG^{-1} \cE)_{N M} \tilde a^M_{(c)  n}
}
.
\label{UsuaBouF}
\end{eqnarray}

With our choices (\ref{fftilde}) and (\ref{Mc0ff})
we have both
$\frac{\hat h f^2}{L}= \frac{f}{L} (-1+\tilde f)\equiv -\frac{f}{L}~~mod~2$ and 
$e^{\frac{i}{2} \Phi_c(k,k)}=1$.  
Hence we can reproduce the boundary
already found in (\cite{DiVecchia:2007dh},\cite{Duo:2007he})\footnote{
Notice that the previous constraints are sufficient but not necessary since we
need  only to impose the equality of the the sign 
in eq. (\ref{UsuaBouF}) and 
the one obtained in the computation 
$|B(F)\rangle= tr( P e^{i \oint A}) |B(F=0)\rangle$
which implies
\begin{equation}
\left.
e^{-\oh i \Phi_{(c)}(k,k)+i \pi \frac{\hat h f^2}{L} m^1 m^2}
\right|_{n=\hat  F m; m=L s}
=
\left.
e^{-\oh i \Phi_{(c)}(k,k)+i \pi \frac{ f}{L} m^1 m^2}\right|_{n=0; m=L s}
.
\end{equation}
This equation has more solutions than the one we have chosen but many
of them are not T-duality invariant and this is the reason why we have
singled out the previous one.
}.

\subsection{$N$ open tachyons and 1 closed tachyon string amplitude on the disk.}
\label{ssec:NOp1Cl}

We can now compute the amplitude given in
fig. \ref{figure:1open-1closed} which is necessary to show the proper
factorization of the non planar amplitude in the closed channel in the
next section. 

The complete amplitude for one closed tachyon and $N$ open tachyon is
given by the sum of all non cyclically equivalent permutations of
the external legs, i.e the sum of all the possible permutations $P$ of the
$N-1$ open string vertices other than ${\cal V}_{(0)T}(x_1)$
\begin{equation}
\cA(t_1,\dots t_N,t_c)
=\sum_P A(t_{1},\dots t_{P(N)},t_c)
\end{equation}
A piece\footnote{
The other pieces are obtained by moving some vertices on the
$\sigma=\pi$ boundary while keeping the cyclical ordering,
} of the partial amplitude associated with the ordering
$1,2,\dots N$  and all vertices at $\sigma=0$ is given by
\begin{eqnarray}
&&
{ A}(N T_o,1 T_c)
=
\Cz ~ \Nt~ \Nop^N
\nonumber\\
&&
~\int \frac{\prod_{r=1}^N d x_r~ d z~d \bar z}{d V_{Killing}}
\langle 0| ~{\cal V}_{(0)T}(x_1;k_1) \dots ~{\cal V}_{(0)T}(x_N;k_N)
{\cal W}_{T_c}(z,\bar z;k_L,k_R) 
|0\rangle
\nonumber\\
\label{NTo1Tc}
\end{eqnarray}
where  $\Nop$ is the
normalization of the open string vertices which is dependent of both the
open and the closed string background,
$d V_{Killing}$ is the volume of the $SL(2,\R)$ gauge invariance which
we discuss later.
We notice that the order of the open vertices w.r.t the closed one 
is not important since they commute when both the operatorial 
and the Chan-Paton parts are considered.

An easy computation gives
\begin{eqnarray}
{ A}(N T_o,1 T_c)&=&
\Cz ~ \Nt~ \Nop^N
~e^{\frac{i}{2}\Phi(k,k)}
~e^{i m^T G \frac{y_0}{\sqa}}
\nonumber\\
&&
\times
~t_{u_1}(k_N)\dots t_{u_1}(k_N)
~tr(T_{u_1}\dots T_{u_N})
~e^{i \pi  \hat h \left( f (n_1 m^1+ n_2 m^2)
- L n_1 n_2 \right) }
\nonumber\\
&&
\times
\left(\frac{1}{\sqrt{L}}\right)^{N-1}
e^{-i \pi \alpha' \sum_{r<s} k_r^T (\Theta+\Theta_{C P}) k_s}
\nonumber\\
&&
\times
~(2\pi)^{D-d}\delta(\sum_r k_{r \mu}+k_\mu)
~(2\pi\sqa)^d \delta_{\sum_r k_r+\cE^T G^{-1} k_L+\cE G^{-1} k_R,0}
\nonumber\\
&&
\times
\int \frac{\prod_{r=1}^N d x_r~ d z~d \bar z}{d V_{Killing}}
\prod_{r<s} (x_r-x_s)^{2\alpha' k_r^T \cG^{-1}  k_s}
\nonumber\\
&&
\times
\prod_r 
\left[ (x_r-z )^{2\alpha' k_L^T \cE^{-T}  k_r}
(x_r -\bar z)^{2\alpha' k_R^T \cE^{-1}  k_r}
\right]
|z-\bar z|^{2\alpha' k_R^T \cE^{-1} \cG \cE^{-1} k_L}
\nonumber\\\end{eqnarray}
where we have used the non operatorial cocycle
in the closed string vertex to write the modulus of the difference
$z-\bar z$ and momentum conservation to simplify the phase coming
from the trace of the $\Lambda$ matrices. In particular the sign 
$~e^{i \pi  \hat h \left( f (n_1 m^1+ n_2 m^2)
- L n_1 n_2 \right) }$ is due to the interaction
among the open string Chan-Paton factors and the would-be closed one.
\COMMENTOm{
When considering amplitudes with more than 1 closed string state then
the would-be closed string Chan-Paton give not only a momentum
dependent sign but a $\Z_N$ momentum dependent phase......
}
We need now the full amplitude since we later want to compare it
with the result of the factorization of the non planar amplitude.
Very roughly  it is convenient to proceed
as follows to obtain the full amplitude (details are given in appendix). 
We fix the $SL(2,\R)$ gauge invariance so that
$z=i$ and $x_N=0$.  Then we can change variable of the integral
to $w= \frac{z-i}{z+i}$ so that the point $z=i$ is mapped into $w=0$
and real axis gets mapped into the circle $|w|=1$. 
In particular for the real number we have $w= e^{i \phi}$ 
with $\phi=\arccos \frac{x}{\sqrt{1+x^2}}$
so that the positive real axis gets mapped into the lower semicircle
in clockwise direction. 
Differently from the gauge fixing we use with pure open string
amplitudes we have only fixed one open string at $x_N=0$ therefore we
must also consider the partial amplitudes with open string emitted on the
$\sigma=\pi$ boundary. 
For example given the same ordering on the unit circle $|w|=1$ as in
eq. (\ref{NTo1Tc}) we have to consider the amplitudes 
\begin{eqnarray}
\Cz ~ \Nt~ \Nop^N
&&
~\int \frac{\prod_{r=1}^{l-1} d y_r~\prod_{r=l}^N d x_r~ d z~d \bar z}{d V_{Killing}}
\nonumber\\
&&\langle 0| R\bigg[
~{\cal V}_{(0)T}(x_k;k_k) \dots ~{\cal V}_{(0)T}(x_l;k_l)
\nonumber\\
&&~~~~
~{\cal V}_{(0)T}(y_{k-1};k_{k-1}) \dots ~{\cal V}_{(0)T}(y_{1};k_{1})
\nonumber\\
&&~~~~
~{\cal V}_{(0)T}(y_N;k_N) \dots ~{\cal V}_{(0)T}(y_{l+1};k_{l+1})
~{\cal W}_{T_c}(z,\bar z;k_L,k_R)
\bigg] 
|0\rangle
\nonumber\\
\end{eqnarray}
with $x_k>\dots> x_l$ and $|y_{k-1}|> \dots |y_1|>|y_{N}|>\dots |y_{l+1}|$
for all $l$s and $k$s ($k<l$) and where $R$ is the radial ordering.
Because the vertices on the two boundaries commute summing over all
the possible $l$s and $k$s and possible radial orderings  amounts to 
integrate over the whole $|w|=1$ circle with all $0<\phi<2\pi$ and 
$\phi_i > \phi_{i+1}$. Details on how the different pieces join
together are given in appendix (\ref{app:NOpen1Closed}): it is
nevertheless noteworthy that the whole procedure works because of some
phase contributed by the would-be closed Chan-Paton factors.

Summing all the previous amplitudes allows 
to extend the $\phi_r$ integrations to the
full range $[0,2\pi]$ we can therefore write
\begin{eqnarray}
{\cal A}^{same~ordering}(N T_o,1 T_c)&=&
\Cz ~ \Nt~ \Nop^N
~e^{\frac{i}{2}\Phi(k,k)}
~e^{i m^T G \frac{y_0}{\sqa}}
\nonumber\\
&&
\times
~t_{u_1}(k_1)\dots t_{u_N}(k_N)
~tr(T_{u_1}\dots T_{u_N})
~e^{i \pi  \hat h \left( f (n_1 m^1+ n_2 m^2)
- L n_1 n_2 \right) }
\nonumber\\
&&
\times
\left(\frac{1}{\sqrt{L}}\right)^{N-2}
e^{-i \pi \alpha' \sum_{r<s} k_r^T (\Theta+\Theta_{C P}) k_s}
\nonumber\\
&&
\times
~(2\pi)^{D-d}\delta(\sum_r k_{r \mu}+k_\mu)
~(2\pi\sqa)^d \delta_{\sum_r k_r+\cE^T G^{-1} k_L+\cE G^{-1} k_R,0}
\nonumber\\
&&
\times
\int_0^{2\pi} \prod_{r=1}^{N-1} d \phi_r ~\theta(\phi_r-\phi_{r+1})
~\prod_{1\le r<s \le N} (2\sin \frac{\phi_r-\phi_s}{2})^{2\alpha' k_r^T \cG^{-1}  k_s}
\nonumber\\
&&
\times
\prod_{r=1}^{N-1} e^{i \phi_r ~\alpha' k_r^T ( \cE^{-1} k_L - \cE^{-T} k_R) } 
\label{NOp1ClFin}
\end{eqnarray}
with $\phi_N=0$.

\subsection{The 1 loop non planar amplitude.}
\label{ssec:1loopNP}
Finally we want to compute the non planar 1 loop amplitude with $N_0$ tachyons
on one border and $N_\pi$ ones on the other.
The full amplitude is given by the sum over all cyclically
nonequivalent amplitudes. 
This can be also be written as
\begin{equation}
\cA \left( t_1,\dots t_{N_0}; t_{N_0+1} t_{N_0+N_\pi} \right)
= 
\sum_{P_0} \sum_{P_\pi} 
\cA ( P_0(1),\dots P_0(N_0); P_\pi(N_0+1), \dots P_\pi(N_0+N_\pi) )
\label{Full1LoopNP}
\end{equation}
where $P_0$ is any of the permutations of vertices at $\sigma=0$,
$P_\pi$ is any of the permutations of vertices at $\sigma=\pi$ which keeps
the index $N_0+N_\pi$ fixed and 
\begin{eqnarray}
&&\cA ( P_0(1),\dots P_0(N_0); P_\pi(N_0+1), \dots P_\pi(N_0+N_\pi) )
= 
\nonumber\\
&&=
\hat\cA ( P_0(1),\dots P_0(N_0),P_\pi(N_0+1),\dots P_\pi(N_0+N_\pi))
\nonumber\\
&&~
+
\hat\cA ( P_0(1),\dots N_0+1,P_0(N_0),\dots P_\pi(N_0+N_\pi))
+
\nonumber\\
&&~
\dots
+\hat\cA ( P_\pi(N_0+1),\dots P_\pi(N_0+N_\pi-1),P_0(1),\dots P_0(N_0),N_0+N_\pi)
\label{PartialFull1LoopNP}
\end{eqnarray}
is the sum over the permutations $Q$ 
of vertices at $\sigma=0$ relative to those at $\sigma=\pi$
which keep fixed the ordering of vertices on both boundaries.

Let us consider the orderings $1\dots N_0$ at $\sigma=0$ boundary and
$N_0+1 \dots N_0+N_\pi$ at $\sigma=\pi$.
For these orderings
we compute first the amplitude corresponding to the simplest relative
ordering on the two boundaries and then we discuss the effect of
summing over all the other relative orderings obtained by applying any
permutation $Q$. 
We use the old formalism and therefore we compute
\begin{eqnarray}
\hat\cA
&&
\hspace{-1em}
( 1,\dots N_0,N_0+1,\dots N_0+N_\pi)
\nonumber\\
&&
\hspace{-2em}
=
\Cu ~\Nop^{ N_0+N_\pi}
~Tr\Big(
\Delta~ \cV_{(0)T}(1;k_1)\dots \cV_{(0)T}(1;k_{N_0})
\Delta \cV_{(\pi)T}(-1;k_{N_0+1})\dots \cV_{(\pi)T}(-1;k_{N_0+N_\pi})
\Big)
\nonumber\\
&&
\hspace{-2em}
=
\Cu ~\Nop^{ N_0+N_\pi}
~t_{u_1}(k_1)\dots t_{u_{N_0+N_\pi}}(k_{N_0+N_\pi})
\nonumber\\
&&
\hspace{-2em}
\times
~\int \frac{d^{D-d} k_\mu}{(2\pi)^{D-d}}
~\frac{1}{(2\pi\sqa)^d}
\sum_{n_1,\dots n_d}
~\int_0^1 d x_1\dots\int_0^1 d x_{ N_0+N_\pi}
\nonumber\\
&&
\hspace{-2em}
\times
tr\left(\Lambda^\dagger(\frac{L^{-1} n}{\sqa}) 
\Lambda(k_1)\dots \Lambda(k_{N_0})
\Lambda(\frac{L^{-1} n}{\sqa})
\Lambda(k_{ N_0+N_\pi})\dots \Lambda(k_{N_0+1})
\right)
~ 
\nonumber\\
&&
\hspace{-2em}
\times
tr\left( T_{u_1}\dots T_{u_{N_0+1}}\right)
~tr\left( T_{u_{N_0+N_\pi}}\dots T_{u_{N_0+1}}\right)
\nonumber\\
&&
\hspace{-2em}
\times
\langle k_\mu,\frac{(L^{-1} n)_i}{\sqa}|
~Tr_{n z m}\Big(
x_1^{L_0-2} 
~:e^{i k_{1 M} X^M(1)}:
\dots
\nonumber\\
&&
\hspace{4em}
x_{ N_0+N_\pi}^{L_0-2} 
e^{i \pi \alpha' k_{ N_0+N_\pi N} \cG^{N M} k_{ N_0+N_\pi M}}
~:e^{i k_{ N_0+N_\pi M} X^M(-1)}:
|k_\mu,\frac{(L^{-1} n)_i}{\sqa}\rangle
\end{eqnarray}
where $\Delta=(L_0-1)^{-1}$ is the open string propator. 
With a straightforward computation, after including the ghost contribution
and  performing a Poisson resummation on the discrete momenta 
we find the following result
\begin{eqnarray}
\hat \cA&&\hspace{-2em}(1,\dots N_0, N_0+1,\dots N_0+N_\pi)
=
\nonumber\\
&=&
\Cu ~(\Nop)^{ N_0+N_\pi} 
~\left[ \det G_{\mu\nu} ~  \det(L \cG_{} L)_{i j}\right]^{\oh}
\nonumber\\
&&
~t_{1 u_1}(k_1)\dots t_{N_0+N_\pi u_{N_0+N_\pi}}(k_{N_0+N_\pi})
~\left( \frac{1}{\sqrt{L}} \right)^{N_0+N_\pi}
~tr\left( T_{u_1}\dots T_{u_{N_0+1}}\right)
~tr\left( T_{u_{N_0+N_\pi}}\dots T_{u_{N_0+1}}\right)
\nonumber\\
&&
~\delta^{D-d}\left(\sqa \sum k_{r \mu}\right)
~\delta_{\sum k_{r i}}
\nonumber\\
&&
\prod_{1\le r < s \le N_0} 
e^{-i \pi \alpha' k_{r i}  \Theta_{tot}^{i    j} k_{s j}}
~
\prod_{N_0+1\le r < s \le N_0+N_\pi} 
e^{+i \pi \alpha' k_{r i} \Theta_{tot}^{i    j} k_{s j}}
\nonumber\\
&&
\int_0^1 \frac{d w}{w^2}
~
\int_0^1 \prod_{r=1}^{N_o+N_\pi-1} \frac{d \rho_r}{\rho_r}
\theta(\rho_r -\rho_{r+1})
~\left[ \frac{-\ln w}{\pi} \right]^{-D/2}
~\left[ \frac{1}{\prod_{n=1}^\infty (1 -w^n)} \right]^{D-2}
\nonumber\\
&&
\prod_{1\le r < s\le  N_0} 
e^{2\alpha' k_{r M} \cG^{M N} k_{s N}~ \ln \psi_{r s} }
~
\prod_{N_0+1\le r < s \le N_0+N_\pi} 
e^{2\alpha' k_{r M} \cG^{M N} k_{s N}~  \ln \psi_{r s} }
~
\nonumber\\
&&
\prod_{1\le r \le N_0< s \le N_0+N_\pi} e^{2\alpha' k_{r M} \cG^{M N}
  k_{s N}~ \ln \psi^T_{r s} }
\nonumber\\
&&
\sum_{(m_0^i)\in \Z^d} 
e^{\frac{\pi^2}{\ln w} 
~\alpha'
\left( \frac {L m_0}{\sqa} + \Theta_{tot} \sum_{r=1}^{N_0} k_{r i} \right)^i 
\cG_{i j} 
\left( \frac {L m_0}{\sqa} + \Theta_{tot} \sum_{s=1}^{N_0} k_{s i} \right)^j
}
~e^{ -i 2\pi \alpha' \sum_{r=1}^{N_0+N_\pi} \frac{\ln \rho_r}{\ln w}~ k_{r i} 
\left( \frac {L m_0}{\sqa} + \Theta_{tot} \sum_{s=1}^{N_0} k_{s } \right)^i
 }
\nonumber\\
\label{Final-1-loop}
\end{eqnarray}
where we have defined  the new integration variables
$\rho_r= x_1 \dots x_r$, $w=\rho_{N_0+N_\pi}$ 
and, for $s>r$, the ratios $c_{s r}= \rho_s/ \rho_r= x_{r+1} \dots
x_s$.
We have also defined the quantities $\psi_{r s}=\psi(c_{s r},w)$
given by exponential of the annulus propagators as
\begin{eqnarray}
\psi(c,w)
&=& 
c^{-\oh} 
\exp{ \left(\frac{(\ln c)^2}{2 \ln w} \right) }
\exp{ \left(- 
\sum_{n=1}^\infty \frac{c^m + ({w}/{c})^m - 2    w^m }{m (1-w^n)}  \right)}
\nonumber\\
\psi^T(c,w)
&=& 
c^{-\oh} 
\exp{\left( \frac{(\ln c)^2}{2 \ln w} \right)}
\exp{\left(
- \sum_{n=1}^\infty \frac{(-c)^m + (-{w}/{c})^m - 2 w^m }{m (1-w^n)} 
\right)}
\end{eqnarray}

In eq. (\ref{Final-1-loop})  the contribution $~\left[ \det G_{\mu\nu}
  ~  \det(L \cG_{} L)_{i j}\right]^{\oh}$ in the first line comes from
Poisson resummation and is fundamental in fixing the normalization of
the amplitudes as we discuss later.

If we now compute any other partial amplitude with a different
relative ordering of the vertices on the two boundary we get almost the same
result as before because the vertices on the boundaries commute: the
only difference is given by the relative ordering of the $\rho$ of the
vertices on the two boundaries.
Therefore when we sum over all possible relative ordering with $|y_{N_0+N_\pi}|$
less than all the other $y$ and $x$ as shown in eq. (\ref{Full1LoopNP})
we get the same result
as above where the ordering of the $\rho$ on the two boundaries are
independent, explicitly
\begin{eqnarray}
\int_0^1 \prod_{r=1}^{N_o+N_\pi-1} \frac{d \rho_r}{\rho_r}
\theta(\rho_r -\rho_{r+1})
\Rightarrow
&&
\int_0^1 \prod_{r=1}^{N_o-1} \frac{d \rho_r}{\rho_r}
\theta(\rho_r -\rho_{r+1})
\frac{d \rho_{N_0}}{\rho_{N_0}}
\theta(\rho_r -\rho_{N_0+N_\pi})
\nonumber\\
&&
~\times
\int_0^1 \prod_{r=N_0+1}^{N_o+N_\pi-1} \frac{d \rho_r}{\rho_r}
\theta(\rho_r -\rho_{r+1})
\end{eqnarray}
since the vertex located in $\rho_{N_0+N_\pi}=w$ is always  the last  
and we keep the ordering on the two boundaries fixed.

We can now perform a modular transformation in order to express the
amplitude using the closed string
variables $q$ and $\nu_r$ ($r=1,\dots N_0+N_\pi-1$) defined as
\begin{eqnarray}
\ln q = \frac{2 \pi^2}{ \ln w},~~~~
\nu_r = \frac{\ln \rho_r}{ \ln w}
\label{ClOpvariables}
\end{eqnarray}
so that the quantities which enter the amplitude become
\begin{eqnarray}
\psi(\nu,q)
&=&
\psi(\rho,w)
=
\frac{2\pi}{-\ln q}
~\sin(\pi\nu)
~\prod_{n=1}^\infty
\frac{(1-e^{i 2\pi \nu} q^{2 n}) (1-e^{-i 2\pi \nu} q^{2 n})
}{
(1-q^{2 n})^2
}
\nonumber\\
\psi^T(\nu,q)
&=&
\psi^T(\rho,w)
=
\frac{\pi}{-\ln q}
~q^{-1/4}
~\prod_{n=1}^\infty
\frac{(1-e^{i 2\pi \nu} q^{2 n-1}) (1-e^{-i 2\pi \nu} q^{2 n-1})
}{
(1-q^{2 n})^2
}
\nonumber\\
\prod_{n=1}^\infty (1 -w^n)
&=&
e^{-\frac{\pi^2}{12 \log q} }
~\left( \frac{-\log q}{\pi} \right)^\oh
~q^{\frac{1}{12}}
\prod_{n=1}^\infty (1 -q^n)
\label{ModTransf}
\end{eqnarray}
The partial amplitude (\ref{PartialFull1LoopNP}) 
obtained by summing over the subclass
of permutations $Q$ can be rewritten for $D=26$ as\footnote{
To make formlae more compact we define $m^M=\Theta^{M N}=0$ when
$M,N\ne i,j$.
}
\begin{eqnarray}
\cA&&\hspace{-2em}(1,\dots N_0; N_0+1,\dots N_0+N_\pi)
=
\nonumber\\
&=&
2^{1-\frac{D}{2}} (2\pi)^{N_0+N_\pi-1}~\Cu ~(\Nop)^{ N_0+N_\pi} 
~\left[ \det G_{\mu\nu} ~  \det(L \cG_{} L)_{i j}\right]^{\oh}
\nonumber\\
&&
~t_{1 u_1}(k_1)\dots t_{N_0+N_\pi u_{N_0+N_\pi}}(k_{N_0+N_\pi})
~\left( \frac{1}{\sqrt{L}} \right)^{N_0+N_\pi}
~tr\left( T_{u_1}\dots T_{u_{N_0+1}}\right)
~tr\left( T_{u_{N_0+N_\pi}}\dots T_{u_{N_0+1}}\right)
\nonumber\\
&&
~\delta^{D-d}\left(\sqa \sum k_{r \mu}\right)
~\delta_{\sum k_{r i}}
\nonumber\\
&&
\prod_{1\le r < s \le N_0} 
e^{-i \pi \alpha' k_{r i}  \Theta_{tot}^{i    j} k_{s j}}
~
\prod_{N_0+1\le r < s \le N_0+N_\pi} 
e^{+i \pi \alpha' k_{r i} \Theta_{tot}^{i    j} k_{s j}}
\nonumber\\
&&
\sum_{(m_0^i)\in \Z^d}
\int_0^1 dq~ 
q^{-3
   + \oh \alpha' \left\{
   (\sum_{r=1}^{N_0} k_r)^T \cG^{-1} (\sum_{s=1}^{N_0} k_s) 
  +\left( \frac {L m_0}{\sqa} + \Theta_{tot} \sum_{r=1}^{N_0} k_{r} \right)^T 
   \cG
   \left( \frac {L m_0}{\sqa} + \Theta_{tot} \sum_{s=1}^{N_0} k_{s} \right)
\right\}
}
\nonumber\\
&&
~
\left[ \prod (1-q^{2 n})^2\right] ^{-24 + 2 (N_0+N_\pi)}
\nonumber\\
%
%
&&
\int_0^1 \prod_{r=1}^{N_0-1} d \nu_r
~\theta(\nu_{r+1}-\nu_r)
~d \nu_{N_0}
~\prod_{r=1}^{N_0}
e^{ -i 2\pi \nu_r \alpha' ~ k_{r }^T
\left( \frac {L m_0}{\sqa} + \Theta_{tot} \sum_{s=1}^{N_0} k_{s } \right)
 }
\nonumber\\
%
%
&&
\hspace{4em}
\prod_{1\le r < s \le N_0} \left[ 2 \sin \pi\nu_{s r} 
~\prod_{n=1}^\infty
\frac{(1-e^{i 2\pi \nu_{s r}} q^{2 n}) (1-e^{-i 2\pi \nu_{s r}} q^{2 n})
}{
(1-q^{2 n})^2
}
\right]^{2\alpha' k_{r} \cG^{-1} k_{s }}
\nonumber\\
%
%
&&
\int_0^1 \prod_{r=N_0+1}^{N_0+N_\pi-1} d \nu_r
~\theta(\nu_{r+1}-\nu_r)
~
\prod_{r=N_0+1}^{N_0+N_\pi}
e^{ -i 2\pi \nu_r \alpha' ~ k_{r }^T
\left( \frac {L m_0}{\sqa} + \Theta_{tot} \sum_{s=1}^{N_0} k_{s } \right)
 }
\nonumber\\
&&
\hspace{4em}
\prod_{1\le r < s \le N_0} \left[ 2 \sin \pi\nu_{s r} 
~\prod_{n=1}^\infty
\frac{(1-e^{i 2\pi \nu_{s r}} q^{2 n}) (1-e^{-i 2\pi \nu_{s r}} q^{2 n})
}{
(1-q^{2 n})^2
}
\right]^{2\alpha' k_{r} \cG^{-1} k_{s }}
\nonumber\\
&&
\prod_{1\le r \le N_0< s} \left[ 
~\prod_{n=1}^\infty
\frac{(1-e^{i 2\pi \nu_{s r}} q^{2 n-1}) (1-e^{-i 2\pi \nu_{s r}} q^{2 n-1})
}{
(1-q^{2 n})^2
}
\right]^{2\alpha' k_{r} \cG^{-1} k_{s }}
\label{DualFinal-1-loop}
\end{eqnarray}
where we have set $\nu_{N_0+N_\pi}=1$.

Summing over all the cyclically equivalent configurations on the
$\sigma=0$ boundary\footnote{
These permutations are a subset of the $P_0$ permutations which are
non trivial since the $P_0$ permutations are not required to keep
$N_0$ fixed.
} and redefining in a proper way the integration
variables as explained in appendix \ref{app:NONPlanar} we finally get
\begin{eqnarray}
\sum_{k=1}^{N_0}&&\hspace{-2em}
 \cA(k,\dots N_0,1,\dots k-1; N_0+1,\dots N_0+N_\pi)
=
\nonumber\\
&=&
2\pi~2^{1-\frac{D}{2}} 
~\Cu ~(\Nop)^{ N_0+N_\pi} 
~\left[ \det G_{\mu\nu} ~  \det(L \cG_{} L)_{i j}\right]^{\oh}
\nonumber\\
&&
~t_{1 u_1}(k_1)\dots t_{N_0+N_\pi u_{N_0+N_\pi}}(k_{N_0+N_\pi})
~\left( \frac{1}{\sqrt{L}} \right)^{N_0+N_\pi}
~tr\left( T_{u_1}\dots T_{u_{N_0+1}}\right)
~tr\left( T_{u_{N_0+N_\pi}}\dots T_{u_{N_0+1}}\right)
\nonumber\\
&&
~\delta^{D-d}\left(\sqa \sum k_{r \mu}\right)
~\delta_{\sum k_{r i}}
\nonumber\\
&&
\prod_{1\le r < s \le N_0} 
e^{-i \pi \alpha' k_{r i}  \Theta_{tot}^{i    j} k_{s j}}
~
\prod_{N_0+1\le r < s \le N_0+N_\pi} 
e^{+i \pi \alpha' k_{r i} \Theta_{tot}^{i    j} k_{s j}}
\nonumber\\
&&
\sum_{(m_0^i)\in \Z^d}
\int_0^1 dq~ 
q^{-3
   + \oh \alpha' \left\{
   (\sum_{r=1}^{N_0} k_r)^T \cG^{-1} (\sum_{s=1}^{N_0} k_s) 
  +\left( \frac {L m_0}{\sqa} + \Theta_{tot} \sum_{r=1}^{N_0} k_{r} \right)^T 
   \cG
   \left( \frac {L m_0}{\sqa} + \Theta_{tot} \sum_{s=1}^{N_0} k_{s} \right)
\right\}
}
\nonumber\\
&&
~
\left[ \prod (1-q^{2 n})^2\right] ^{-24 + 2 (N_0+N_\pi)}
\nonumber\\
%
%
&&
~\int_0^1 d \nu_{N_0}
~e^{ -i 2\pi \nu_{N_0} ~\alpha' \sum_{r=1}^{N_0} k_{r}^T \frac{L m_0}{\sqa} }
\nonumber\\
&&
~\int_0^{2\pi} \prod_{r=1}^{N_0-1} d \phi_r
~\theta(\phi_{r}-\phi_{r+1})
~\prod_{r=1}^{N_0}
e^{ +i  \phi_r \alpha' ~ k_{r }^T
\left( \frac {L m_0}{\sqa} + \Theta_{tot} \sum_{s=1}^{N_0} k_{s } \right)
 }
\nonumber\\
%
%
&&
\hspace{4em}
\prod_{1\le r < s \le N_0} 
\left[ 2 \sin \frac {\phi_{r s}}{2} 
~\prod_{n=1}^\infty
\frac{(1-e^{i  \phi_{r s}} q^{2 n}) (1-e^{-i  \phi_{r s}} q^{2 n})
}{
(1-q^{2 n})^2
}
\right]^{2\alpha' k_{r} \cG^{-1} k_{s }}
\nonumber\\
%
%
&&
\int_0^{2\pi} \prod_{r=N_0+1}^{N_0+N_\pi-1} d \phi_r
~\theta(\phi_{r+1}-\phi_{r})
~
\prod_{r=N_0+1}^{N_0+N_\pi}
e^{ - i \phi_r \alpha' ~ k_{r }^T
\left( \frac {L m_0}{\sqa} + \Theta_{tot} \sum_{s=1}^{N_0} k_{s } \right)
 }
\nonumber\\
&&
\hspace{4em}
\prod_{N_0+1\le r < s \le N_0+N_\pi} 
\left[ 2 \sin \frac {\phi_{s r}}{2} 
~\prod_{n=1}^\infty
\frac{(1-e^{i \phi_{s r}} q^{2 n}) (1-e^{-i \phi_{s r}} q^{2 n})
}{
(1-q^{2 n})^2
}
\right]^{2\alpha' k_{r} \cG^{-1} k_{s }}
\nonumber\\
&&
\prod_{1\le r \le N_0< s} \left[ 
~\prod_{n=1}^\infty
\frac{(1-e^{i ( \phi_{r}+\phi_r - 2 \pi \nu_{N_0})} q^{2 n-1}) 
(1-e^{-i ( \phi_{s} + \phi_r - 2 \pi \nu_{N_0}) } q^{2 n-1})
}{
(1-q^{2 n})^2
}
\right]^{2\alpha' k_{r} \cG^{-1} k_{s }}
\label{SumDualFinal-1-loop}
\end{eqnarray}
with $\phi_{N_0}=0$, $\phi_{N_0+N_\pi}=2\pi$.

We can now expand in powers of $q$, 
integrate over $\nu_{N_0}$, shift 
$\bar \phi_r=\phi_r+2\pi-\phi_{N_0+1}$ for $r> N_0$ so that $\bar
\phi_{N_0+1}=0$
and then we can compare with the $N$ open tachyons - 1
closed tachyon amplitude. 
If we consider in particular the terms 
$e^{i \phi_r \alpha' ~ k_{r }^T
\left( \frac {L m_0}{\sqa} + \Theta_{tot} \sum_{s=1}^{N_0} k_{s } \right)
 }$ we must identify
\begin{equation}
 \left(\frac {L m_0}{\sqa} + \Theta_{tot} \sum_{r=1}^{N_0} k_{r} \right)
= ( \cE^{-1} k_L - \cE^{-T} k_R)
\end{equation}
which implies
\begin{equation}
m_0= \hat h ~\epsilon n + \tilde f m
\end{equation}
so that we have
\begin{eqnarray}
\sum_{k=1}^{N_0}&&\hspace{-2em}
 \cA(k,\dots N_0,1,\dots k-1; N_0+1,\dots N_0+N_\pi)
\sim
\nonumber\\
&&
\int \frac{d^{D-d}k_C}{ (2\pi)^{D-d} ~(2\pi\sqa)^d}
\sum_{n_C}
~\cA(1,\dots N_0; C) ~
\cA(N_0+N_\pi,\dots N_0+1; -C)
~\frac{1}{k_C^T \cG^{-1} k_C -\frac{4}{\alpha'}}
\nonumber\\
\end{eqnarray}
where $C$ stands for the closed string tachyon appearing in the mixed amplitude
(\ref{NOp1ClFin}), $-C$ the closed string tachyon with opposite momentum
and 
the disk amplitude associated to the $\sigma=\pi$ boundary is
run in the opposite direction w.r.t. the one associated to the
$\sigma=0$ boundary as a simple picture shows it is the case.

Making the previous equation more precise and comparing the overall coefficients we can then write
\begin{equation}
(\Cz ~\Nop)^2
~(2\pi\sqa)^{D}
=
2\pi~2^{1-\frac{D}{2}}
~\cC_1
~\left[ \det G_{\mu\nu} ~  \det \cG_{i j}\right]^{\oh}
~\frac{2}{\alpha'}
\label{NormC0Nt0}
\end{equation}
which matches the result from the annulus given in appendix
(\ref{app:annulus}) and together the sewing relations
(\cite{DiVecchia:1995iy} \footnote{
In \cite{DiVecchia:1995iy} the open string normalization was $\Cz
~\Nop^2 ~\alpha'$ because the gauge generators were normalized as
$tr(T_u T_v)= \oh \delta_{u v}$ while here they are normalized as
$tr(T_u T_v)= \delta_{u v}$.
})
\begin{equation}
\Cz ~\Nop^2 ~\alpha'= \tilde{\cal C}_0(E) ~\Nt^2 ~\frac{\alpha'}{2}=1
\end{equation}
allow to fix the normalization.
\section{ T-duality action on vertices.}
\label{sect:T-duality}
In this section we would like to show that all the previous amplitudes
for a theory of $N=N_1 L$ wrapped $D25$ branes with magnetic field
$\hF_{12}=\frac{f}{L}$  which breaks the gauge group to $U(N_1)$ 
can be obtained by a T-duality transformation of the 
amplitudes for a theory of $N_1$ $D25$ branes with vanishing magnetic field.

In doing so we prove the equivalence of the two theories also when
gravitational interactions are considered.
This happens because the phases depend on momenta only and we can use
sewing techniques to argue that all amplitudes are invariant
once we we have shown that pure open string amplitudes and mixed amplitudes
with one closed string are invariant.

We can always choose the abelian field strength block diagonal
in space-time and, because of that, in the previous sections 
we have considered the case
where the magnetic field is turned on only in two directions within a
non factorized torus.  
We consider therefore only T-duality transformations which act on
$x^1$ and $x^2$ whose non trivial part of the $\Lambda$ matrix is
given by
\begin{eqnarray}
\Lambda &\supset& \left( \begin{array}{cc} 
\tilde f \uno_2 & \hat h \epsilon \\ 
-f\epsilon & L \uno_2
\end{array}\right) \in O(2,2,\Z)
~~~~
L \tilde f- f \hat h =1
~~~~
f\tilde f \in 2 \Z
\end{eqnarray}
This T-duality transformation acts on zero modes as
\begin{eqnarray}
 \left( \begin{array}{c} 
m^1 \\ n_2
\end{array}
\right)
=
 \left( \begin{array}{cc} 
L  & -\hat h  \\ 
-f & \tilde f
\end{array}\right)
 \left( \begin{array}{c} 
m^{t 1} \\ n_2^t
\end{array}
\right)
~~~~
 \left( \begin{array}{c} 
m^2 \\ n_1
\end{array}
\right)
=
 \left( \begin{array}{cc} 
L  & \hat h  \\ 
f & \tilde f
\end{array}\right)
 \left( \begin{array}{c} 
m^{t 2} \\ n_1^t
\end{array}
\right)
\label{nmTrans}
\end{eqnarray}
The previous T-duality can be understood as the result of a T-duality,
followed by a rotation and then by another T-duality as shown in
eq. (\ref{Gura-Rango})
therefore the closed string vertices must be
transformed in order to have invariant amplitudes 
as shown in (\cite{Hellerman:2006tx}, \cite{Duo:2007he}) as
\begin{eqnarray}
\cW^{(c)}_{\beta_L, \beta_R}(n,m;[M_{c 0}])
\rightarrow
e^{i\pi m^2 n_2}
~e^{i\pi m^{t 2} n_2^t}
~\cW^{(c)}_{\beta_L^t, \beta_R^t}(n^t,m^t;[M_{c 0}])
\label{TdualTrasnClo}
\end{eqnarray}
where we have written the explicit dependence on the cocycle through
the equivalence class of the matrix $M_{c 0}$ (\ref{Mc0}).
It is then possible to check that the phase $e^{i \Phi_{(c)}(k,l)}$ 
in eq. (\ref{OPE-closed}) is the same of the one obtained by computing
the product
\begin{equation}
e^{i\pi m^2_k n_{k 2}}
~e^{i\pi m^{t 2}_k n_{2 k}^t}
{\cal W}^{(c)}_{\beta_L^t,\beta_R^t}( k^t_{L },k^t_{R }; [M_{c 0}])
\times
e^{i\pi m^2_l n_{l 2}}
~e^{i\pi m^{t 2}_l n_{2 l}^t}
{\cal W}^{(c)}_{\alpha_L^t,\alpha_R^t}(l^t_{L},l^t_{R }; [M_{c 0}])
\end{equation}
upon the use of the transformations (\ref{nmTrans}).
Notice that the same result holds if we perform the T-dualities along
the $x$-axes.

We can now consider what happens in the open sector.
In \cite{DiVecchia:2007dh} we discussed how the T-duality
transformations can be implemented in open string formalism.
The starting point was to impose the same transformations  in
eq. (\ref{TdualTrasnClo}) on the closed string vertices 
written in open string formalism, i.e.
\begin{eqnarray}
\cW_{\beta_L, \beta_R}(n,m;[M_{c 0}])
\rightarrow
e^{i\pi m^2 n_2}
~e^{i\pi m^{t 2} n_2^t}
~\cW_{\beta_L^t, \beta_R^t}(n^t,m^t;[M_{c 0}])
\label{TdualTrasnOpe}
\end{eqnarray}
where the closed string vertices are the generalization of the one in
eq. (\ref{ClosedStringVertex}) to an arbitrary closed string state
$(\beta_L, \beta_R)$.
In our particular case
the first vertex has winding dependent ``Chan-Paton'' factors 
while the second has not (since we have only a stack of branes with
equal Wilson lines).
The results of the discussion in \cite{DiVecchia:2007dh} 
can be summarized in the following transformation rules
for the open string quantities
\begin{eqnarray}
k^t= T^{-T}(F)~k
&~~~~&
\theta= T^T(F)~\theta^t
\nonumber\\
\cG= T^T(F)~\cG^t~ T(F)
&~~~~&
\Theta^t=T(F)~\Theta~T^T(F)+ \B T^T(F)
\nonumber\\
\hat X_{L(0)}(z)=\hat X_{L(0)}^t(z)
&~~~~&
\hat X_{R(0)}(\bar z)=\hat X_{R(0)}^t(\bar z)
\end{eqnarray}
where we have defined $T(F)=\A+\B \hat F$.
In the case at hand where $\hat F^t=0$ it was shown that
\begin{equation}
T(F)=\D^{-T}= L^{-1} \uno
\end{equation}

We are now in the position of showing that the product of two open string
vertices in eq. (\ref{ProdOpVer},\ref{OpenOPEsigma=pi}) is invariant. 
This amounts to show that
\begin{eqnarray}
k^T \cG^{-1} l
&=&
k^{t T} \cG^{t -1} l^t
\nonumber\\
e^{-i \pi \alpha' k^T (\Theta + \Theta_{CP} ) l}
&=&
e^{-i \pi \alpha' k^{t T} \Theta^t  l^t}
\end{eqnarray}
This first equation is trivially satisfied. 
We can now use the explicit expression for $\B=\hat h \epsilon$ 
and $\Theta_{CP}= L \hat h \epsilon$ to explicitly show that 
the second one holds:
\begin{equation}
 \Theta^t
=
T~(\Theta + T^{-1} \hat h \epsilon)~T^T
=
T~(\Theta + \Theta_{CP} )~T^T
\end{equation}

Next we can consider the $N$ open tachyons - $1$ closed tachyon
amplitude given in eq. (\ref{NOp1ClFin}).  
To show that it is invariant we notice that the previous closed string vertex
transformation (\ref{TdualTrasnOpe}) can be written as
\begin{eqnarray}
\cW_{\beta_L, \beta_R}(n,m;[M_{c 0}])
\rightarrow
e^{i\pi \hat h f(m^1 n_1+m^2 n_2)}
~e^{i\pi \hat h L n_1 n_2}
~\cW_{\beta_L, \beta_R}(n^t,m^t;[M_{c 0}])
\end{eqnarray}
The sign in this expression reproduces exactly the one from the
would-be closed string Chan-Paton in the second line of eq. (\ref{NOp1ClFin}). 
Momentum conservation and all products of two momenta can be easily
checked to be invariant. 
It is then immediate to check that this amplitude is T-duality invariant
when the normalization satisfies
\begin{equation}
\Cz ~\Nop^N ~\Nt=
\cC_0(E^t,F^t) ~\cN_0(E^t,F^t)^N ~\tilde\cN_0(E^t)
\end{equation} 
but this is was verified in \cite{DiVecchia:2007dh}.



\appendix
\section{Conventions.}
\label{app-conventions}
\begin{itemize}
\item Indices:\\
Compact $i,j,\dots= 1,\dots d$;
non compact $\mu,\nu,\dots= 0,d+1\dots D$;
general $M,N,\dots=0,\dots D$;
\\
Color $a,b,\dots$;

\item $\delta^{[N]}_{m,n}$ means $m\equiv n~~~mod~N$;

\item Given a matrix $Q$, we use $[Q]_S$, $[Q]_A$ to mean respectively
  the symmetric and the antisymmetric part and $[Q]_>$ to denote the
  upper diagonal part, i.e. the matrix where we set $Q_{i j}=0$ when
  $i<j$, and similarly for $[Q]_<$;

\item 't Hooft matrices $P_N$ and $Q_N$: 
$Q_N P_N= e^{- 2\pi i \frac{1}{N}}P_N Q_N$.

\item Background matrices:\\
\begin{eqnarray}
E&=& \parallel E_{i j} \parallel = G+B
\nonumber\\
{\cal E} &=& \parallel \cE_{i j} \parallel=E^T + 2\pi \alpha' q_0 F
= G -{\cal B}
\end{eqnarray}
and
\begin{eqnarray}
\hat F &=& 2\pi \alpha' q_0 F
\nonumber\\
{\cal B} &=& B - 2\pi \alpha' q_0 F = B -\hat F
\nonumber\\
{\cal E}^{-1} &=& {\cal G}^{-1} -\Theta
\end{eqnarray}
from which we deduce that
\begin{eqnarray}
&&{\cal E}{\cal G}^{-1} {\cal E}^T
={\cal E}^T{\cal G}^{-1} {\cal  E}
=G
\nonumber\\
&&
\Theta= \frac{1}{2}\left({\cal E}^{-T}-{\cal E}^{-1} \right)
=- {\cal E}^{-1} {\cal B} {\cal E}^{-T}
\end{eqnarray}
Moreover we can extend all the previous quantities to both compact and
non compact indices by setting:
\begin{equation}
G_{i 0}=B_{i 0}= F_{i 0}=0.
\end{equation}


\end{itemize}


\section{Vertices and cocycles for dipole strings.}
\label{app-cocyles}

\subsection{Useful formula.}
Since we are using the open string formalism we define the logarithm as
\begin{equation}
\ln(z-w) =\ln(z) - \sum_1^\infty \frac{1}{n}
\left(\frac{w}{z}\right)^n
~~~~
|w|<|z|,
0\le arg(z),arg(w)<\pi
\label{DefLog}
\end{equation}
as suggested by the operatorial formalism and then we analytically continue it to the whole complex plane.
Given the previous range for the arguments we find
($arg(\bar z) =-arg(z)$)
\begin{eqnarray}
\ln(z-\bar w) &=& \ln(\bar w -z) +i \pi
\nonumber\\
\ln(z- w) &=& \ln( w -z)
+i \pi ~sgn\left(arg\left(\frac{z}{w}\right)\right)
\nonumber\\
&\Rightarrow&
\ln(\bar z- \bar w)
=
\ln(\bar w -\bar z)
+i \pi ~sgn\left(arg\left(\frac{\bar z}{\bar w}\right)\right)
=
\ln(\bar w -\bar z)
-i \pi ~sgn\left(arg\left(\frac{z}{w}\right)\right)
\nonumber\\
~~
\label{UseLogForm}
\end{eqnarray}
but if we write $y=|y| e^{i \pi}$ then we are out of our range and we
actually find
\begin{equation}
\ln(y-\bar w) = \ln(\bar w -y) +i \pi
\end{equation}

In the following we need also the following expressions.
If we fix $z= |z| e^{i \zeta}$ with $0<\zeta <\pi$, we get for $x>0>y$
\begin{eqnarray}
\ln(x-z)= \ln|x-z| + i\psi
&,~~~~&
\ln(x-\bar z)= \ln|x-z| - i\psi,~~~~
-\pi+\zeta< \psi< 0
\nonumber\\
\ln(y-z)= \ln|y-z| + i\psi
&,~~~~&
\ln(y-\bar z)= \ln|y-z| + i(2\pi-\psi),~~~~
\pi< \psi< \pi+\zeta
\nonumber\\
\label{LogRanges}
\end{eqnarray}
The ranges can be obtained by comparing the $x,|y| \to \infty$  and
$x=|y|=|z|$  values with the definition given in eq. (\ref{DefLog})
which is valid exactly for $x,|y|> |z|$.
The expression for $\ln(y-\bar z)$ is obtained since 
$\ln(y-\bar z)= i2\pi+ (\ln(y- z) )^*$ because $y= |y| e^{i\pi}=
y^* e^{2i\pi}$.

\subsection{Cocycles for closed string vertexes in presence of a $B$
  background. }
\label{Closed-Cocycles}
Before computing the closed vertices OPEs in open string formalism
 we need to determine the closed string cocycles in closed string
 formalism so
that any two closed string vertices commute and then compute their
 OPEs which we must reproduce in open string formalism.
The compact part of the closed string vertex for the emission
of a state with right (left)
quantum number $\beta_L$ ($\beta_R$)
 we consider is of the form
\begin{eqnarray}
{\cal W}_{\beta_L \beta_R}(z,\bar z; k_L, k_R)
&=&
c(k_L, k_R;p) W_{\beta_L \beta_R}(z,\bar z; k_L, k_R)
\label{clos-vert-with-coc}
\\
c(k_L, k_R;p)
&=&
e^{i\pi \left(
n_i A_c^{i j} \hat n_j
+n_i B_{c~j}^i \hat m^j
+m^i C_{c~ i}^{~~j} \hat n_j
+m^i D_{c~i j} \hat m^j
\right)}
\label{coc-of-clos-str0}
\end{eqnarray}
where $W_{\beta_L \beta_R}(z,\bar z; k_L, k_R)=V_{\beta_L}(z,k_L)
\,
\tilde V_{\beta_R}(\bar z, k_R)$
is the usual closed string vertex
without cocycle, $n$, $m$ are the momentum and the winding associated
to $k_L, k_R$
and  $\hat n$, $\hat m$ are the operator given by
\begin{eqnarray*}
\hat n = \parallel \hat n_i \parallel =
\sqrt{\alpha'} \left( E p_L + E^T p_R \right)
&&
\hat m = \parallel \hat m^i \parallel
= \sqrt{\alpha'}  \left( p_L - p_R \right)
\end{eqnarray*}
and we want to fix the constant matrices $A_c, B_c, C_c, D_c$.

To compute these matrices we compare
the analytic continuation from $|z|>|w|$ to   $|z|<|w|$
of the expression
$\left[ {\cal W}(z,\bar z; k_L, k_R) {\cal W}(w,\bar w; l_L, l_R)\right]_{an.cont} $
with the other expression
${\cal W}(w,\bar w; l_L, l_R) {\cal W}(z,\bar z; k_L, k_R)$:
\begin{eqnarray*}
{\cal W}(z,\bar z; k_L, k_R) {\cal W}(w,\bar w; l_L, l_R)
&=&
e^{i \Phi_c(k,l) }
c(k_L+l_L, k_R+l_R ; p )
\nonumber\\
&&\hspace{-5em}
V_L(z,k_L) V_L(w,l_L)\, V_R(\bar z, k_R) V_R(\bar w, l_R)
~~~~~
|z|>|w|
\end{eqnarray*}
with
\begin{eqnarray}
e^{i \Phi_c(k,l) }
&=&
e^{-i\pi\left[
n^T_l A_{c}  n_k
+n^T_l B_c  m_k
+m^T_l C_c  n_k
+m^T_l D_{c}  m_k
\right]}
\nonumber\\
&=&
e^{-i\pi \alpha' \left[
l_L^T G^{-1}( E^T A_{c } E + E^T B_c + C_c E + D_{c }) G^{-1} k_L
+l_R^T G^{-1}( E A_{c} E^T - E B_c - C_c E^T + D_{c }) G^{-1} k_R
\right]}
\nonumber\\
&&\times
e^{-i\pi \alpha' \left[
l_L^T G^{-1}( E^T A_{c } E^T - E^T B_c + C_c E^T - D_{c })G^{-1} k_R
+l_R^T G^{-1}( E A_{c} E  + E B_c - C_c E - D_{c }) G^{-1} k_L
\right]}
\nonumber\\
\label{OPE-Phase-closed00}
\end{eqnarray}
\COMMENTO{
\begin{eqnarray}
\Phi_c(k,l)
&=&
- \pi \left(
n^T_l A_c  n_k
+n^T_l B_c  m_k
+m^T_l C_c  n_k
+m^T_l D_c  m_k
\right)
\end{eqnarray}
}
where  we have used $k_L=\frac{1}{2 \sqrt{\alpha'}}\left(n_k +E^T m_k\right)$,
$k_R=\frac{1}{2 \sqrt{\alpha'}}\left(n_k -E m_k\right)$
and similarly for $l$.
This expressions shows that the matrices $A_c$ and so on are defined
up to arbitrary even integer valued matrices, i.e. for example
$ A_c \equiv A_c + 2 Z_A$ with $Z_A$ an arbitrary integer valued matrix.

Now we commute the vertexes and we make use of
$[ \ln(z-w) ]_{an.cont}= ln(w-z)+ i\pi S$ and
$[ \ln(\bar z-\bar w) ]_{an.cont}= ln(\bar w-\bar z)- i\pi S$
with $S\in 2\Z +1$\footnote{
We have that
$S= 2(n_z-n_w)+ sgn([\phi_z]-[\phi_w])$
with $arg(z)=[\phi_z]+2\pi n_z$, $-\pi<[\phi_z]\le \pi$.
}
where the opposite sign is due to the fact that $\ln(\bar z-\bar w) =
\overline{\ln(z-w)} $  as follows from the operatorial method where
$\ln(z-w)=\ln(z)-\sum_{n=1}^\infty\frac{1}{n}\left(\frac{w}{z}\right)^n$
and $\ln(z)=\overline{\ln(z)}$, hence the equation we need solving is
\begin{eqnarray}
e^{i \Phi_c(k,l)}
e^{i 2\pi\alpha' S ( k_L^T G^{-1} l_L -k_R^T G^{-1} l_R ) }
&=&
e^{i \Phi_c(l,k)}
\end{eqnarray}
\COMMENTO{
\begin{eqnarray*}
\Phi_c(k,l)
+2\pi\alpha' S ( k_L^T G^{-1} l_L -k_R^T G^{-1} l_R )
&=&
\Phi_c(l,k)
+2\pi n
\end{eqnarray*}
with an arbitrary integer $n$.
}
The previous expression can be rewritten 
as
\begin{eqnarray}
e^{-i \pi (
n^T_l (A_c-A_c^T)  n_k
+n^T_l (B_c-C_c^T-S \uno)  m_k
+m^T_l (C_c-B_c^T-S \uno)  n_k
+m^T_l (D_c-D_c^T)  m_k )}=1
\Rightarrow
\nonumber\\
n^T_l (A_c-A_c^T)  n_k
+n^T_l (B_c-C_c^T-S \uno)  m_k
+m^T_l (C_c-B_c^T-S \uno)  n_k
+m^T_l (D_c-D_c^T)  m_k
&\equiv&
0 ~~~~ mod~ 2
\nonumber\\
\label{closed-comm-const}
\end{eqnarray}
and immediately solved as
\begin{eqnarray}
A_c&=&[A_{c}]_S+  Z_{A~A}
~~~
[A_c]_A\equiv Z_{A~A}= - Z_{A~A}^T
\nonumber\\
D_c&=&[D_c]_{S}+ Z_{D~A}
~~~
[D_c]_A\equiv Z_{D~A}= - Z_{D~A}^T
\nonumber\\
C_c&=&B^T_c-\uno+2 Z_C
\end{eqnarray}
by choosing respectively $n_{k,l}\ne 0, m_{k,l}=0$, $m_{k,l}\ne 0,
n_{k,l}=0$, 
and $n_k,m_l\ne 0, m_k=n_l=0$
where $[A_c]_{S}, [D_c]_{S}$ are arbitrary symmetric complex matrices, 
$B_c$ is an arbitrary complex matrix,
$Z_{A~A}, Z_{D~A}$ are arbitrary antisymmetric integer valued matrices and
$Z_C$ is an arbitrary integer  valued matrix.

There is actually another constraint. 
It comes from the request of having hermitian vertices. 
If we compute using the vertex in eq. (\ref{clos-vert-with-coc})
and we suppose $A_c, B_c, C_c$ and $D_c$ are real , we get
\begin{eqnarray}
({\cal W}_{\beta_L \beta_R}(z, \bar z; k_L , k_R))^\dagger
&=&
\frac{1}{|z|^4 }
 W_{\beta_L \beta_R}(\frac{1}{\bar z},\frac{1}{z}; -k_L, -k_R)
c(-k_L, -k_R;p)
\nonumber\\
&=&
e^{i \Phi_c(k,k)}
\frac{1}{|z|^4 }
{\cal W}_{\beta_L \beta_R}(\frac{1}{\bar z},\frac{1}{z}; -k_L, -k_R)
\end{eqnarray}
hence we get the constraint
\begin{eqnarray}
e^{i \Phi_c(k,k)}
&=&
e^{-i \pi n^T [A_c]_{ S} n}
~e^{-i \pi m^T [D_c]_{S} m}
~e^{-i \pi m^T (2 B_{c }^T -\uno +2 Z_C) n}
=1
\label{cl-vertex-herm}
\end{eqnarray}
which can be solved by setting  
$[A_c]_{S}=  Z_{A 0}+ Z_{A 0}^T$ 
i.e. choosing a symmetric matrix with even integer 
diagonal entries and arbitrary integer entries otherwise,
$[D_c]_{S}= Z_{D 0}+ Z_{D 0}^T$ and
$B_{c}= \frac{1}{2} \uno +Z_{B}$ where 
all these $Z$ matrices are integer valued.
The final solution for the matrices $A_c, B_c, C_c$ and $D_c$ is
\begin{eqnarray}
A_c &=& Z_{A~A} + 2[ Z_{A 0}]_S+ 2 Z_A,
~~~~
Z_{A~A}^T=-Z_{A~A},
\nonumber\\
D_c &=& Z_{D~A} + 2 [Z_{D 0}]_S+2 Z_D,
~~~~
Z_{D~A}^T=-Z_{D~A},
\nonumber\\
B_c&=& \frac{1}{2} \uno +Z_{B}
\nonumber\\
C_c&=&B^T_c-\uno+2 Z_C =  -\frac{1}{2} \uno +Z_{B}^T + 2 Z_C
\label{gen-sol-coc.closed1}
\end{eqnarray}
where all matrices are integer valued and all matrices $A_c, B_c, C_c$
and $D_c$ are in the same class of equivalence for any choice of the
$Z_A, Z_D$ and $Z_C$ matrices, i.e. they yield the same phase (\ref{OPE-Phase-closed00}). 

Four points are worth noticing:
\begin{itemize}
\item
a change in $Z_B$ gives rise to a different phase;
\item
the previous equations (\ref{gen-sol-coc.closed1}) and (\ref{OPE-Phase-closed00})
 mean that all the entries are actually equivalent either to $0$ or $1$;
\item
we can always set $Z_{A A}=Z_{D A}=0$ by choosing
$Z_A= -[Z_{A A}]_<$ so that $A_c=[ [Z_{A A}]_>]_S+ 2 [Z_{A 0}]_S=
[A_c]_S $ 
and this
possible ``gauge'' is important for solving the conditions which allow
the product of two closed string vertices in open string formalism to
reproduce the same product in closed string formalism;

\item
more generally we can always choose $Z_A$ so that $[A_c]_A=\hat Z_A$ where
$\hat Z_A$ is an arbitrary antisymmetric integer valued matrix by
choosing $Z_A= [\hat Z_A-Z_{A A}]_<$ and similarly for $Z_D$.

\end{itemize}

We could try to fix the undetermined matrices requiring the phase
$e^{i \Phi_c(k,l) }$ to be invariant under T-duality,
but from the transformation properties
$m \rightarrow  n$ and $ n \rightarrow  n$
we see that the best we can do is to set
$A_c=D_c=C_c=0~~mod~2$ and $B_c= 1~~mod~2$
which nevertheless does not give neither a T-duality
invariant phase nor a proper hermitian conjugation.
\COMMENTO{
\begin{eqnarray}
e^{i \Phi_c(k,l) }
&=&
e^{-i\pi n^T_l m_k }
\nonumber\\
&=&
e^{-i \pi \alpha' \left[
l_L^T G^{-1} E^T G^{-1} k_L
-l_R^T G^{-1}  E  G^{-1} k_R
-l_L^T G^{-1} E^T G^{-1} k_R
+l_R^T G^{-1} E  G^{-1} k_L
\right]}
\nonumber\\
&&\times
e^{-i 2\pi \alpha' \left[
l_L^T G^{-1}( E^T Z_A E + E^T Z_B + Z_C E + Z_D ) G^{-1} k_L
+l_R^T G^{-1}( E Z_A E^T - E Z_B - Z_C E^T + Z_D ) G^{-1} k_R
\right]}
\nonumber\\
&&\times
e^{-i 2 \pi \alpha' \left[
l_L^T G^{-1}( E^T Z_A E^T - E^T Z_B + Z_C E^T - Z_D)G^{-1} k_R
+l_R^T G^{-1}( E Z_A E  + E Z_B - Z_C E - Z_D) G^{-1} k_L
\right]}
\nonumber\\
\label{OPE-Phase-closed}
\end{eqnarray}
where all matrices $Z_{A,B,C,D}$ are arbitrary integer valued.
}
The simplest choice compatible with hermitian conjugation is
\begin{equation}
A_c=D_c\equiv 0,~~~
B_c=-C_c \equiv \frac{1}{2} \uno
~~~~~
mod~2
\label{clo-coc0}
\end{equation}
so that
\begin{eqnarray}
e^{i \Phi_c(k,l) }
&=&
e^{-i\frac{1}{2}\pi ( n^T_l m_k - m^T_l n_k) }
\nonumber\\
&=&
e^{i \pi \alpha' \left[
l_L^T G^{-1} B G^{-1} k_L
+l_R^T G^{-1}  B G^{-1} k_R
+l_L^T G^{-1} k_R
-l_R^T G^{-1} k_L
\right]}
\nonumber\\
&&\times
e^{-i 2\pi \alpha' \left[
l_L^T G^{-1}( E^T Z_A E + E^T Z_B + Z_C E + Z_D ) G^{-1} k_L
+l_R^T G^{-1}( E Z_A E^T - E Z_B - Z_C E^T + Z_D ) G^{-1} k_R
\right]}
\nonumber\\
&&\times
e^{-i 2 \pi \alpha' \left[
l_L^T G^{-1}( E^T Z_A E^T - E^T Z_B + Z_C E^T - Z_D)G^{-1} k_R
+l_R^T G^{-1}( E Z_A E  + E Z_B - Z_C E - Z_D) G^{-1} k_L
\right]}
\nonumber\\
\label{OPE-Phase-closed0}
\end{eqnarray}

As in the case $B=0$  \cite{Pesando:2003ww}
the existence of different possible cocycles
does not mean that we have different theories since we get an
overall common phase to a given amplitudes independently of the genus
of the Riemann surface they are computed on.

\subsection{Dipole vertices.}

In this subsection we work on $R\otimes S^{D-1}$
and we use the usual $X_{L(0)}(z)$ expansion
which contains the commuting $x_0^i$ instead of 
$x^i=x^i_0- \pi\alpha' \Theta^{i l} {\cal  G}_{l m} p^m$ as given in
eq. (\ref{hatXExp}) of the main text, i.e. in this subsection we use
\begin{equation}
X_{L(0)}^i
= x_0^i-i 2\alpha' p^i \ln(z)+ n.z.m.
= \hat X_{L(0)}^i  + \pi\alpha' \Theta^{i l} {\cal  G}_{l m} p^m
\end{equation}
with OPEs
\begin{eqnarray}
X_{L (0)}(z) X^{T}_{L (0)}(w) &\sim& -2 \alpha' \ln(z-w) {\cal G}^{-1}
\nonumber\\
X_{L (0)}(z) X^{T}_{R (0)}(\bar w) &\sim& -2 \alpha' \ln(z-\bar w) {\cal G}^{-1}
\nonumber\\
X_{R (0)}(\bar z) X^{T}_{R (0)}(\bar w) &\sim& -2 \alpha' \ln(\bar z-\bar
w) {\cal G}^{-1}
\label{openOPEs-X0}
\end{eqnarray}
We expect therefore a non vanishing $p$ dependent cocycle due to
the shift by $p$ in the relation between $x$ and $x_0$.

\COMMENTO{
or using ${\cal E}{\cal G}^{-1} {\cal E}^T
={\cal E}^T{\cal G}^{-1} {\cal  E} =G^{-1}$
\begin{eqnarray}
\hat X_{L }(z) \hat X^{T}_{L }(w) &=& -2 \alpha' \ln(z-w) {G}^{-1}
\nonumber\\
\hat X_{L }(z) \hat X^{T}_{R }(\bar w)
&=&
-2 \alpha' \ln(z-\bar w)
G^{-1} {\cal E} {\cal G}^{-1} {\cal E} G^{-1}
=
-2 \alpha' \ln(z-\bar w)
{\cal E}^{-T} {\cal E} G^{-1}
\nonumber\\
\hat X_{R }(\bar z) \hat X^{T}_{R }(\bar w) &=& -2 \alpha' \ln(\bar z-\bar
w) {G}^{-1}
\label{openOPEs-X}
\end{eqnarray}
}

We consider the bundle given in 
eqs. (\ref{gen-A}) and (\ref{item-more-general-bundle}) 
where there are also Wilson lines in the part of the group without
magnetic field so that the surviving group is $U(1)^N$ 
hence we have to take into account that 
the momenta do depend on the the Wilson lines as
described in footnote \ref{foot:GenBundMom}, i.e.
the spectrum of the momentum $k_{M; I J}$  which can be obtained imposing
the periodicity of the vertices as given in eq. (\ref{DefTranInvOp})
is given by
\begin{equation}
\sqa  k_{i; I J}= \frac{n_i}{L}+ \theta_i^I -\theta_i^J
\label{Speck}
\end{equation}
with $0\le \theta_i^I< \frac{1}{L}$ as discussed around eq.(\ref{theta-range}).

In order to simplify the notation we consider as in the main text the
simplest non trivial case where $r=2$, i.e. the magnetic field is
turned on in $U(L)\subset U(L)\otimes U(\Lrpu)\subset U(L \Lrpu)$ therefore
the color indices have two components e.g. $I=(I_1,I_2)$.
Notice however that when $r=2$ the momentum $k_{i; I J}$ depends only on $I_2$
and $J_2$ and not on $I_1,J_1$ since their dependence  cancels 

Since  all problems associated with phases are captured by
tachyonic vertex operators we consider only the vertices for the
emission of an open tachyon from $\sigma=0$, $\sigma=\pi$ boundaries and
of a closed string tachyon.in eq. (\ref{Speck}).
The most general form of these vertices is 
\newcommand{\NN}{{N}}
\newcommand{\MM}{{M}}
\begin{eqnarray}
{\cal V}_{(0) T}(x; k)
&=&
t_{I_2 J_2}(k)
~
e^{ i 2\alpha'\left( k_{\NN; I J} D_0^{\NN \MM} k_{\MM; I J} 
+ k_{\MM; I J} (C_0\cG)_{~\NN}^{\MM} p^\NN\right) }
~
e^{i k_{\MM; I J} H^\MM_{0. \NN} y_0^\NN}
\nonumber\\
&&\times
:e^{ i k_{\MM; I J} 
X^\MM_{L (0)}(x) 
} :
\nonumber\\
&&
\times
\Lambda_{L;I_1 J_1}(n_1,n_2) 
|(I_1, I_2)\rangle_{\sigma=0}~{}_{\sigma=0}\langle (J_1, J_2)|
\nonumber\\
{\cal V}_{(\pi) T}(y; k)
&=&
t_{J_2 I_2}(k)
~
e^{ i 2\alpha'\left( k_{\MM; I J} D_\pi^{\MM \NN} k_{\NN; I J} + k_{\MM; I J} (C_\pi \cG)_{~\NN}^\MM p^\NN\right) }
~
e^{i k_{\MM; I J} H^\MM_{\pi. \NN} y_0^\NN}
\nonumber\\
&&\times
: e^{ i k_{\MM; I J} 
X^\MM_{L (0)}(y) 
} :
\nonumber\\
&&
\times
\Lambda_{L;J_1 I_1}(n_1,n_2) 
|(I_1, I_2)\rangle_{\sigma=\pi}~ {}_{\sigma=\pi}\langle (J_1, J_2)|
\nonumber\\
{\cal W}_{T_c}(z,\bar z; k_L,k_R,y_0)
&=&
t_{(c)}(k)
~
e^{ i \left( n_{ \MM}  A_{o}^{\MM \NN} n_{ \NN}
+ m^\MM D_{o~\MM \NN} m^\NN + m^\MM C_{o~ \MM}^{~~. \NN} n_\NN\right) }
~
e^{i \sqa \left( n_\MM (B_1\cG)^\MM_\NN + m^\MM (B_2\cG)_{\MM \NN}\right) p^\NN}
\nonumber\\
&&\times
e^{i (n_\MM (I_1)^\MM_{ \NN}+ m^\MM (I_2)_{\MM \NN}) \frac{y_0^\NN} {\sqa} } 
~
: e^{ i k_{L \MM} (G^{-1} \cE)^\MM_{.\NN} X_{L (0)}^\NN(z)} :
~
: e^{ i k_{R \MM} (G^{-1} \cE^T)^\MM_{.\NN}   X_{R (0)}^\NN(\bar z)} :
\nonumber\\
&&\times
\Lambda_{L;I_1 J_1}(-f m_2, f m_1)\otimes \uno_{ I_2 J_2}
|(I_1, I_2), K\rangle~ \langle(J_1, I_2),K|
\label{OpenStringVertices}
\end{eqnarray}
where  $|(I_1, I_2)\rangle_{\sigma=0}$  ($_{\sigma=0}\langle (I_1,I_2)|$)  
is the ket (bra)  for the state with 
color $(I_1,I_2)$ at $\sigma=0$,
similarly for $|(I_1, I_2)\rangle_{\sigma=\pi}$
and 
$|(I_1, I_2), K\rangle=|(I_1, I_2)\rangle_{\sigma=0} 
~|(K_1, K_2)\rangle_{\sigma=\pi}$.
We have also used
$x=|x| e^{i 0}$, $y=|y| e^{i \pi}$.

It is worth stressing that the implicit sum over $I_2$ and $J_2$ which also
label the momenta  $k_\MM =\parallel k_{\MM; I J} \parallel 
=\parallel k_{\MM; (I_1,I_2) (J_1,J_2)} \parallel$ gives
the vertices a non trivial, i.e. non factorized in operatorial and
color space, matricial structure.

These vertices contain cocycles which are determined by the constant
matrices $C_{0,\pi}$, $D_{0,\pi}$,  $H_{0,\pi}$ and  
 $A_o, D_o, C_o$, $B_{1,2}$, $I_{1,2}$
which we want to determine in order to satisfy the
requirements  needed for a proper CFT 
as discussed in section \ref{sec-coc-vert-open-str}\footnote{
We could also have expressed the vertices using $\hat X$ 
(for the compact part of the closed string
emission vertex) as
\begin{eqnarray}
{\cal W}_{T_c, compact}(z,\bar z; k_L,k_R,y_0)
&=&
e^{ i \left( n_{ i}  \tilde A_{o}^{i j} n_{ j}
+ m^i \tilde D_{o~i j} m^j
+ m^i \tilde  C_{o~ i}^{~. j} n_j\right) }
e^{i \sqa \left( n_i (B_1\cG)^i_j + m^j (B_2\cG)_{i j}\right) p^j}
\nonumber\\
&&\times
e^{i (n_i (\tilde I_1)^i_{ j}+ m^i (\tilde I_2)_{i  j}) \frac{y_0^j} {\sqa} } 
: e^{ i k_{L i} \hat X_{L}^i(z)} :
: e^{ i k_{R i} \hat X_{R}^i(\bar z)} :
\end{eqnarray}
but this amounts to a redefinition of the matrices:
\begin{eqnarray}
A_o= \tilde A_o,~~~~
D_o&=& \tilde D_o+ [(I_2+\frac{1}{2})\theta G]_S,~~~~
C_o= \tilde C_o- G\theta (I_1-\frac{1}{2} )~~~~
\nonumber\\
I_1= \tilde I_1,~~&~&~
I_2= \tilde I_2 +G
\end{eqnarray}
where $\theta$ is defined in eq. (\ref{Commy0y0}).
}.
For the non compact direction our conventions are
\begin{equation}
k_0=\frac{1}{2}k_{L 0}=\frac{1}{2}k_{R 0}=\frac{n_0}{\sqa},~~~~
m_0=0
\end{equation}
along with the hypothesis that mixed compact - non compact components
of all matrices vanish, as for example 
\begin{equation}
\parallel C_0^{\mu\nu} \parallel=
\left(
\begin{array}{cc}
C_0^{0 0} & \\
& C_0
\end{array}
\right),~~~~
C_0=\parallel C_0^{i j} \parallel
\end{equation}
and similarly for all the other matrices. 
We take also
\begin{equation}
D_o^{0 0}=C_o^{0 0}=0.
\end{equation}

In the following subsection we will also assume that
\begin{equation}
[y_0^i,y_0^j]= i~2\pi \alpha'~ \theta^{i j} 
\label{Commy0y0}
\end{equation}
even if our aim is to stick as close as possible with the naive
form of the vertices by choosing
\begin{equation}
D_0=C_0=H_0=0,~~~~
H_\pi=0,~~~~
I_1=0,~~~~ I_2=G,~~~~
\theta^{i j}=0
\label{imposed-constraints-on-cocycles}
\end{equation}

Generalizing previous results  obtained in
(\cite{Pesando:2003ww},\cite{Billo:2002ff},\cite{Fre:1996js} ) to this case
we assume also that $y_0$ gets shifted after the
emission of an open string with momentum $k$ from the $\sigma=0$ boundary
as 
\begin{eqnarray}
\Delta_0(k_{ I J}) y_{0}^i &=& 2\sqa  \nu^{i j}_{0 }
\Theta_j( \sqa k_{I J})
\label{Delta-shifts}
\end{eqnarray}
where $k_{ I J}$ is the momentum for a string starting at $\sigma=0$ with color
$(I_1,I_2)$ and ending at $\sigma=\pi$ with color $(J_1 J_2)$ and we
have defined 
\begin{equation}
\sqa k_{i; I J}= \frac{n_i}{L}+ \theta_i^I -\theta_i^J
\Rightarrow
I_i(\sqa k_{I J})= \frac{n_i}{L} \delta_{I,J},~~~~
\Theta_i( \sqa k_{I J})=\theta^I_i -\theta^J_i
\end{equation}

For use in the computations we recall that the product of two momentum dependent Chan-Paton is
\begin{equation}
\Lambda_L(n_1,n_2) \Lambda_L(m_1,m_2)
= 
e^{-i \pi  \alpha' (\frac{n_1}{\sqa L},\frac{n_2}{\sqa L}) \Theta_{C P} 
(\frac{m_1}{\sqa L},\frac{m_2}{\sqa L})^T
}
\frac{1}{\sqrt{L}}
\Lambda_L(n_1+m_1,n_2+m_2)
\end{equation}
where we have defined for later convenience
\begin{equation}
\Theta_{C P}
= L {\hat h} \epsilon
= L {\hat h} 
\left(\begin{array}{cc}
0& +1 \\
-1 & 0
\end{array}\right)
\label{DefThetCP}
\end{equation}
with ${\hat h} f\equiv -1~~~~mod~L$.

\subsection{Requirements for a proper CFT.}
\label{app-requ-prop-CFT}
In order to have a well defined CFT we want all operators to be
relatively local, i.e. we do not want them to create cuts in the
surface where they are inserted
therefore we check the following properties
\begin{enumerate}
\item
The open string vertices on the opposite boundaries commute;
\item
The open string emission vertex from $\sigma=0$ commutes with the
closed string vertex;
\item
In a similar way the open string emission vertex from $\sigma=\pi$
commutes with the closed string vertex;
\item
The product of two closed string vertices in open string formalism
must reproduce the result in the closed string formalism.
\item 
Proper behavior under Hermitian conjugation.
\end{enumerate}

In details the constraints listed above 
become:
\begin{enumerate}
\item
The open string vertices on the opposite boundaries commute:
\begin{eqnarray}
&&
\hspace{-4ex}
\left[ {\cal V}_{(0) T}(x; k,y_0+\Delta_\pi(l) y_0)
~{\cal V}_{(\pi) T}(y; l,y_0) \right]_{an.~cont.}
\nonumber\\
&&=
{\cal V}_{(\pi) T}(y; l,y_0+\Delta_0(k) y_0) ~{\cal V}_{(0) T}(x; k,y_0)
\end{eqnarray}
where $an.~cont.$ means analytically continued from the region 
$|x|>|y!$ to $|x|<|y|$ and
we have introduced a shift in the $y_0$ as in eq. (\ref{Delta-shifts})
\begin{equation}
\Delta_0(k) y_0 = 2\alpha' \nu_0 \Theta(k)
\end{equation}
and we have also introduced for generality the analogous shifts in the $y_0$
for the emission from the $\sigma=\pi$ boundary:
\begin{equation}
\Delta_\pi(l) y_0 = 2\alpha' \nu_\pi \Theta(l).
\end{equation}
even if it can be argued to be zero.

The constraint we get is {\bf independent} of the Chan-Paton factors 
since when we commute the two vertices on the two borders 
we do not change the matrix ordering but only the radial ordering.
If we denote the $\sigma=0$ indices as $I$ and $J$ and
$\hat I$ and $\hat J$ the $\sigma=\pi$ ones
we get the constraint
\begin{eqnarray}
&&e^{ i 2\alpha' k_0 \left( - C_\pi^T
+ C_0 
-  \pi {\cal G}^{-1}
\right)^{0 0} l_0 }
\nonumber\\
&\times&
e^{ i 2\alpha' k^T_{I J} ~H_{0} \nu_\pi ~\Theta(l_{\hat I \hat J})}
\nonumber\\
&\times&
e^{ - i 2\alpha' k^T_{I J}  ~C_\pi^T ~l_{\hat I \hat J}}
\nonumber\\
&\times&
e^{ - i 2\pi \alpha' k^T_{I J}  ~H_{0} \theta  H_{\pi}^T ~l_{\hat I \hat J} }
\nonumber\\
&\times&
e^{ - i 2\pi \alpha' k^T_{I J}  ~\cG ~l_{\hat I \hat J} }
e^{  i 2\alpha' k^T_{I J}  ~C_0^T ~l_{\hat I \hat J}}
\nonumber\\
&=&
e^{ i 2\alpha' l^T_{\hat I \hat J} ~H_{\pi} \nu_0 ~\Theta(k_{I J})}
\end{eqnarray}
where the second contribution is from the $y_0$ shift, the third one
from the $p$ cocycle, the fourth one from the $y_0$ cocycle,
and the fifth from the usual operatorial vertex.
The previous expression can be rewritten as
\begin{eqnarray}
e^{ i 2\alpha' \left[
k^T_{I J} ~\left( - C_\pi^T
+ C_0 
-  \pi {\cal G}^{-1}
-\pi H_{0} \theta H_{\pi}^T
\right) ~l_{\hat I \hat J}
+ k^T_{I J} ~H_{0} \nu_\pi ~\Theta(l_{\hat I \hat J})
- \Theta^T(k_{I J}) ~\nu_0^TH_{\pi}^T ~l_{\hat I \hat J}
\right]
}
=1
\label{V0Vpi}
\end{eqnarray}
for the compact directions and 
for the time direction as
\begin{equation}
e^{ i 2\alpha' k_0 \left( - C_\pi^{0 0}
+ C_0^{0 0} -  \pi { G}^{0 0} \right)^{0 0} l_0 }=1.
\end{equation}
%
%
%
\item
The open string emission vertex from $\sigma=0$ commutes with the
closed string vertex:
\begin{eqnarray}
&&\hspace{-4ex}
\left[ {\cal V}_{(0) T}(x; k,y_0
) ~{\cal W}_{ T_c}(z,\bar z; k_L,k_R,y_0) \right]_{an.~cont.}
\nonumber\\
&&=
{\cal W}_{ T_c}(z,\bar z; k_L,k_R,y_0+\Delta_0(k) y_0) 
~{\cal V}_{(0) T}(x; k,y_0)
\end{eqnarray}
where the emission of a closed string does not change the ``carrier''
string.

As before we denote the $\sigma=0$ indices as $I$ and $J$ then
we find the constraint for the compact directions
\begin{eqnarray}
&&e^{-2 i \pi ~\II{\sqa k_{I J}}^T  ~\Theta_{C P} ~(-1)\hF m}
\nonumber\\
&& e^{-i \sqa ~k^T_{I J} ~(B_1^T n+ B_2^T m)}
\nonumber\\
&&  e^{-i 2\pi \sqa ~k^T_{I J} ~H_{0} \theta~( I_1^T n + I_2^T m)}
\nonumber\\
&& e^{-i 2\alpha' \pi ~k^T_{I J} ~\cG^{-1} \cE^T G^{-1} ~k_L} 
   ~e^{i 2\alpha'  ~k^T_{I J} ~C_0 \cE^T G^{-1} ~k_L}
\nonumber\\
&& e^{i 2\alpha' \pi ~k^T_{I J} ~\cG^{-1} \cE G^{-1} ~k_R} 
   ~e^{i 2\alpha'  k^T_{I J} ~C_0  \cE G^{-1} ~k_R}
\nonumber\\
&=& e^{i (n^T I_1 + m^T I_2) \frac{1}{\sqa}\Delta_0(k_{I J}) y_0}
\end{eqnarray}
where the first line is due to the commuting of the Chan-Paton
factors\footnote{
We have written $~\II{\sqa k_{I J}}$ since the open Chan-Paton matrices
$\Lambda(k)$ in eq.(\ref{LambdaDef}) depend on the ``integer'' part of
$k_{I J}$  given in eq. (\ref{Speck}).
},
the second one to the commuting of the $p$ dependent closed string
cocycle,
the third one to $X_L(z)$, 
the fourth one to $X_R(\bar z)$
and the last one to the shift in $y_0$ induced by the emission of an
open string.

This constraint can be rewritten as
\begin{eqnarray}
&&
e^{i \sqa ~k^T_{I J} \left( -B_1^T +2\pi \Theta + 2 C_0
- 2 \pi H_{0} \theta I_1^T  
-2 \nu_0^T I_1^T \right) n
}
~e^{i \sqa ~\Theta(k_{I J})^T \left(-2 \nu_0^T I_1^T \right) n}
\nonumber\\
&&
e^{i \sqa k^T_{I J} 
\left(-B^T -2\pi \uno - 2\pi \Theta  \hat F -2 C_0 \hat F 
-2 \pi H_{0} \theta I_2^T  
\right) m }
~e^{i \sqa ~\Theta(k_{I J})^T \left(-2 \nu_0^T I_2^T \right) m}
\nonumber\\
&&
e^{i \sqa \II{k_{I J}}^T \left(2\pi \Theta_{C P}\hF  \right) m}
=1
\nonumber\\
\label{V0W}
\end{eqnarray}
where we have used ${\cal E}^{-T}=G^{-1} {\cal E} {\cal G}^{-1} $,
${\cal E}^{-1}=G^{-1} {\cal E}^T {\cal G}^{-1} $
and
\begin{eqnarray}
\cE^{-1} E^T + \cE^{-T} E &=& 2 \uno +2 \Theta \hF
\nonumber\\
\cE^{-1} E^T - \cE^{-T} E &=& -2 \cG^{-1} \hF
\end{eqnarray}
For the non compact direction the constraint in eq. (\ref{V0W})
becomes
\begin{equation}
e^{i \alpha' k_0 (-B_1 ^{ 0 } + 2 C_0^{0 0}) k_0^{(c)}} =1
\label{V0W-noncomp}
\end{equation}
\item
In a similar way the open string emission vertex from $\sigma=\pi$
commutes with the closed string vertex:
\begin{eqnarray}
&&\hspace{-4ex}
\left[ {\cal V}_{(\pi) T}(y; k,y_0
) 
{\cal W}_{ T_c}(z,\bar z; k_L,k_R,y_0) \right]_{an.~cont.}
\nonumber\\
&&=
{\cal W}_{ T_c}(z,\bar z; k_L,k_R,y_0+\Delta_\pi(k) y_0) 
~{\cal V}_{(\pi) T}(y; k,y_0)
\end{eqnarray}
We find therefore the constraint for the compact directions
\begin{eqnarray}
&& e^{-i \sqa ~k^T_{\hat I \hat J} ~(B_1^T n+ B_2^T n)}
\nonumber\\
&&  e^{-i 2\pi \sqa ~k^T_{\hat I \hat J} ~H_{\pi} \theta ~( I_1^T n + I_2^T m)}
\nonumber\\
&& e^{i 2\alpha' \pi ~k^T_{\hat I \hat J} ~\cG^{-1} \cE^T G^{-1} ~k_L} 
   e^{i 2\alpha'  ~k^T_{\hat I \hat J} ~C_\pi  \cE^T G^{-1} ~k_L}
\nonumber\\
&& e^{i 2\alpha' \pi ~k^T_{\hat I \hat J} ~\cG^{-1} \cE G^{-1} ~k_R} 
   e^{i 2\alpha' ~k^T_{\hat I \hat J} ~C_\pi \cE G^{-1} ~k_R}
\nonumber\\
&=& e^{i (n^T I_1 + m^T I_2 )\frac{1}{\sqa}\Delta_\pi(k_{\hat I \hat J}) y_0}
\end{eqnarray}
It is worth noticing that the previous expression is independent on
$\Theta_{CP}$ since the closed string vertex has a ``Chan-Paton
factor'' which acts on $\sigma=0$.
In performing this computation it is also necessary to be careful in
computing the phases since $y= |y| e^{i \pi}$ and $y^*= e^{-2i \pi} y$,
explicitly we have
\begin{equation}
\ln(y-z)= \ln(z-y) + i\pi,~~~~
\ln(y-\bar z)= \ln(\bar z-y) + i\pi
\end{equation}

which can be rewritten as
\begin{eqnarray}
&&
e^{i \sqa ~k^T_{\hat I \hat J} ~\left( -B_1^T +2\pi \cG^{-1} + 2 C_\pi
- 2 \pi H_{\pi} \theta I_1^T  
 -2 \nu_\pi^T I_1^T
\right) ~n
}
\nonumber\\
&&
e^{i \sqa ~k^T_{\hat I \hat J}
~\left(- B_2^T - 2\pi \cG^{-1} \hat F -2 C_\pi \hat F 
- 2 \pi H_{\pi} \theta I_2^T  
-2\nu_\pi^T I_2^T
\right) ~m  }
=1
\nonumber\\
\label{VpiW}
\end{eqnarray}
while the time constraint reads
\begin{equation}
e^{i \alpha' ~k_0 ~(-B_1 ^{ 0 } + 2 \pi G^{0 0} + 2 C_\pi^{0 0}) ~k_0^{(c)}} =1
\label{VpiW-noncomp}
\end{equation}
\item
The product of two closed string vertices in open string formalism
must reproduce the result in the closed string formalism.
An easy computation yields
\begin{eqnarray}
\hspace{-2em}
{\cal W}_{T_c}(z,\bar z; k_L,k_R,y_0
)
&\times&
{\cal W}_{T_c}(w,\bar w; l_L,l_R,y_0)
\nonumber\\
&=&
e^{i \Phi_{o}(k,l)}
~
(z-w)^{2\alpha'~ k_L^T  G^{-1} l_L }
(\bar z-\bar w)^{2\alpha'~ k_R^T  G^{-1} l_R }
\nonumber\\
&&\times
N_o\left[ {\cal W}_{T_c}(z,\bar z;y_0) {\cal W}_{T_c}(w,\bar w;y_0) \right]
\nonumber\\
~
\end{eqnarray}
In this expression the phase is given by
\begin{eqnarray}
e^{i \Phi_{o}(k,l)}
&=&
e^{ i \left[
-2 n_l^T A_o n_k -2 m_l^T D_o m_k -m_l^T C_o n_k -n_l^T C_o^T m_k
\right]}
\nonumber\\
&&
e^{-i \sqa \left( k_L^T G^{-1} \cE +k_R^T G^{-1} \cE^T\right)
\left( B_1^T n_l +B_2^T m_l\right)
}
\nonumber\\
&&e^{-i \pi \left(n_k^T I_1 + m_k^T I_2\right) \theta 
           \left(I^T_1 n_l +I^T_2 m_l\right) }
\nonumber\\
&&e^{-i \pi 2\alpha' k_R^T \cE^{-1} \cE^T G^{-1} l_L}
\nonumber\\
&& 
e^{-i \pi m_k^T \hF^T \Theta_{C P} \hF m_l}
\nonumber\\
&&
e^{i \alpha' l_0 (-2 A^{0 0}_0 - B_1^{0 0} - \frac{\pi}{2} G^{0 0}) k_0 }
\label{OPE-open-Phase-closed00}
\end{eqnarray}
where the first line is due to the numerical cocycles,
the second one to the $p$ dependent cocycle,
the third one to the $y_0$ dependent cocycle,
the fourth one to the commuting of $X_{(0)L}(w)$ with $X_{(0)R}(\bar
z)$
and the last but one to the ``Chan-Paton'' factors.

The open string equivalent of the closed string normal ordering 
$:{\cal W}_{T_c}(z,\bar z) {\cal W}_{T_c}(w,\bar w) :$
is then given by
\begin{eqnarray}
N_o\left[ {\cal W}_{T_c}(z,\bar z) {\cal W}_{T_c}(w,\bar w) \right]
&=&
e^{ i \left( (n_k+n_m)^T  A_{o} (n_{k}+n_m)
+ (m_k+m_l)^T D_{o} (m_k+m_l)
+ (m_k+m_l)^T C_{o~ }(n_k+n_l)\right) }
\nonumber\\
&&e^{\sqa \left( (n_k+n_l)^T B_1 + (m_k+m_l)^T B_2\right) \cG p}
~e^{i \left( (n_k+n_m)^T I_1 + (m_k+m_l)^T I_2 \right) \frac{y_0} {\sqa} } 
\nonumber\\
&&
: e^{ i k_{L }^T G^{-1}\cE X_{L (0)}(z) + i l_{L }^T G^{-1}\cE X_{L (0)}(w) } :
\nonumber\\
&&
: e^{ i k_{R} G^{-1}\cE^T X_{R (0)}(\bar z) + i l_{R} G^{-1}\cE^T X_{R (0)}(\bar w)
} :
\Lambda_L(-L~\hF(m_k+m_l))
\nonumber\\
\end{eqnarray}

We can now explicitly write down the constraint by equating the phases
given in 
eq. (\ref{OPE-open-Phase-closed00}) and eq. (\ref{OPE-Phase-closed00})
\begin{eqnarray}
e^{i \Phi_{c}(k,l)}
&=&
e^{i \Phi_{o}(k,l)}
\nonumber\\
&=&
e^{i n_l^T \left( -2 A_o + \pi I_1\theta I_1^T 
         -\frac{\pi}{2} \cE^{-T} \cG \cE^{-T} -B_1\right) n_k}
\nonumber\\
&&
e^{i n_l^T \left( - C_o^T + \pi I_1 \theta I_2^T  
         +\frac{\pi}{2} \cE^{-T} \cG \cE^{-T} E 
         +B_1 \hat F \right) m_k}
\nonumber\\
&&
e^{i m_l^T \left( - C_o + \pi I_2\theta I_1^T 
         -\frac{\pi}{2} E \cE^{-T} \cG \cE^{-T}  
         -B_2\right) n_k}
\nonumber\\
&&
e^{i m_l^T \left( - 2D_o + \pi I_2 \theta I_2^T  
         +\frac{\pi}{2} E \cE^{-T} \cG \cE^{-T} E 
          +B_2 \hat F 
          -\pi  \hF^T \Theta_{C P} \hF
          \right) m_k }
\nonumber\\
\label{cl-open-phase-equal}
\end{eqnarray}
along with  the constraint for the non compact direction
\begin{equation}
e^{i \alpha' k_0 (-B_1 ^{ 0 } + 2 \pi G^{0 0} + 2 C_\pi^{0 0}) k_0^{(c)}} =1
\end{equation}

\item Proper behavior under Hermitian conjugation.

We start computing the Hermitian of the open string vertices
under the assumption that all Chan-Paton factors are {\sl
  Hermitian}\footnote{
For example 
in the case of twisted bundle with degeneration this is not true and
we have 
$\left[\Lambda_{C P}(k)\right]^\dagger = e^{i h \cdot k}\Lambda_{C
  P}(-k) $
where $h$ is a vector which does not depend on the momentum $k$.
In this case it is not possible to redefine $\Lambda_{C P}$ to
reabsorb the phase $e^{i h \cdot k}$ which shows also up in the 
Zamolodchikov metric. This phase cannot hence be removed from
the Hermitian conjugate of a vertex, i.e.
$\left[ {\cal V}_{(0) T}(x; k) \right]^\dagger
\propto e^{i h \cdot k} {\cal V}_{(0) T}(\frac{1}{x^*}; -k)$.
}
\begin{equation}
\left[\Lambda_{C P}(k)\right]^\dagger = \Lambda_{C P}(-k)
\end{equation}
then  we easily get
\begin{eqnarray}
\left[ {\cal V}_{(0) T}(x; k) \right]^\dagger
&=&
e^{ -i 2\alpha' ~ k^T ( D_0 + D_0^\dagger  + C_0^\dagger ) k}
e^{ i 2\alpha'~ k^T (C_0- C_0^*) \cG p}
\nonumber\\
&& \times
e^{-i k^T(H_0^*-H_0) y_0} e^{-i \sqa \pi k^T H_0^* \theta H_0^T k^T} 
\frac{ {\cal V}_{(0) T}(\frac{1}{x^*}; -k) }
     { x^{* ~ 2\alpha' k^T    {\cal G}^{-1} k } }
\nonumber\\
&&
\Rightarrow
C_0=C_0^*,~~~~
H_0=H_0^*,~~~~
\left[ D_0 + D_0^\dagger + C_0^\dagger \right]_S = \pi n_0 \cG^{-1}
\nonumber\\
\left[ {\cal V}_{(\pi) T}(y; k) \right]^\dagger
&=&
e^{ -i 2\alpha' ~ k^T ( D_\pi + D_\pi^\dagger  +C_\pi^\dagger ) k}
e^{ i 2\alpha'~ k^T (C_\pi- C_\pi^*) \cG p}
\nonumber\\
&&\times
e^{-i k^T(H_\pi^*-H_\pi) y_0} e^{-i \sqa \pi k^T H_\pi^* \theta H_\pi^T k^T} 
\frac{ {\cal V}_{(\pi) T}(\frac{1}{y^*}; -k) }
     { y^{* ~ 2\alpha' k^T    {\cal G}^{-1} k } }
\nonumber\\
&&
\Rightarrow
C_\pi=C_\pi^*,~~~~
H_\pi=H_\pi^*,~~~~
\left[ D_\pi + D_\pi^\dagger + C_\pi^\dagger \right]_S = \pi n_\pi \cG^{-1}
\nonumber\\
\label{herm-open-const}
\end{eqnarray}
where we have used the fact that $\alpha' k^T \cG^{-1} k \in \Z$,
$[Q]_S$ means the symmetric part of the matrix $Q$ and
it is necessary to write $y^*$ because the need to keep track
of the phase due to the logarithm.
Despite the fact that $C_0$ contains an antisymmetric part the
symmetrization of the previous expression is such that
fortunately we can satisfy the constraints with $D_0$ and $D_\pi$.
This can be seen decomposing $D_0$ into the real and imaginary part as
 $D_0= D_0^T = D_{0~R}+i D_{0~I}$, then the (Hermitian conjugate of the)
constraint can be formally solved as
\begin{equation}
D_{0~R}  
= -\frac{1}{2} \left[ C_0  \right]_S +\oh \pi n_0 \cG^{-1}
\end{equation}
and similarly for $D_\pi$:
\begin{equation}
D_{\pi~R}  = -\frac{1}{2} \left[ C_\pi \right]_S
= \frac{n_\pi+1}{2} \pi \cG^{-1} -\frac{1}{2}  \left[ C_0  \right]_S
\end{equation}
where we have already used eq. (\ref{SolOpStrCons}).
In the following we choose
\begin{equation}
n_0=0,~~~~ n_\pi=-1
\end{equation}
Once $n_0=0$ has been chosen then  $n_\pi$ is also fixed by the request
of the cyclicity of amplitudes which, for example, requires that the
four open strings amplitude can be computed either by fixing
$x_4=0,x_2=1,x_1=+\infty$ and integrating over $0\le x_3 \le 1$ or by
fixing $x_3=0,x_2=1,x_1=+\infty$ and integrating over $x_4 \le 0$.

We can now exam what happens to the closed string vertex where with
the help of $arg\left(\frac{1}{z^*}\right)=arg(z)$ and
$arg\left(\frac{1}{\bar z^*}\right)=arg(\bar z)$ we find
\begin{eqnarray}
\left[ {\cal W}_{T_c}(z,\bar z; k_L,k_R,y_0) \right]^\dagger
&=&
e^{ i \left[
- n_k^T (A_o^*+A_o) n_k - m_k^T (D_o^*+D_o) m_k - m_k^T (C_o^*+C_o) n_k 
\right]}
\nonumber\\
&&
e^{-i \sqa \left( k_L^T G^{-1} \cE +k_R^T G^{-1} \cE^T\right)
\left( B_1^{* T} n_k +B_2^{* T} m_k\right)
}
e^{i \sqa [ n^T (B_1-B_1^*)+ m^T (B_2 -B_2^*)] \cG p }
\nonumber\\
&&
e^{i (n^T (I_1 -I_1^*)+ m^T (I_2-I_2^*)) \frac{y_0} {\sqa} } 
e^{-i \pi (n^T (I_1 -I_1^*)+ m^T (I_2-I_2^*)) \theta (I_1^T n + I_2^T m) } 
\nonumber\\
&&e^{-i \pi 2\alpha' k_R^T \cE^{-1} \cE^T G^{-1} k_L}
\nonumber\\
&&
\times
\frac{1}{z^{*~ 2\alpha' k_L^T G^{-1} k_L} }
\frac{1}{\bar z^{*~ 2\alpha' k_R^T G^{-1} k_R} }
{\cal W}_{T_c}(\frac{1}{\bar z^*}, \frac{1}{z^*}; -k_L,-k_R,y_0)
\nonumber\\
\end{eqnarray}
where the first contribution  is due to the non operatorial cocycle,
the second one to the $p$ dependent cocycle,
the third one to to the $y_0$ dependent cocycle
and the last ones to the reordering of the naive vertices.
From the operatorial part we get the constraints
\begin{equation}
B_{1,2}^*=B_{1,2},
~~~~
I_{1,2}^*=I_{1,2}
\end{equation}
and if we assume that
\begin{equation}
A_o^*=A_o,~~~~
D_o^*=D_o,~~~~
C_o^*=C_o,~~~~
\end{equation}
we can immediately write
\begin{eqnarray}
\nonumber\\
\left[ {\cal W}_{T_c}(z,\bar z; k_L,k_R,y_0) \right]^\dagger
&=&
e^{i \Phi_o(k,k)} 
\nonumber\\
&&\times
\frac{1}{z^{*~ 2\alpha' k_L^T G^{-1} k_L} }
\frac{1}{\bar z^{*~ 2\alpha' k_R^T G^{-1} k_R} }
{\cal W}_{T_c}(\frac{1}{\bar z^*}, \frac{1}{z^*}; -k_L,-k_R,y_0)
\nonumber\\
\end{eqnarray}
Because of the constraint in eq. (\ref{cl-open-phase-equal}) and because
$e^{i\Phi_c(k,k)}=1$ as in eq. (\ref{cl-vertex-herm})
we do not get any further constraints.

\end{enumerate}

\subsection{The solution of the constraints.}
We now impose the constraints (\ref{imposed-constraints-on-cocycles})
to keep the solution as close as possible to the naive vertices.

\subsubsection{ Solving open-closed string constraints.}
The constraints for the non compact direction can be easily solved as
\begin{equation}
B_1^{0 0}= C_0^{0 0}=0,~~~~
C_o^{0 0}=G^{0 0},~~~~
A_o^{0 0}=-\frac{\pi}{4} G^{0 0}
\end{equation}
as the case of the trivial background.

Since we consider the general background given in
eq. (\ref{item-more-general-bundle}) with Wilson lines in the part of
the group without magnetic fields we generically have that 
$\Theta(k)\ne 0$ therefore we get
\begin{equation}
\nu_0= -\pi G^{-1}
\end{equation}
in agreement with the naive expectation.
Eq.s (\ref{V0W}) and (\ref{VpiW}) together with  the constraint
(\ref{V0Vpi}) can be easily solved to yield
\begin{eqnarray}
C_0 &=& [C_0]_S -\frac{\pi}{2} \Theta
\nonumber\\
~C_\pi &=& [C_0]_S -\pi \cG^{-1} + \frac{\pi}{2} \Theta
\nonumber\\
B_1 &=& 2 [C_0]_S -\pi \Theta
\nonumber\\
B_2 &=& \hF ~B_1= \hF~(  2 [C_0]_S -\pi \Theta)
\label{SolOpStrCons}
\end{eqnarray}

A closer look to eq. (\ref{V0W}) reveals that  
we are left with a term proportional $\II{k}$
after using the previous equations.
This term reads
\begin{eqnarray}
&&e^{i \sqa \II{k}^T 
\left(-B_2^T -2\pi \uno - 2\pi \Theta  \hat F -2 C_0 \hat F 
-2 \pi H_{0} \theta I_2^T  
\right) m }
~e^{i \sqa \II{k}^T \left(2\pi \Theta_{C P}\hF  \right) m}
\nonumber\\
&=&
e^{i \sqa \II{k}^T \left(2\pi \Theta_{C P}\hF +2 \nu_0^T I_2^T \right) m}
=
e^{i \frac{1}{L} n_k^T\left(2\pi \Theta_{C P}\hF +2 \nu_0^T I_2^T \right) m}
=1
\end{eqnarray}
when we use eq. (\ref{DefThetCP}) we can write
\begin{equation}
\Theta_{CP} \hF = (1- \tilde f L) \uno
\label{DefTildef}
\end{equation}
and check that it does not imply any further constraint.

\subsubsection{Consistency of the closed string constraints.}
We can now write explicitly the constraint
(\ref{cl-open-phase-equal}) using eq. (\ref{OPE-Phase-closed00}) as
\begin{eqnarray}
- \pi A_c 
&=&
     -2 A_o 
     -\frac{\pi}{2} \cE^{-T} \cG \cE^{-T} 
     -B_1
\nonumber\\
-\pi B_c
&=&
- C_o^T 
         +\frac{\pi}{2} \cE^{-T} \cG \cE^{-T} E 
         +B_1 \hat F 
\nonumber\\
-\pi C_c
&=&
- C_o 
         -\frac{\pi}{2} E \cE^{-T} \cG \cE^{-T}  
         -B_2
\nonumber\\
-\pi D_c
&=&
 - 2D_o 
         +\frac{\pi}{2} E \cE^{-T} \cG \cE^{-T} E 
          +B_2 \hat F
-\pi \hF \Theta_{C P} \hF
\nonumber\\
\label{ADC-open}
\end{eqnarray}

One could think these constraints just give the matrices $A_o$, $D_o$
and $C_o$: this is not so because 
\begin{equation}
A_o=A_o^T,~~~~D_o=D_o^T,
\end{equation}
moreover  $C_o$ enters two equations
and we have to remember eq.s (\ref{gen-sol-coc.closed}).

Because both $A_o$ and $D_o$ are symmetric we have to consider what
happens
to the antisymmetric part of the first and fourth equation; the first
equation becomes
\begin{eqnarray}
-\pi [A_c]_A
&=&
-\pi ( Z_{ A A} + 2 [Z_A]_A )
\nonumber\\
&=& 
- \frac{\pi}{2} [\cE^{-T} \cG \cE^{-T}]_A - [B_1]_A
\nonumber\\
&=&
-\pi \Theta -[B_1]_A
=0
\end{eqnarray}
where in the last line we have used eq. (\ref{SolOpStrCons}) 
from the previous section and
\begin{equation}
[\cE^{-T} \cG \cE^{-T}]_A
=
\frac{1}{2} \left( (\cE^{-1} + 2 \Theta)  \cG \cE^{-T} - \cE^{-1}  \cG
\cE^{-1}\right)
=
2 \Theta
\end{equation}
This constraint is nothing else but a choice of ``gauge'': $[A_c]_A=0$.

The antisymmetric part of the fourth equation can be now written as
\begin{eqnarray}
-\pi [D_c]_A
&=&
-\pi ( Z_{ D A} + 2 [Z_D]_A )
\nonumber\\
&=& 
+ \frac{\pi}{2} [E \cE^{-T} \cG \cE^{-T} E]_A 
+  [B_2 \hat F ]_A
-\pi \hF \Theta_{C P} \hF
\nonumber\\
&=&
+\pi \hat F + \pi \hF \Theta \hF +  [B_2 \hat F ]_A
-\pi \hF \Theta_{C P} \hF
\nonumber\\
&=& -\pi~L {\tilde f} \hF 
\end{eqnarray}
where we have used eq. (\ref{DefTildef}) and
\begin{eqnarray}
 [E \cE^{-T} \cG \cE^{-T} E]_A
&=&
 [(\cE^T+ \hat F) \cE^{-T} \cG \cE^{-T} (\cE^T+ \hat F)]_A
= 2 \hat F + 2 \hat F \Theta \hat F
\end{eqnarray}
This constraint is again nothing else but a choice of ``gauge'' on the
antisymmetric part: $[D_c]_A=\tilde f  ~L \hF\in \Z$.

Finally summing the opposite of the transpose of the second equation 
with the third we eliminate $C_o$ and we get
\begin{eqnarray}
\pi ( B_c^T -C_c)
&=&
\pi( \uno -2 Z_C)
\nonumber\\
&=&
-\frac{\pi}{2} E^T \cE^{-1} \cG \cE^{-1}
-\frac{\pi}{2} E \cE^{-T} \cG \cE^{-T}
+\hF B_1^T -B_2
\nonumber\\
&=&
-\pi \uno
-2 \pi \hF \Theta 
+\hF B_1^T -B_2
=-\pi \uno
\end{eqnarray}
where we have used
\begin{eqnarray}
 E^T \cE^{-1} \cG \cE^{-1}+ E \cE^{-T} \cG \cE^{-T}
&=&
(\cE -\hF) \cE^{-1} \cG \cE^{-1}+ (\cE^T +\hF) \cE^{-T} \cG \cE^{-T}
\nonumber\\
&=&
2 \uno - \hF (  \cE^{-1} \cE^T -  \cE^{-T} \cE ) G^{-1}
\nonumber\\
&=& 
2 \uno - \hF (  (\cE^{-T}- 2\Theta) \cE^T -  (\cE^{-1}+ 2\Theta) \cE
) G^{-1}
= 2\uno +4\hF \Theta
\nonumber\\
\end{eqnarray}
which is again a choice of ``gauge'' ($Z_C=\uno$).

\subsubsection{Solving the closed string constraints.}

When we have solved the previous constraints we can then determine the
remaining matrices $A_o$, $D_o$ and $C_o$ as
\begin{eqnarray}
A_o 
&=& 
\frac{\pi}{2} [A_c]_S
-\frac{\pi}{4}[ \cE^{-T} \cG \cE^{-T} ]_S 
-\frac{1}{2}[B_1]_S
\nonumber\\
D_o
&=&
\frac{\pi}{2} [D_c]_S
+\frac{\pi}{4} [ E \cE^{-T} \cG \cE^{-T} E ]_S 
+\frac{1}{2}[B_2 \hF]_S
\nonumber\\
C_o
&=&
\frac{\pi}{2} (C_c + B_c^T)
+\frac{\pi}{4}( E^T \cE^{-1} \cG \cE^{-1}- E \cE^{-T} \cG \cE^{-T} )
-\frac{1}{2}(B_2+\hF B_1^T)
\end{eqnarray}
where the first and second equations are obtained by taking the
symmetric part of the first and last equations in (\ref{ADC-open}),
while the last is obtained by summing the transpose of the second with
the third one.

Notice that the matrices $[Z_A]_S$,  $[Z_D]_S$ and  $Z_B$ are
completely arbitrary but integers and hence the the numerical cocycle of closed
string vertices in open string formalism will be determined up to
powers of $i$, actually we can write
\COMMENTO{
\begin{eqnarray}
&&e^{ i \left( n_{ i}  A_{o}^{i j} n_{ j}
+ m^i D_{o~i j} m^j + m^i C_{o~ i}^{~. j} n_j\right) }
=
\nonumber\\
&&=
e^{-i \pi \alpha' k_L^T \cE^{-T} \cG \cE^{-T} k_R}
~e^{-i \frac{1}{2} (n^T B_1 + m^T B_2) (n-\hff m)}
\nonumber\\
&&\times
e^{i \pi (n^T I_1 +m^T I_2) I^{-1} (\Zu n +\Zd m) }
~e^{i \frac{\pi}{2} ( n^T A_c n + n^T B_c m + m^T C_c n+m^T D_c m)}
\nonumber\\
\label{non-op-coc}
\end{eqnarray}
If $H_0=H_\pi=\nu_\pi=0$ we can simplify further the previous
expression and write
}
\begin{eqnarray}
&&e^{ i \left( n_{ i}  A_{o}^{i j} n_{ j}
+ m^i D_{o~i j} m^j + m^i C_{o~ i}^{~. j} n_j\right) }
e^{i k_0 A^{0 0}_o k_0}
=
\nonumber\\
&&=
e^{-i \pi \alpha' k_L^T \cE^{-T} \cG \cE^{-T} k_R}
~e^{-i  (n - \hF m)^T 
[C_\pi]_S
(n-\hF m)}
\nonumber\\
&&\times
~e^{i \frac{\pi}{2} ( n^T A_c n + n^T B_c m + m^T C_c n+m^T D_c m)}
\nonumber\\
\label{non-op-coc1}
\end{eqnarray}

Notice that 
\begin{enumerate}
\item
the dependence on  $A_c,B_c,\dots$ is like a square root of the
closed string dependence but it can be fixed to give a trivial result
as done in eq. (\ref{halfPhi1}).
\item
there is also another arbitrary matrix $[C_0]_S$ which can be and will
be fixed to the trivial value $[C_0]_S=0$.
\end{enumerate}

Finally we can summarize our discussion by writing
\begin{eqnarray}
{\cal V}_{(0) T}(x; k)
&=&
t_{I_2 J_2}(k)
~
e^{ -i \alpha' k_{\NN; I J} [C_0]_S^{\NN \MM} k_{\MM; I J} }
~
e^{ i 2\alpha' k_{\MM; I J} 
\left([C_0]_S \cG -\frac{\pi}{2}\Theta\cG\right)_{~\NN}^{\MM} p^\NN }
\nonumber\\
&&\times
: e^{ i k_{\MM; I J} X^\MM_{L (0)}(x) }: 
\nonumber\\
&&
\times
\Lambda_{L;I_1 J_1}(n_1,n_2)  
~ |(I_1,I_2)\rangle_{\sigma=0}
~~{}_{\sigma=0}\langle (J_1 ,J_2)|
\nonumber\\
&=&
t_{I_2 J_2}(k)
~
e^{ -i \alpha' k_{\NN; I J} [C_0]_S^{\NN \MM} k_{\MM; I J} }
~
e^{ i 2\alpha' k_{\MM; I J} \left([C_0]_S\cG\right)_{~\NN}^{\MM} p^\NN }
\nonumber\\
&&\times
: e^{ i k_{\MM; I J} \hat X^\MM_{L (0)}(x) }: 
\nonumber\\
&&
\times
\Lambda_{L;I_1 J_1}(n_1,n_2) 
~ |(I_1, I_2)\rangle_{\sigma=0}
~~{}_{\sigma=0}\langle (J_1, J_2)|
\nonumber\\
\nonumber\\
{\cal V}_{(\pi) T}(y; k)
&=&
t_{J_2 I_2}(k)
~
e^{ i \alpha' k_{\MM; I J} (
- [C_0]_S)^{M N} k_{\NN; I J}  }
e^{ i 2\alpha'  k_{\MM; I J} \left(- \pi \uno +  [C_0]_S\cG +\frac{\pi}{2}\Theta \cG\right)_{~\NN}^\MM p^\NN }
\nonumber\\
&&\times
: e^{ i k_{\MM; I J} X^\MM_{L (0)}(y) } :
\nonumber\\
&&
\times
~\Lambda_{L;J_1 I_1}(n_1,n_2) 
~| (I_1, I_2 )\rangle_{\sigma=\pi}
~~{}_{\sigma=\pi} \langle (J_1,J_2)|
\nonumber\\
&=&
t_{J_2 I_2}(k)
~
e^{ i \alpha' k_{\MM; I J} (
- [C_0]_S)^{M N} k_{\NN; I J}  }
~
e^{ i 2\alpha'  k_{\MM; I J} (-\pi \uno +\pi \Theta \cG +[C_0]_S \cG)_{~\NN}^\MM p^\NN }
\nonumber\\
&&\times
: e^{ i k_{\MM; I J} \hat X^\MM_{L (0)}(y) } :
\nonumber\\
&&
\times
~\Lambda_{L;J_1 I_1}(n_1,n_2) 
~| (I_1, I_2 )\rangle _{\sigma=\pi}
~{}_{\sigma=\pi}\langle (J_1, J_2)|
\nonumber\\
&=&
t_{J_2 I_2}(k)
~
e^{ -i \alpha' k_{\MM; I J} (
+ [C_0]_S)^{M N} k_{\NN; I J}  }
~
e^{ i 2\alpha'  k_{\MM; I J} ([C_0]_S \cG)_{~\NN}^\MM p^\NN }
\times
: e^{ i k_{\MM; I J}  X^\MM(y) } :
\nonumber\\
&&
\times
~\Lambda_{L;J_1 I_1}(n_1,n_2) 
~| (I_1, I_2 )\rangle _{\sigma=\pi}
~{}_{\sigma=\pi}\langle (J_1, J_2)|
\nonumber\\
\nonumber\\
{\cal W}_{T_c}(z,\bar z; k_L,k_R,y_0)
&=&
e^{\frac{i}{2} \Phi_{(c)}(k,k)}
~e^{-i\pi \alpha' k_{L;M} (\cE^{-T}\cG\cE^{-T})^{M N} k_{R;N}}
e^{-i \pi \sqa \left( n_\MM  + m^\NN \hF_{\NN \MM}\right) (\Theta\cG)^\MM_{~L} p^L}
\nonumber\\
&&\times
e^{i m^\MM G_{\MM \NN} \frac{y_0^\NN} {\sqa} } 
: e^{ i k_{L \MM} (G^{-1} \cE)^\MM_{.\NN} X_{L (0)}^\NN(z)} :
: e^{ i k_{R \MM} (G^{-1} \cE^T)^\MM_{.\NN}   X_{R (0)}^\NN(\bar z)} :
\nonumber\\
&&\times
\sqrt{L}~\Lambda_{L;I_1 J_1}(-f m_2, f m_1)\otimes \uno_{ I_2 J_2}
~ |(I_1, I_2), K\rangle
~\langle (J_1,J_2), K|
\nonumber\\
&=&
e^{\frac{i}{2} \Phi_{(c)}(k,k)}
~e^{i\pi \alpha' k_{L;M} (G^{-1} \cE^{T} \Theta \cE^{T} G^{-1})^{M N} k_{R;N}}
\nonumber\\
&&\times
: e^{ i k_{L \MM}  X_{L }^\MM(z)} :
: e^{ i k_{R \MM}   X_{R }^\MM(\bar z)} :
\nonumber\\
&&\times
\sqrt{L}~\Lambda_{L;I_1 J_1}(-f m_2, f m_1)\otimes \uno_{ I_2 J_2}
~ |(I_1, I_2), K\rangle
~\langle (J_1,J_2), K|
\nonumber\\
\label{OpenStringVertices1}
\end{eqnarray}
where the different expressions are obtained using
\begin{eqnarray}
X(x,x)&=& \hat X_{L (0)}(x)= X_{L (0)}(x)- \pi \alpha' \Theta \cG p, 
\nonumber\\
X(y,y)&=& \hat X_{L (0)}(y) + 2 \pi \alpha' (-\uno + \Theta \cG ) p 
=  X_{L (0)}(y) +  2 \pi \alpha' (-\uno + \oh \Theta \cG ) p 
\end{eqnarray}
where the $\hat X_{L (0)}$ fields are whose expansion contains 
the non commuting $x$ and
the $X_{L,R}$ in the case of the closed string.

\section{Details on different amplitude computations.}
\subsection{Open OPEs with Wilson lines.}
\label{appssec:OpenOPEs}
It is easy to check that the OPEs of two open string vertices in
presence of Wilson lines are given by
\begin{eqnarray}
{\cal V}_{(0) T}(x_1; \parallel k_{I L} \parallel)
~{\cal V}_{(0) T}(x_2; \parallel l_{L J} \parallel)
&=&
\frac{1}{\sqrt{L}} 
e^{ - i \pi \alpha' ~ \II{k_{M; I L}} \Theta_{C P}^{ M N} 
\II{l_{N; L J}}  }
\nonumber\\
&&
\sum_L
e^{ - i \pi \alpha' ~ k_{M; I L} \Theta ^{ M N} l_{N; L J} }
(x_1 - x_2)^{2 \alpha' k_{M; I L} \cG^{ M N} l_{N; L J}}
\nonumber\\
&& 
: e^{ i   k_{M; I L} X^M(x_1) + i   l_{M; L J} X^M(x_2) } :
\nonumber\\
&&
~\Lambda_{L; I_1 J_1}(k+l)\otimes (T_u T_v)_{I_2 J_2}
~ |(I_1, I_2)\rangle_{\sigma=0}
~~{}_{\sigma=0}\langle (J_1, J_2)|
\nonumber\\
\end{eqnarray}
when we set $[C_0]_S=0$ and where we have used the fact that $\parallel k_{I L} \parallel$ depends
only on $I_2$ and $L_2$, and that $\II{ k_{M; I L} }$ does not depend on
$I,L$ at all.

\subsection{Details on the boundary derivation from reggeon vertices.}
\label{app:BounRegg}
In this appendix we want to give some details on the computations
performed for obtaining the boundary state from the reggeon formalism.
In particular we want to start from eq. (\ref{BouGen})
\begin{eqnarray}
\langle B(F); V_L,V_R|
&=&
\frac{ \Cz ~ \Nt}{2\pi}
\nonumber\\
&&
\langle x_L=x_R=0; 0_{a_{(c)}}, 0_{\tilde a_{(c)}} |
~e^{-\frac{i}{2 } \Phi_{(c)}(G p_{(c)},G p_{(c)})}
~e^{-i \pi \alpha'~ p_{R}^M (\cE^{T} \cG^{-1} \cE^{T})_{M N} p_{L}^N}
\nonumber\\
&&
\times
Tr\left( \sqrt{L}
\Lambda_L\left(-L \hat F m\right) 
\otimes \uno_{\Lrpu}
\right)
~\times
{}_p\langle 0| \cS_L(z;V_L) ~\cS_R(\bar z;V_R) |0\rangle_p
\nonumber\\
\label{BouGen1}
\end{eqnarray}
where we have introduced the state $\langle x_L=x_R=0|$ normalized as
$\langle x_L=x_R=0| k_L, k_R\rangle=1 $
and the Sciuto-Della Selva-Saito vertices as
discussed in (\cite{DiVecchia:1988cy}, \cite{DiVecchia:1986jv}, 
\cite{Pesando:1999ex})
\begin{eqnarray}
\cS_L(z;V_L) 
&=&
:\myexp{
-\frac{1}{2\alpha'} \oint_{u=0} \frac{d u}{2\pi i}
\partial X_{L (c)}^M(u) ~G_{M N} ~X^N_L(V_L(u;z))
}:
\nonumber\\
&&\times
:\myexp{
\frac{i}{2 \sqrt{2\alpha'} } \oint_{u=0} \frac{d u}{2\pi i}
\partial X_{L (c)}^M(u) ~G_{M N} ~\alpha^N_{(c) 0}
~\log \frac{d V_L}{d u}
}:
\nonumber\\
&=&
:\myexp{
\frac{i}{ \sqrt{2 \alpha'}} \sum_{n=0}^\infty \alpha^M_{(c) n} 
 ~\cE_{M N} \frac{1}{n!}~\partial^n_u  \hat X^N_L(V_L(u;z)) |_{u=0}
}:
~e^{
\frac{i}{ \sqrt{2 \alpha'}} \alpha^M_{(c) 0} ~G_{M N} ~y^N_0} 
\nonumber\\
&&\times
:\myexp{ \frac{1}{2} \sum_{n=0}^\infty \alpha^M_{(c) n}
~G_{M N} ~\alpha^N_{(c) 0}
~\frac{1}{n!}~\partial^n_u \log \frac{d V_L}{d u} |_{u=0}
}:
\end{eqnarray}
and
\begin{eqnarray}
\cS_R(\bar z;V_R) 
&=&
:\myexp{
-\frac{1}{2\alpha'} \oint_{u=0} \frac{d u}{2\pi i}
\partial X_{R (c)}^M(u) ~G_{M N} ~X^N_R(V_R(u;\bar z))
}:
\nonumber\\
&&\times
:\myexp{
\frac{i}{2 \sqrt{2\alpha'} } \oint_{u=0} \frac{d u}{2\pi i}
\partial X_{R (c)}^M(u) ~G_{M N} ~\tilde \alpha^N_{(c) 0}
~\log \frac{d V_R}{d u}
}:
\nonumber\\
&=&
:\myexp{
\frac{i}{ \sqrt{2 \alpha'}} \sum_{n=0}^\infty 
\tilde \alpha^M_{(c) n} ~(\cE^T)_{M N} ~ \frac{1}{n!}\partial^n_u  \hat
X^N_R(V_R(u;\bar z)) |_{u=0}
}:
~e^{-
\frac{i}{ \sqrt{2 \alpha'}} \tilde \alpha^M_{(c) 0} ~G_{M N} ~y^N_0} 
\nonumber\\
&&\times
:\myexp{ \frac{1}{2} \sum_{n=0}^\infty \tilde \alpha^M_{(c) n}
~G_{M N} ~\tilde \alpha^N_{(c) 0}
~\frac{1}{n!}~\partial^n_u \log \frac{d V_R}{d u} |_{u=0}
}:
\end{eqnarray}
where the zero modes are defined as 
$a^M_{(c) 0}=\alpha^M_{(c) 0}=\sqrt{2\alpha'} p^M_L$, 
$\sqrt{n} a^M_{(c) n}=\alpha^M_{(c) 0}$ ($n>0$) and similarly for the
right moving operators.
These vertices are given for arbitrary $SL(2,\C)$ local 
coordinates $V_L(u;z)$ and $V_R(u;\bar z)$:
\begin{eqnarray}
V_L(u;z)
&=&
\frac{a_L u + b_L z}{c_L u +b_L},
~~~~
b_L(a_L - c_L z)=1,
~~~~
V_L(0;z)=z
\label{defVL1}
\end{eqnarray}
and similarly for $V_R$.

Performing explicitly the computation we get therefore\footnote{
Notice that the phase
$~e^{-i \pi \alpha'~ p_{R}^M (\cE^{T} \cG^{-1} \cE^{T})_{M N}
  p_{L}^N}$ which enters the definition of the closed string vertex
(\ref{ClosedStringVertex}) 
and is present in eq. (\ref{SLSRder}) is not anymore present in
eq. (\ref{SLSRwithD}).
The reason is that the phase 
$~e^{-i \pi \alpha'~ p_{R}^M (\cE^{T}  \cG^{-1} \cE^{T})_{M N}  p_{L}^N}$ 
is required in order to write the expression (\ref{SLSRwithD}) 
with the use $D_{0 0}$ given in eq.s (\ref{defD})
since
$\frac{d}{d u} ( U_L V_R) |_{u=0}
= \frac{ e^{-i \pi} }{ |z-\bar z|^2} 
\frac{d}{d u} ( V_L) |_{u=0}
\frac{d}{d u} ( V_R) |_{u=0}$
where we have used the first of eq.s (\ref{UseLogForm}) to fix the phase. 
Another way to check this is to plug the special choice
(\ref{SpecVLVR}) 
in order to get the final result (\ref{UsuaBouF}).
}
\begin{eqnarray}
\hspace{5em}&&\hspace{-5em}
~e^{-i \oh \pi \alpha^N_{(c) 0} (\cE \cG^{-1} \cE)_{N M} ~ \tilde \alpha^N_{(c) 0}}
{}_p\langle0|  \cS_L(z;V_L) ~\cS_R(\bar z;V_R) |0\rangle_p
\nonumber\\
&=&
\myexp{\sum_{n,m=0}^\infty \alpha^N_{(c) n} (\cE \cG^{-1} \cE)_{N M} \tilde \alpha^M_{(c)    m}
~\frac{\partial^n_u}{n!} \frac{\partial^m_v}{m!}
\log( V_L(u;z)-V_R(v;\bar z))|_{u=v=0}
}
\nonumber\\
&&
\myexp{ \frac{1}{2} \sum_{n=0}^\infty \alpha^M_{(c) n}
~G_{M N} ~\alpha^N_{(c) 0}
~\frac{1}{n!}~\partial^n_u \log \frac{d V_L}{d u} |_{u=0}
}
\nonumber\\
&&
\myexp{ \frac{1}{2} \sum_{n=0}^\infty \tilde \alpha^M_{(c) n}
~G_{M N} ~\tilde \alpha^N_{(c) 0}
~\frac{1}{n!}~\partial^n_u \log \frac{d V_R}{d u} |_{u=0}
}
\nonumber\\
&& 
\myexp{i \oh \pi  ~ \alpha^N_{(c) 0} (\cE \Theta \cE)_{N M} \tilde \alpha^M_{(c) 0}}
~\myexp{-i \oh \pi \alpha^N_{(c) 0} (\cE \cG^{-1} \cE)_{N M} ~ \tilde \alpha^N_{(c) 0}}
\nonumber\\
&&
\myexp{\frac{i}{ \sqrt{2 \alpha'}} (\alpha^M_{(c) 0}-
  \tilde\alpha^M_{(c) 0}) ~G_{M N} ~y^N_0} 
\nonumber\\
&&
~(2\pi\sqa)^{D-d}\delta^{D-d}(\alpha_{0 \mu})
~(2\pi\sqa)^d \delta_{\cE^T \alpha_{(c) 0} +\cE \tilde \alpha_{(c) 0},0}
\nonumber\\
\label{SLSRder}
\end{eqnarray}
\begin{eqnarray}
\hspace{5em}&=&
\myexp{- \sum_{n,m=0}^\infty a^N_{(c) n} (\cE \cG^{-1} \cE)_{N M} \tilde a^M_{(c) m}
D_{n m}(U_L V_R)
}
\nonumber\\
&& \myexp{-i {\pi\over 2}  ~ a^N_{(c) 0} (\cE \Theta \cE^T)_{N M} \tilde a^M_{(c) 0}}
~~\myexp{\frac{i}{ \sqrt{2 \alpha'}} ( a^M_{(c) 0}- \tilde a^M_{(c) 0}) ~G_{M N} ~y^N_0} 
\nonumber\\
&&(2\pi)^{D-d}\delta^{D-d}(k_\mu)
~(2\pi\sqa)^d \delta_{\cE^T  a_{(c) 0} +\cE \tilde a_{(c) 0},0}
\label{SLSRwithD}
\end{eqnarray}
where $U_L(u)=\Gamma V^{-1}_L(u)= \frac{1} { V^{-1}_L(u) }$.
In the previous equation we have also introduced
$D_{n m}(\gamma)$ which is a representation of the $SL(2,\R)$
group given by
\begin{eqnarray}
D_{n m}(\gamma) = \sqrt{\frac{m}{n}} \frac{1}{m!} \partial^m
[\gamma^n(u)] |_{u=0},
~~~~&&~~~~
D_{00}(\gamma) = \oh \log \gamma'(0)
\nonumber\\
D_{n 0}(\gamma) = \sqrt{\frac{1}{n}} [\gamma^n(u)] |_{u=0},
~~~~&&~~~~
D_{0 n}(\gamma) = \oh\sqrt{m} \frac{1}{m!} \partial^m \log \gamma'(0)
\label{defD}
\end{eqnarray}
with $n,m>0$.

The amplitude (\ref{BouGen1}) can then be written as
\begin{eqnarray}
\langle B(F);V_L, V_R | 
&=&
N~ \frac{ \Cz ~ \Nt}{2\pi}
\nonumber\\
&&\hspace{-5em}
~\langle k_\mu=0|
~\sum_{s\in\Z^d} \frac{1}{(2\pi \sqa)^d} \langle n=L~ \hF ~s, m= L~ s|
e^{-\oh i \Phi_{(c)}(k,k) +i \pi \frac{\hat h f^2}{L} m^1 m^2}
~e^{i ~m^M ~G_{M N} ~\frac{y^N_0}{\sqa} } 
\nonumber\\
&\times&
\langle 0_a, 0_{\tilde a} |~
\myexp{- \sum_{n,m=0}^\infty a^N_{(c) n} (\cE \cG^{-1} \cE)_{N M} \tilde a^M_{(c) m}
D_{n m}(U_L V_R)
}
\label{1cl-from-SDS1}
\end{eqnarray}
where we used the definition of $\langle x_L=x_R=0|$ 
and the momentum conservation. In particular the term with $\Theta$.
is canceled due to momentum conservation.

This expression allows to compute any 
one point closed string amplitude as given in eq. (\ref{1ptGenAmp}).

\subsection{Details on $N$ open - $1$ closed tachyon amplitudes.}
\label{app:NOpen1Closed}
We want to compute the amplitudes with strings on the $\sigma=\pi$
border and show that they are necessary to complete the integration
over the whole circle. In doing so a lot of care must be used in the
treatment of phases, in particular eq.s (\ref{LogRanges}) must be used.
All these problems arises because we have been working with
coordinates in closure of the upper complex plane where there is a
special point $\infty$ for the ordering of the vertices; this would
not happen with a disk formulation where the cyclically equivalent
orderings are obvious.

We start therefore from our formulation in the closure of the upper
complex plane and want to perform the change of variable
\begin{equation}
x=\frac{ \bar z~ e^{i\phi}-z}{ e^{i\phi}-1}
\label{changxfi}
\end{equation}
which maps the circle $0\le \phi < 2\pi $ into the real axis.
The key point is to correctly connect $\phi$ with the phase $\psi$ given in eq.s
(\ref{LogRanges}).
This can be done comparing the phase of $\ln(x-z)$, explicitly we have
\begin{equation}
x-z = e^{i (\pi+\oh \phi + 2\pi k)} \frac{Im ~z}{ \sin \oh \phi}
\rightarrow
\ln(x-z ) = \ln \bigg| \frac{Im ~z}{ \sin \oh \phi} \bigg|
+ i (\pi+\oh \phi)+ i 2\pi k 
\end{equation}
where we have used the modulus even if $Im~z,~\sin \oh \phi >0$.
Comparing with the ranges given in eq.s (\ref{LogRanges}) we get
\begin{eqnarray}
\oh \phi=
\left\{ \begin{array}{ccc}
\psi+\pi, & x>0, & \zeta<\oh \phi < \pi \\
\psi-\pi, & x<0, & 0<\oh \phi < \zeta \\
\end{array}\right.
\end{eqnarray}

We have therefore to consider the cyclically equivalent correlators.
For $k>l$ we have
\begin{eqnarray}
A_{k l}&=&\langle 0| R\bigg[
~{\cal V}_{(0)T}(x_k;k_k) \dots ~{\cal V}_{(0)T}(x_l;k_l)
\nonumber\\
&&~~~~
~{\cal V}_{(0)T}(y_{k-1};k_{k-1}) \dots ~{\cal V}_{(0)T}(y_{1};k_{1})
\nonumber\\
&&~~~~
~{\cal V}_{(0)T}(y_N;k_N) \dots ~{\cal V}_{(0)T}(y_{l+1};k_{l+1})
~{\cal W}_{T_c}(z,\bar z;k_L,k_R)
\bigg] 
|0\rangle
\nonumber\\
\end{eqnarray} 
where $R[..]$  is the radial ordering and we have chosen
$x_k>\dots> x_l$ and $|y_{k-1}|> \dots |y_1|>|y_{N}|>\dots |y_{l+1}|$
in order to consider cyclically equivalent configurations
and for $k<l$ we write
\begin{eqnarray}
A_{k l}&=&\langle 0| R\bigg[
~{\cal V}_{(0)T}(x_k;k_k) \dots ~{\cal V}_{(0)T}(x_1;k_1)
\nonumber\\
&&~~~~
~{\cal V}_{(0)T}(x_N;k_N) \dots ~{\cal V}_{(0)T}(x_{l};k_{l})
\nonumber\\
&&~~~~
~{\cal V}_{(0)T}(y_{l-1};k_{l-1}) \dots ~{\cal V}_{(0)T}(y_{k+1};k_{k+1})
~{\cal W}_{T_c}(z,\bar z;k_L,k_R)
\bigg] 
|0\rangle
\nonumber\\
\end{eqnarray} 
with $x_l> \dots > x_1>x_N> \dots x_k$ and $ |y_{k-1}|> |y_k|> \dots
|y_{k-1}|$ in order to keep the desired cyclical ordering.

We start compute the cyclically equivalent correlators $A_{1 l}$ and
then we discuss the others in order to 
show that they cover different ranges of the circle which together
all the others $A_{k l}$ correlators cover all the circle.
Explicitly we have 
\begin{eqnarray}
A_{1 l}&=&\langle 0| 
~{\cal V}_{(0)T}(x_1;k_1) \dots ~{\cal V}_{(0)T}(x_l;k_l)
\nonumber\\
&&
~{\cal V}_{(0)T}(y_N;k_N) \dots ~{\cal V}_{(0)T}(y_{l+1};k_{l+1})
~{\cal W}_{T_c}(z,\bar z;k_L,k_R) 
|0\rangle
\nonumber\\
\end{eqnarray} 
for all\footnote{
$A_{1 0}$ is the correlator where all the vertices are on the
  $\sigma=\pi$ border.
} $0\le l\le N$ with $x_k>\dots x_l> |y_N|>\dots |y_{l+1}|> |z|$
which a simple computation shows to be
\begin{eqnarray}
A_{1 l}&=&
~
e^{i 2\pi \alpha' (\sum_{u=l+1}^N k_u) ^T ~\Theta ~(\hat k_L + \hat k_R)} 
~
e^{i 2\pi \alpha' 
(\sum_{u=l+1}^N k_u) ^T ~\Theta_{CP} ~(- \frac{\hF m}{\sqa})} 
~
e^{- i \pi \hat h L 
(\sum_r k_r- \frac{\hF m}{\sqa})_1
(\sum_r k_r- \frac{\hF m}{\sqa})_2
}
\nonumber\\
&&~
e^{-i 2\pi \alpha' \sum_{l+1 \le u<v} k_u \cdot k_v}
~
e^{-i 2\pi \alpha' \sum_{u=l+1}^N k_u \cdot (\hat k_L + \hat k_R)}
\nonumber\\
&&
~e^{\frac{i}{2}\Phi(k,k)}
~e^{i m^T G \frac{y_0}{\sqa}}
~e^{-i \pi \alpha' \sum_{1\le r <s \le N} k_r^T (\Theta+\Theta_{CP}) k_s} 
\nonumber\\
&&
\left(\frac{1}{\sqrt{L}}\right)^{N-2}
~t_{u_1}(k_N)\dots t_{u_1}(k_N)
~tr(T_{u_1}\dots T_{u_N})
\nonumber\\
&&
\prod_{1\le r <s \le l} (x_{r }-x_s)^{2\alpha' k_r\cdot k_s} 
~\prod_{r < \le l<u} (x_{r }-y_u)^{2\alpha' k_r\cdot k_u} 
~\prod_{l+1\le u <v \le l} (y_v-y_u)^{2\alpha' k_v\cdot k_u} 
\nonumber\\
&&
\prod_{1\le r \le l} 
\left[ (x_{r }-z)^{2\alpha' k_r\cdot \hat k_L} 
~(x_r -\bar  z)^{2\alpha' k_r\cdot \hat k_R}\right]
~
\prod_{l+1\le u \le N} 
\left[ (y_u-z)^{2\alpha' k_u\cdot \hat k_L} 
~(y_u -\bar  z)^{2\alpha' k_u\cdot \hat k_R}\right]
\nonumber\\
&&
|z-\bar z| ^{2\alpha' \hat k_L \cdot \hat k_R}
\times
~(2\pi)^{D-d}\delta(\sum_r k_{r \mu}+k_\mu)
~(2\pi\sqa)^d \delta_{\sum_r k_r+\hat k_L+\hat k_R,0}
\label{A1lstep1}
\end{eqnarray}
where we have written in the first and second line all the terms from cocycles
and Chan-Paton which change with $l$.
We have also defined $\hat k_L=\cE^T \cG^{-1} k_L$, $\hat k_R=\cE \cG^{-1} k_R$
and $k \cdot l = k^T \cG^{-1} l$.
By writing $y_v-y_u= |y_v-y_u| e^{i \pi}$ we cancel the first term of
the second line.
Writing $y_u-\bar z= |y_u-  z| e^{i (\pi -\oh \phi_u)}$,
$y_u-z= |y_u-  z| e^{i (\pi +\oh \phi_u)}$ with $0< \phi_u < \zeta$
and $x_r-z= (x_r-  \bar z)^* =|x_r-  z| e^{i (-\pi +\oh \phi_r)}$ with
$\zeta<\phi_r<2 \pi$ we cancel all the remaining terms in the first two
lines with the help of eq. (\ref{ThetaF}).
The final result is
\begin{eqnarray}
A_{1l}&=&
~e^{\frac{i}{2}\Phi(k,k)}
~e^{i m^T G \frac{y_0}{\sqa}}
~e^{-i \pi \alpha' \sum_{1\le r <s \le N} k_r^T (\Theta+\Theta_{CP}) k_s} 
~e^{i \pi  \hat h \left( f (n_1 m^1+ n_2 m^2)
- L n_1 n_2 \right) }
\nonumber\\
&&
\left(\frac{1}{\sqrt{L}}\right)^{N-2}
~t_{u_1}(k_N)\dots t_{u_1}(k_N)
~tr(T_{u_1}\dots T_{u_N})
\nonumber\\
&&
\prod_{1\le r <s \le N} |x_{r }-x_s|^{2\alpha' k_r\cdot k_s} 
~
\prod_{1\le r \le N} |x_{r }-z|^{2\alpha' k_r\cdot (\hat k_L+\hat k_R) } 
~
\prod_{1\le r \le N} e^{i \phi_r \alpha' k_r\cdot (\hat k_L-\hat k_R) } 
\nonumber\\
&&
|z-\bar z| ^{2\alpha' \hat k_L \cdot \hat k_R}
\times
~(2\pi)^{D-d}\delta(\sum_r k_{r \mu}+k_\mu)
~(2\pi\sqa)^d \delta_{\sum_r k_r+\hat k_L+\hat k_R,0}
\nonumber\\
\end{eqnarray}
where $\phi_r$s are such that $\phi_r> \phi_{r+1}$  and 
have two different ranges according whether $r\le l$
or $l<r$ as discussed above, i.e.
$ 2\pi> \phi_1 > \dots \frac{}{}_l > \zeta> \phi_{l+1}> \dots \phi_N>0$.

To get the amplitude we change variables according to
eq. (\ref{changxfi}), we multiply for the measure
\begin{equation}
\prod_{r=1}^N d x_r ~d z~d \bar z=
(\oh Im~z)^N
\prod_{r=1}^N \frac{d \phi_r}{ \sin^2 \frac{\phi_r}{2}}
 ~d z~d \bar z
\end{equation}
and divide for the gauge fixing
\begin{equation}
d V_{Killing} = \frac{1}{4} (Im~z)^{-2} d \phi_N  ~d z~d \bar z
\end{equation}
which can simply be obtained by requiring that the $N=1$ amplitude is
independent on $z$ and $\phi$.

Finally fixing $\phi_N=\alpha$ we can sum the $A_{1l}$ correlators
as $ \sum_{l=0}^N A_{1l}$ to eliminate the dependence on $\zeta$ and then
we get a partial amplitude given by
\begin{eqnarray}
\cA_{1}&=&
\Cz ~\Nop^N~\Nt
~e^{\frac{i}{2}\Phi(k,k)}
~e^{i m^T G \frac{y_0}{\sqa}}
~e^{-i \pi \alpha' \sum_{1\le r <s \le N} k_r^T (\Theta+\Theta_{CP}) k_s} 
\nonumber\\
&&
\left(\frac{1}{\sqrt{L}}\right)^{N-2}
~t_{u_1}(k_N)\dots t_{u_1}(k_N)
~tr(T_{u_1}\dots T_{u_N})
~e^{i \pi  \hat h \left( f (n_1 m^1+ n_2 m^2)
- L n_1 n_2 \right) }
\nonumber\\
&&
~(2\pi)^{D-d}\delta(\sum_r k_{r \mu}+k_\mu)
~(2\pi\sqa)^d \delta_{\sum_r k_r+\hat k_L+\hat k_R,0}
\nonumber\\
&&
\int^{2\pi}_\alpha 
\prod_{r=1}^{N-1} 
{d \phi_r}
\theta(\phi_r-\phi_{r+1})
\prod_{1\le r <s \le N} (2 \sin \frac{\phi_r-\phi_s}{2})^{2\alpha' k_r\cdot k_s} 
~
\prod_{1\le r \le N} e^{i \phi_r \alpha' k_r\cdot (\hat k_L-\hat k_R) } 
\nonumber\\
\label{calA1}
\end{eqnarray}
where we still have
a dependence on $\phi_N=\alpha$. To get rid of it we need consider all
the others amplitudes.

Let us now exam all the other correlators $A_{k l}$.
The previous discussion shows that all $A_{1 l}$
correlators are the same of $A_{1 N}$ up to the ranges where the
$\phi_r$ are defined, i.e. we can freely move all the vertices from
the $\sigma=\pi$ boundary to the $\sigma=0$ one while keeping the
cyclical order.
We deduce therefore that all $A_{k l}$ are the analytic
continuation of the correlators
\begin{eqnarray}
A_{k ~k-1}&=&
\langle 0| R\bigg[
~{\cal V}_{(0)T}(x_k;k_k) \dots ~{\cal V}_{(0)T}(x_N;k_N)
\nonumber\\
&&~~~~
~{\cal V}_{(0)T}(x_{1};k_{1}) \dots ~{\cal V}_{(0)T}(x_{k-1};k_{k-1})
~{\cal W}_{T_c}(z,\bar z;k_L,k_R)
\bigg] 
|0\rangle
\nonumber\\
\end{eqnarray} 
If we compare the explicit expression (\ref{A1lstep1}) for the usual
ordering $A_{1 N}$ with the corresponding one for the previous correlators 
$A_{k k-1}$ which can be obtained mutata mutandis we find that
\begin{eqnarray}
A_{k l}
&=& 
e^{2 i \pi \alpha' \sum_{1\le r \le k-1 } k_r^T (\Theta+\Theta_{CP})  
\sum_{k \le s \le N }k_s }
~
e^{-2 \pi i  \alpha' \sum_{1\le r \le k-1 } k_r\cdot (\hat k_L-\hat
  k_R) }
~
A_{1 N}
\nonumber\\
&=& 
e^{2 i \pi  L \sqa \sum_{1\le r \le k-1 } k_r^T ( \hat h \epsilon n
  + \tilde f m) }
~
A_{1 N}
=
A_{1 N}
\end{eqnarray}
where the first phase is due to the difference  with the overall phase
in the first line of eq (\ref{calA1}) and
depends crucially on the presence of the would-be closed string Chan-Paton,
the second one is due to the shift of the range  $\phi_r\rightarrow \phi_r-2\pi$
for $1\le r \le k-1$ so that the new $\phi_r$ are  ordered as in the $A_{1
  l}$ amplitudes even if with a different range
$\phi_{N}< \dots \phi_k< 2\pi < \phi_{k-1}<\dots \phi_1 < 2\pi+ \phi_N$.
 To perform the computations 
we have used the momentum conservation, the definition of
$\Theta_{CP}$ and the fact that $L \sqa k_r\in \Z$.

Notice that there is not a contribution from $\sin
\frac{\phi_r-\phi_s}{2}$ since it can be written as $|\sin
\frac{\phi_r-\phi_s}{2}|$ because of the ordering $\theta(\phi_r-\phi_{r+1})$.

Changing variables, multiplying for the measure, dividing for the
gauge group and fixing $\phi_N=\alpha$ we can sum all $A_{kl}$ correlators
with fixed $k$
to get a partial amplitude given by
\begin{eqnarray}
\cA_{k}&=&
\Cz ~\Nop^N~\Nt
~e^{\frac{i}{2}\Phi(k,k)}
~e^{i m^T G \frac{y_0}{\sqa}}
~e^{-i \pi \alpha' \sum_{1\le r <s \le N} k_r^T (\Theta+\Theta_{CP}) k_s} 
\nonumber\\
&&
\left(\frac{1}{\sqrt{L}}\right)^{N-2}
~t_{u_1}(k_N)\dots t_{u_1}(k_N)
~tr(T_{u_1}\dots T_{u_N})
~e^{i \pi  \hat h \left( f (n_1 m^1+ n_2 m^2)
- L n_1 n_2 \right) }
\nonumber\\
&&
~(2\pi)^{D-d}\delta(\sum_r k_{r \mu}+k_\mu)
~(2\pi\sqa)^d \delta_{\sum_r k_r+\hat k_L+\hat k_R,0}
\nonumber\\
&&
\int^{2\pi}_\alpha 
\prod_{r=k}^{N-1} 
{d \phi_r}
~\int_{2\pi}^{2\pi+\alpha} 
\prod_{r=1}^{k-1} 
{d \phi_r}
~
\theta(\phi_r-\phi_{r+1})
\prod_{1\le r <s \le N} (2 \sin \frac{\phi_r-\phi_s}{2})^{2\alpha' k_r\cdot k_s} 
~
\prod_{1\le r \le N} e^{i \phi_r \alpha' k_r\cdot (\hat k_L-\hat k_R) } 
\nonumber\\
\end{eqnarray}

Finally summing over all $k$ we get the previous amplitude with
integration range $[\alpha,2\pi+\alpha]$ for all $\phi_r$ ($1\le r \le
N-1$) variables, we can then shift all the variables as
$\phi\rightarrow \phi+\alpha$ to get final result described in the
main text.
Notice that the shift is allowed since using the momentum
conservation we get
\begin{eqnarray}
e^{i \alpha \alpha' \sum_r k_r\cdot (\hat k_L-\hat k_R) }
=
e^{-i \alpha \alpha' (\hat k_L+\hat k_R)\cdot (\hat k_L-\hat k_R) }
=
e^{-i \alpha \alpha' (k_L^T G^{-1} k_L -k_R^T G^{-1} k_LR ) }
=1.
\end{eqnarray}
as long as $k_L^T G^{-1} k_L =k_R^T G^{-1} k_R $ which is granted the
the $\sigma$  rotational invariance of closed string.

\subsection{Details on factorization of non planar one loop amplitudes.}
\label{app:NONPlanar}
In this section we want to describe the details of the computations
necessary to derive the final expression in eq. (\ref{SumDualFinal-1-loop}).
This amounts essentially to follow the steps as in
(\cite{Cremmer:1973ig}) taking care of some more factors.

We start showing that
\begin{eqnarray}
&& \cA(k,\dots N_0,1,\dots k-1; N_0+1,\dots N_0+N_\pi)
\nonumber\\
&&''=''
\prod_{k< r < s \le N_0} 
e^{i 2\pi \alpha' k_{r i}  \Theta_{tot}^{i    j} k_{s j}} 
~\prod_{k< r < s \le N_0} 
e^{-i 2\pi \alpha' k_{r i}  \Theta_{tot}^{i    j} k_{s j}} 
 ~
\cA(1,\dots N_0; N_0+1,\dots N_0+N_\pi)
\nonumber\\
&&=\cA(1,\dots N_0; N_0+1,\dots N_0+N_\pi)
\end{eqnarray}
where we have written $''=''$ since the lhs and the rhs differ in the
range of integration variables only.
The first phase in the second line 
is obtained while rewriting the non commutative phase
in a canonical form, i.e. as in the $\cA(1,\dots N_0; N_0+1,\dots
N_0+N_\pi)$ amplitude.
The second contribution arises because we redefine the integration
variables from $\nu$ to $\bar \nu$ as ($k>1$)
\begin{equation}
\nu_r
=
\left\{ \begin{array}{cc}
\bar \nu_r & r=1,\dots k-1 \\
\bar \nu_r -1 & r=k,\dots N_0-1
\end{array}
\right.
\end{equation}
which gives a contribution
\begin{eqnarray}
\prod_{s=k}^{N_0} e^{ i 2\pi ~\alpha' k_s^T
(\frac{L m_0}{\sqa} +\Theta_{tot} \sum_{r=1}^{N_0} k_r)}
=
\prod_{k< r < s \le N_0} 
e^{-i 2\pi \alpha' k_{r i}  \Theta_{tot}^{i    j} k_{s j}} 
\end{eqnarray}
when we use that $L \sqa k_r\in \Z$.
The reason why we redefine the integration variables is that the
original integration variables $\nu$ in the $\cA(k,\dots N_0,1,\dots k-1;
N_0+1,\dots N_0+N_\pi)$ amplitude have range
\begin{equation}
0< \nu_k< \dots \nu_{N_0} < \nu_1 < \dots \nu_{k-1} < 1
\end{equation}
and a different ordering w.r.t. the $\cA(1,\dots N_0; N_0+1,\dots
N_0+N_\pi)$ amplitude
while the new ones $\bar \nu$ have range
\begin{equation}
\nu_{N_0} < \bar \nu_1 < \dots \bar \nu_{k-1} < 1 < 
\bar \nu_k < \dots \bar \nu_{N_0-1} < 1+\nu_{N_0}
\end{equation}
but the same ordering as in the amplitude 
$\cA(1,\dots N_0; N_0+1,\dots N_0+N_\pi)$.
If we now redefine the integration variables 
$\nu_r=  \bar \nu_r -1$ for $ r=1,\dots N_0$ in the
$k=1$ amplitude 
$\cA(1,\dots N_0; N_0+1,\dots N_0+N_\pi)$ 
the amplitude is left invariant.
We can then perform the sum 
$
\sum_{k=1}^{N_0}
 \cA(k,\dots N_0,1,\dots k-1; N_0+1,\dots N_0+N_\pi)
$
since the $\bar \nu$ variables cover different ranges of integration
which can be joined together into a bigger range.
The sum has then the same functional expression as
the amplitude $\cA(1,\dots N_0; N_0+1,\dots
N_0+N_\pi)$ but with a different integration range, explicitly
\begin{eqnarray}
\sum_{k=1}^{N_0}&&\hspace{-2em}
 \cA(k,\dots N_0,1,\dots k-1; N_0+1,\dots N_0+N_\pi)
\nonumber\\
&=& 
\dots
~\int_0^1 d \nu_{N_0}
~e^{ -i 2\pi (\nu_{N_0}+1) ~\alpha' k_r^T (\frac{L m_0}{\sqa}
  +\Theta_{tot} \sum_{s=1}^{N_0} k_s)}
\nonumber\\
&&~~~
~\int_{\nu_{N_0}}^{\nu_{N_0}+1} \prod_{r=1}^{N_0-1} d \bar \nu_r
~\theta(\bar \nu_{r}-\bar \nu_{r+1}) 
~e^{ -i 2\pi \bar \nu_{r} ~\alpha' k_r^T (\frac{L m_0}{\sqa} +\Theta_{tot} \sum_{s=1}^{N_0} k_s)}
\dots
\nonumber\\
\end{eqnarray}
where we have written only the pieces which differ from the original amplitude.
Next we can change variables as
\begin{eqnarray}
\phi_r=
\left\{
\begin{array}{cc}
2\pi (1+\nu_{N_0}-\bar\nu_r) & r=1,\dots N_0-1
\\
2\pi \bar \nu_r & r=N_0+1,\dots N_0+N_\pi
\end{array}
\right.
~~~~
\phi_{N_0}\equiv 0
\end{eqnarray}
and get the result given in the main text whose main pieces are
\begin{eqnarray}
\sum_{k=1}^{N_0}&&\hspace{-2em}
 \cA(k,\dots N_0,1,\dots k-1; N_0+1,\dots N_0+N_\pi)
\nonumber\\
&=& 
\dots
\frac{1}{(2\pi)^{N_0+N_\pi-2}}
~\int_0^1 d \nu_{N_0}
~e^{ -i 2\pi \nu_{N_0} ~\alpha' k_r^T \frac{L m_0}{\sqa} }
\nonumber\\
&&~~~
~\int_{0}^{2\pi} \prod_{r=1}^{N_0-1} d \phi_r
~\theta(\phi_{r}-\phi_{r+1}) 
~e^{ +i \phi_{r} ~\alpha' k_r^T (\frac{L m_0}{\sqa} +\Theta_{tot} \sum_{s=1}^{N_0} k_s)}
\dots
\nonumber\\
&&~~~
~\int_{0}^{2\pi} \prod_{r=N_0+1}^{N_0+N_\pi-1} d \phi_r
~\theta(\phi_{r+1}-\phi_{r}) 
~e^{ -i \phi_{r} ~\alpha' k_r^T (\frac{L m_0}{\sqa} +\Theta_{tot} \sum_{s=1}^{N_0} k_s)}
\dots
\nonumber\\
\end{eqnarray}
where we have $\phi_{N_0}=0$
and $\phi_{N_0+N_\pi}=2\pi$.

\subsection{The annulus amplitude.}
\label{app:annulus}
In this section we check the equation  (\ref{NormC0Nt0}) which gives
the normalization of the amplitudes by computing the annulus,
explicitely
we compute
\begin{eqnarray}
Z
&=&
 -2\times\oh Tr\left( \log(L_0-1) \right)
=
\Cu \int_{0}^1 \frac{d w}{\ln w} Tr( w^{L_0-2} ) 
\nonumber\\
&=&
\Cu ~N_1^2
~\frac{ \delta^{D-d}(0)}{(\sqa)^{D-d}}
~\left[ \det G_{\mu\nu} ~  \det(L \cG_{} L)_{i j}\right]^{\oh}
\nonumber\\
&&\times
\int_{0}^1 \frac{d w}{w^2 \ln w}
~\frac{1}{ \left[ \prod_1^\infty (1-w^n) \right]^{D-2}}
~\left(-\frac{\ln w}{\pi }\right)^{D/2}
\sum_{(m^i)\in \Z^d}e^{\frac{\pi^2}{\ln w} m^T L \cG L m }
\end{eqnarray}
Changing variable as in eq. (\ref{ClOpvariables}) 
and using the modular transformations given in eq. (\ref{ModTransf})
we can rewrite the previous amplitude for $D=26$ as
\begin{eqnarray}
Z
&=&
\frac{\Cu}{2\pi} 
~2^{-D/2}
~N^2
~\frac{ \delta^{D-d}(0)}{(\sqa)^{D-d}}
~\left[ \det\cG_{M N}\right]^{\oh}
\int_{0}^1 \frac{d q}{q^3 }
~\frac{1}{ \left[ \prod_1^\infty (1-q^n) \right]^{D-2}}
\sum_{(m^i)\in \Z^d}e^{\oh \ln q~ m^T L \cG L m }
\nonumber\\
\end{eqnarray}
which can be factorized on the tachyons as
\begin{eqnarray}
Z
&\sim&
\frac{\Cu}{2\pi} 
~2^{-D/2}
~N^2
~\frac{ \delta^{D-d}(0)}{(\sqa)^{D-d}}
~\left[ \det\cG_{M N}\right]^{\oh}
\sum_{(m^i)\in \Z^d}
\frac{1}{\oh m^T L \cG L m -2}
\nonumber\\
\end{eqnarray}
When we compare with the expected form 
\begin{eqnarray}
Z\sim&&
\int \frac{d^{D-d}k_C}{ (2\pi)^{D-d} ~(2\pi\sqa)^d}
\sum_{n_C}
~\cA(C) ~
\cA( -C)
~\frac{1}{k_C^T \cG^{-1} k_C -\frac{4}{\alpha'}}
\nonumber\\
\end{eqnarray}
where $C$ stands for the closed string tachyon appearing in the mixed amplitude
(\ref{1ptCloTacFin}), $-C$ the closed string tachyon with opposite momentum,
we get
\begin{equation}
\frac{\Cu}{2\pi} 
~2^{-D/2}
~\left[ \det\cG_{M N}\right]^{\oh}
=
( \frac{\Cz \Nt}{2\pi})^2 
~(2\pi \sqa)^D
\frac{\alpha'}{2}
\end{equation}
which reproduces the result given in the main text.

\end{document}